\documentclass[preprint]{aastex701}

\usepackage{graphicx}
\usepackage{subcaption}
\usepackage{float}
\usepackage{multirow}
\usepackage{amsmath}
\usepackage{soul}
\usepackage{tabularx}
\usepackage{array}
\usepackage{caption}

\begin{document}

\title{CPD--82 291: Multi-sector TESS Photometry and SED Analysis of an Active Eclipsing Binary}

\correspondingauthor{Volkan Bak{\i}\c{s}}

\author[0000-0002-3125-9010]{Volkan Bak{\i}\c{s}}
\affiliation{Department of Space Sciences and Technologies, Faculty of Science, Akdeniz University, 07058, Antalya, T\"{u}rkiye}
\email[show]{volkanbakis@akdeniz.edu.tr}

\author[0000-0002-9846-3788]{G\"{o}khan Y\"{u}cel}
\affiliation{Department of Space Sciences and Technologies, Faculty of Science, Akdeniz University, 07058, Antalya, T\"{u}rkiye}\email{gokhannyucel@gmail.com}

\begin{abstract}

We present a multi-sector \textit{TESS} photometric and spectral
energy distribution analysis of the eclipsing binary CPD--82~291
(RX~J0853.1--8244). Eight \textit{TESS} sectors reveal well-defined
eclipses and substantial, changing out-of-eclipse modulation. The
asymmetries are reproduced by different surface-brightness
configurations at different epochs, consistent with strong and evolving
stellar activity, although the fitted circular regions are not unique
maps of individual starspots.
A detached-binary light-curve model yields an orbital inclination of
$i\simeq82.1^\circ$ and a photometric mass ratio of
$q_{\rm phot}\simeq0.464$. Without a spectroscopic radial-velocity
orbit, the mass ratio and absolute parameters are model-dependent
rather than dynamical measurements. Combining the light-curve geometry
with the SED radius scale and \textit{Gaia} distance gives
$R_1\simeq1.25\,R_\odot$ and $R_2\simeq3.70\,R_\odot$; the
corresponding model-inferred masses remain substantially less secure
than the dimensionless light-curve parameters.
The stellar photospheres reproduce the observed SED from the optical
region through WISE W4, with no statistically significant
mid-infrared excess. The only positive long-wavelength deviation is a
low-significance $60\,\mu$m measurement, while the $90\,\mu$m result
is an upper limit. A cool circumbinary-dust model can reproduce this
tentative signal, but such a component is not required by the
photometry, and its inferred mass and radial extent are non-unique and
strongly model dependent.
Evolutionary tracks place the primary close to the zero-age main
sequence and the secondary well above the main-sequence locus. However,
the absence of radial velocities and spectroscopic youth indicators,
together with \textit{Gaia} astrometry inconsistent with Chamaeleon~I
membership, prevents a secure pre-main-sequence classification.
CPD--82~291 is therefore best regarded as an active eclipsing binary
with uncertain evolutionary status and only a marginal far-infrared
indication that could be reproduced by cool circumstellar material.

\end{abstract}


\keywords{
\uat{Eclipsing binary stars}{444} ---
\uat{Pre-main sequence stars}{1290} ---
\uat{Young stellar objects}{1834} ---
\uat{Circumbinary disks}{235} ---
\uat{Protoplanetary disks}{1300} ---
\uat{Spectral energy distribution}{2129}
}

\section{Introduction}
\label{sec:Intro}

Mass and radius are the fundamental parameters governing stellar structure and evolution, and their accurate determination across a wide range of masses and ages is essential for constraining theoretical models. Detached eclipsing binaries with well-sampled eclipse light curves and double-lined spectroscopic radial-velocity measurements provide the most direct and reliable measurements of these quantities, because they enable largely model-independent determinations of the stellar masses, radii, and orbital scale \citep{Torres2010,Serenelli2021}. For this reason, eclipsing systems
serve as critical benchmarks for testing stellar-evolution theory. In the absence of a spectroscopic orbit, however, the light curve primarily constrains the orbital geometry, fractional radii, temperature ratio, and, under favourable conditions, a
model-dependent photometric mass ratio. The absolute masses and orbital scale must then be inferred using additional assumptions or external
constraints and are therefore not direct dynamical measurements.

Young eclipsing binaries are particularly valuable laboratories for testing stellar evolutionary models, since they probe the pre-main-sequence (PMS) phase during which stellar properties such as radius, luminosity, and effective temperature evolve rapidly. Fundamental parameters derived from eclipsing binaries with both photometric and spectroscopic orbital solutions provide some of the most stringent observational constraints on PMS evolution, especially in the low- and intermediate-mass regime where theoretical models still show significant uncertainties
\citep[e.g.][]{Stassun2014,Gillen2017}.

Despite substantial observational progress in recent years, important discrepancies between theoretical predictions and observed stellar properties remain. In particular, many PMS systems exhibit inflated stellar radii, lower effective temperatures, and luminosities that are difficult to reproduce simultaneously with standard evolutionary models. These discrepancies are often attributed to magnetic activity, starspots, accretion history, or remaining limitations in PMS evolutionary calculations for young low-mass stars \citep[e.g.][]{Stassun2014,Gillen2017}.

The number of known PMS eclipsing binaries has increased considerably over the last decade thanks to wide-field photometric surveys and space-based missions. Recent discoveries include systems such as MML~48 \citep{MML48_2025}, Mon--735 \citep{Gillen2020}, THOR--42 \citep{Murphy2020}, TOI--450 \citep{Tofflemire2023}, and several newly identified PMS eclipsing binaries summarized by \citet{MML48_2025}. Nevertheless, the currently known sample remains relatively limited and does not yet fully cover the parameter space required for robust calibration of PMS evolutionary models, particularly for active low-mass systems, inflated secondaries, and binaries associated with circumstellar or circumbinary material. The value of individual systems as evolutionary benchmarks nevertheless depends strongly on the availability of spectroscopic orbital constraints and on the security of their youth and association membership.

In addition to analyses based on light curves and radial velocities, the spectral energy distribution (SED) provides an independent and complementary diagnostic of stellar systems. In young stellar objects (YSOs), circumstellar material such as discs and envelopes can contribute significantly to the observed flux, particularly at infrared wavelengths, and can also introduce additional extinction components. As a result, extinction estimates derived from large-scale Galactic reddening maps may not accurately represent the true line-of-sight attenuation toward individual systems. SED modelling can therefore provide useful constraints on stellar radii, extinction, and infrared-excess components, although the inferred circumstellar properties remain dependent on the adopted geometry, opacity prescription, distance, and treatment of non-simultaneous photometry

CPD--82 291 (RX J0853.1--8244; ASASSN-V
J085305.34--824360.0) was originally classified as a weak-line T~Tauri star in the direction of the Chamaeleon~I star-forming region \citep{Spangler2001}. However, its youth and association
membership have not been established unambiguously. In the
\textit{Spitzer} Formation and Evolution of Planetary Systems
survey, \citet{Carpenter2009} classified the source as a field
star with an activity-based age estimate of
$\log(\mathrm{age/yr})=8.6$. They found its 16-to-8 and
24-to-8~$\mu$m flux ratios to be consistent with photospheric
emission and reported no significant infrared excess in the
IRAC, IRS, MIPS 24~$\mu$m, or MIPS 70~$\mu$m observations.
The absence of a significant excess in the \textit{Spitzer} and WISE mid-infrared measurements, together with the low-significance \textbf{$60\,\mu$m} flux reported by \citet{Spangler2001}, indicates that the available photometry does not establish a persistent infrared excess. The far-infrared emission is therefore treated in this study as tentative and is explored only through a conditional dust model.
The Gaia DR3 astrometry does not support physical membership in Chamaeleon~I. CPD--82~291 has proper-motion components of $\mu_{\alpha*}=+3.435\pm0.0169$ and
$\mu_{\delta}=-13.006\pm0.014$~mas\,yr$^{-1}$, whereas the northern and southern Chamaeleon~I subclusters have mean proper motions of
approximately $(-22.07,-0.05)$ and
$(-23.13,+1.59)$~mas\,yr$^{-1}$, respectively
\citep{Roccatagliata2018}. The proper-motion vector of CPD--82~291 is
therefore separated from those of the two subclusters by approximately
29--30~mas\,yr$^{-1}$. In addition, its corrected parallax,
$\varpi_{\rm corr}=2.0538\pm0.0134$~mas, is markedly smaller than the mean Chamaeleon~I parallax of approximately $5.25$~mas. The source is thus located roughly 300~pc farther away along the line of sight than the association. Its Gaia DR3 RUWE value of 0.885 does not indicate a poor astrometric fit. We therefore conclude that CPD--82~291 is not a physical member of Chamaeleon~I, although its evolutionary status and possible youth must be assessed independently from spectroscopic and stellar-parameter diagnostics.

The eclipsing-binary nature of CPD--82 291 was previously recognized by \citet{Christy2022} in their study of unusual variables identified through the Citizen ASAS-SN project. Using ASAS-SN photometry, they revised the initially reported period of approximately 20.4~days to $P \simeq 10.21$~days and showed that the phase-folded light curve contains both primary and secondary eclipses. They also attributed the prominent sinusoidal modulation to surface activity and starspots in a nearly synchronized binary. Their treatment was primarily classificatory, however, and did not include detailed eclipse modelling or a physical analysis of the stellar and infrared-emitting components.

In the present study, we extend this earlier identification through the analysis of eight sectors of high-precision \textit{TESS} photometry. We examine the changing out-of-eclipse morphology on a sector-by-sector basis, model the eclipse geometry and evolving surface-brightness distribution, and search for candidate flare events. We further combine the light-curve constraints with broadband photometry and the adopted \textit{Gaia} distance to investigate the stellar scale and the possible origin of the tentative far-infrared emission. Because no spectroscopic radial-velocity orbit or spatially resolved infrared observation is available, the resulting component
masses, orbital scale, and dust-structure parameters are treated as model-inferred quantities rather than as direct dynamical or spatial measurements.

Our principal aim is therefore not to report the discovery of the binary, but to provide a detailed multi-sector photometric characterization of its eclipse morphology and activity, together with a model-dependent investigation of its stellar properties and infrared-emitting environment. The tentative far-infrared emission is explored using a conditional circumbinary-dust model, while alternative explanations such as unresolved contamination, non-simultaneous variability, and different circumstellar geometries are considered when assessing this interpretation.

The structure of the paper is as follows. In Sect.~2, we describe the observational data and the procedures used to compile and prepare them. Sect.~3 presents the eclipse-timing, light-curve, SED, surface-brightness, and flare analyses. In Sect.~4, we present and discuss the binary configuration and model-inferred absolute parameters, the possible evolutionary status of the system, the conditional interpretation of the tentative far-infrared emission,
the surface and flare activity, and the principal model limitations. Sect.~5 summarizes the main conclusions.

\section{Observations and Data Preparation}
\subsection{Broadband Data}
\label{sec:sed_data}

\citet{Spangler2001} observed CPD--82 291 using the ISOPHOT C100 detector \citep{isophot1996} onboard the \textit{Infrared Space Observatory} (ISO), employing the 60~$\mu$m and 90~$\mu$m filters. They reported a flux density of $0.026\pm0.019$~Jy at 60~$\mu$m, corresponding to a significance of only approximately $1.4\sigma$. We therefore regard this value as a marginal measurement rather than a secure far-infrared detection. The 90~$\mu$m measurement was reported as an upper limit, which is listed explicitly in Table~\ref{tab:SED_data} and treated as a one-sided constraint in the SED analysis.

CPD--82 291 was subsequently observed as part of the \textit{Spitzer} Formation and Evolution of Planetary Systems (FEPS) survey. \citet{Carpenter2009} reported normalized flux ratios of $R_{16/8}=0.97\pm0.05$ and $R_{24/8}=1.01\pm0.03$, both of which are consistent with photospheric emission. They found no significant infrared excess in the IRAC, IRS, MIPS 24~$\mu$m, or MIPS 70~$\mu$m observations. Thus, the \textit{Spitzer} measurements do not indicate a persistent mid- or far-infrared excess at the epoch of the FEPS observations.

In addition to the far-infrared measurements reported by \citet{Spangler2001} and the \textit{Spitzer}/FEPS observations presented by \citet{Carpenter2009}, we compiled multi-wavelength photometric data from several surveys using the VizieR photometric data service \citep{vizier2000}. The SED data were retrieved in flux density units, as provided by the VizieR SED tool, which homogenizes measurements from different photometric systems using appropriate zero-point calibrations. However, the original catalog entries were independently checked to verify the source identifiers, coordinates, angular separations, reported uncertainties, and available photometric quality flags.

The dataset includes optical, near-infrared, and mid-infrared measurements from surveys such as \textit{Gaia}, 2MASS, and \textit{WISE}, together with the available \textit{ISO} and \textit{Spitzer} measurements and constraints. Catalog entries with uncertain source associations, unreliable quality flags, or possible contamination were excluded from the modelling.

Because CPD--82 291 is photometrically variable and the observations were obtained at different epochs, measurements from different surveys or observing epochs were not averaged into a single representative flux. Instead, individual measurements were retained separately whenever their observing epochs or instrumental characteristics differed. This is particularly important for the infrared data, since the \textit{Spitzer} and \textit{WISE} observations exhibit different SED profiles and may sample different states of the unresolved infrared-emitting environment.

The broader point-spread function of \textit{WISE}, especially in the W3 and W4 bands, also makes its measurements more susceptible to contamination by nearby sources or diffuse background emission than the corresponding \textit{Spitzer} observations. Differences between the \textit{Spitzer} and \textit{WISE} infrared profiles are therefore interpreted cautiously, since they may represent genuine temporal variability, but may also arise from source confusion, background emission, or survey-dependent aperture effects.

The compiled and processed SED data used in the analysis are listed in
Table~\ref{tab:SED_data}. For each photometric point, the table
provides the effective wavelength, adopted flux density (or upper
limit), photometric uncertainty, bandpass, number of distinct
photometric records contributing to the combined value, corresponding
source catalogues, observing epoch when available, uncertainty treatment, and its treatment in the SED analysis (included, excluded,
or treated as an upper limit). Exact duplicate catalogue entries were removed before independent measurements within the same bandpass were
combined. For records lacking a valid catalogue uncertainty, a wavelength-dependent uncertainty was assigned during the data-cleaning
procedure. The uncertainties listed in Table~\ref{tab:SED_data} are
these adopted photometric uncertainties; the additional 20\% systematic term was added in quadrature separately when evaluating the SED likelihood, as described in Sect.~\ref{sec:sed_analysis}.


\begin{table*}[htbp!]
\centering
\caption{
Photometric data adopted for the SED analysis of CPD--82~291 after
quality-independent catalogue matching, removal of exact duplicate
entries, and combination of distinct photometric records within each
bandpass. Flux densities and their adopted photometric uncertainties
are given in Jy. The column $N$ gives the number of distinct
photometric records contributing to the tabulated flux after exact
duplicate removal. Source codes identify the VizieR catalogue tables
contributing to each combined point and are defined below. Epochs are
the JD (TCB) values supplied by the VizieR SED service, expressed as
JD$-2400000$; a range is given when more than one finite epoch is
available, while ``--'' indicates that no epoch was supplied.
The uncertainty code C denotes points based exclusively on available
catalogue uncertainties, whereas C+I denotes combined points for which
at least one contributing record lacked a valid catalogue uncertainty
and was assigned the wavelength-dependent uncertainty described in
Section~\ref{sec:sed_analysis}. The listed uncertainties do not include
the additional 20\% systematic term added in quadrature in the SED
likelihood. Measurements at $\lambda<0.4\,\mu$m and
$1.5<\lambda<3.0\,\mu$m were excluded from the fit. The
$60\,\mu$m measurement is retained only as a low-significance
($\simeq1.4\sigma$) flux measurement. The $90\,\mu$m entry from
\citet{Spangler2001} is a non-detection with a reported
$1\sigma$ uncertainty of 0.007~Jy and was incorporated as a
one-sided $3\sigma$ upper limit of 0.021~Jy.
}
\label{tab:SED_data}

\scriptsize
\renewcommand{\arraystretch}{1.08}
\setlength{\tabcolsep}{2.4pt}

\begin{tabular}{rccccclccc}
\hline
No. &
$\lambda_{\rm eff}$ &
Flux &
Unc. &
Bandpass &
$N$ &
Sources &
Epoch &
Err. &
Treatment \\
&
($\mu$m) &
(Jy) &
(Jy) &
&
&
&
(JD$-2400000$) &
&
\\
\hline

 1 & 0.3498 & 0.005288 & 0.000100 & SkyMapper $u$ & 5 & S1,S2,S3 & -- & C & Excluded \\
 2 & 0.3519 & 0.015200 & 0.000200 & SDSS $u$ & 1 & S4 & -- & C & Excluded \\
 3 & 0.3871 & 0.015400 & 0.000100 & SkyMapper $v$ & 3 & S1,S2,S3 & -- & C & Excluded \\
 4 & 0.4203 & 0.033292 & 0.007039 & Tycho $B_T$ & 2 & S5,S6 & -- & C & Fit \\
 5 & 0.4442 & 0.042599 & 0.003901 & Johnson $B$ & 6 & S7,S8,S9,S10,S11,S12 & 58072.819 & C+I & Fit \\
 6 & 0.4820 & 0.060000 & 0.007700 & SDSS $g$ & 1 & S4 & -- & C & Fit \\
 7 & 0.4820 & 0.059900 & 0.007700 & SDSS $g'$ & 1 & S9 & -- & C & Fit \\
 8 & 0.4968 & 0.076350 & 0.001906 & SkyMapper $g$ & 4 & S1,S2,S3 & -- & C & Fit \\
 9 & 0.5036 & 0.071700 & 0.000600 & Gaia DR3 $G_{\rm BP}$ & 1 & S13 & -- & C & Fit \\
10 & 0.5046 & 0.070600 & 0.000300 & Gaia DR2 $G_{\rm BP}$ & 1 & S14 & -- & C & Fit \\
11 & 0.5319 & 0.087456 & 0.007114 & Tycho $V_T$ & 3 & S5,S6,S15 & -- & C & Fit \\
12 & 0.5537 & 0.086869 & 0.010102 & Johnson $V$ & 8 & S8,S9,S16,S10,S17,S18,S11,S12 & 58072.819 & C+I & Fit \\
13 & 0.5822 & 0.102000 & 0.010200 & Gaia DR3 $G$ & 1 & S13 & -- & C+I & Fit \\
14 & 0.6003 & 0.084700 & 0.008470 & IA598 & 1 & S12 & -- & C+I & Fit \\
15 & 0.6041 & 0.098138 & 0.000644 & SkyMapper $r$ & 5 & S1,S2,S3 & -- & C & Fit \\
16 & 0.6226 & 0.105000 & 0.010500 & Gaia DR2 $G$ & 1 & S14 & -- & C+I & Fit \\
17 & 0.6247 & 0.089200 & 0.035600 & SDSS $r$ & 1 & S4 & -- & C & Fit \\
18 & 0.6247 & 0.088900 & 0.035500 & SDSS $r'$ & 1 & S9 & -- & C & Fit \\
19 & 0.6730 & 0.095621 & 0.000274 & Gaia $G$ (compiled) & 4 & S19,S16,S20,S21 & 58072.819 & C+I & Fit \\
20 & 0.7621 & 0.147000 & 0.001000 & Gaia DR3 $G_{\rm RP}$ & 1 & S13 & -- & C & Fit \\

\hline
\end{tabular}

\end{table*}


\begin{table*}[htbp!]
\ContinuedFloat

\caption{Continued.}

\centering
\scriptsize
\renewcommand{\arraystretch}{1.08}
\setlength{\tabcolsep}{2.4pt}

\begin{tabular}{rccccclccc}
\hline
No. &
$\lambda_{\rm eff}$ &
Flux &
Unc. &
Bandpass &
$N$ &
Sources &
Epoch &
Err. &
Treatment \\
&
($\mu$m) &
(Jy) &
(Jy) &
&
&
&
(JD$-2400000$) &
&
\\
\hline

21 & 0.7635 & 0.118000 & 0.042000 & SDSS $i'$ & 1 & S9 & -- & C & Fit \\
22 & 0.7635 & 0.119000 & 0.042000 & SDSS $i$ & 1 & S4 & -- & C & Fit \\
23 & 0.7725 & 0.150000 & 0.001000 & Gaia DR2 $G_{\rm RP}$ & 1 & S14 & -- & C & Fit \\
24 & 0.7886 & 0.148970 & 0.008992 & Cousins $I$ & 20 & S22 & -- & C & Fit \\
25 & 0.9018 & 0.166000 & 0.003000 & SDSS $z$ & 1 & S4 & -- & C & Fit \\
26 & 1.2390 & 0.199500 & 0.003571 & 2MASS $J$ & 2 & S23,S16 & 58072.819 & C & Fit \\
27 & 1.2500 & 0.202638 & 0.012049 & Johnson $J$ & 27 & S22,S24,S15 & 51541.725 & C & Fit \\
28 & 1.6300 & 0.214872 & 0.003827 & Johnson $H$ & 3 & S24,S6,S15 & 51541.725 & C & Excluded \\
29 & 1.6495 & 0.216000 & 0.005000 & 2MASS $H$ & 1 & S25 & 58072.819 & C & Excluded \\
30 & 2.1638 & 0.163000 & 0.003000 & 2MASS $K_s$ & 1 & S26 & 58072.819 & C & Excluded \\
31 & 2.1900 & 0.160396 & 0.008703 & Johnson $K$ & 22 & S22,S24 & 51541.725 & C & Excluded \\
32 & 3.3500 & 0.078420 & 0.000933 & WISE W1 & 6 & S27,S28,S29,S30,S10,S12 & 57165.259--58072.819 & C+I & Fit \\
33 & 3.4000 & 0.073800 & 0.001600 & Johnson $L$ & 1 & S6 & -- & C & Fit \\
34 & 4.6000 & 0.043730 & 0.000465 & WISE W2 & 5 & S4,S27,S29,S30,S12 & 57165.259--58072.819 & C+I & Fit \\
35 & 5.0300 & 0.040600 & 0.000700 & Johnson $M$ & 1 & S6 & -- & C & Fit \\
36 & 11.5598 & 0.008055 & 0.000276 & WISE W3 & 3 & S4,S27,S12 & 58072.819 & C+I & Fit \\
37 & 22.0907 & 0.002218 & 0.000795 & WISE W4 & 2 & S27,S31 & 58072.819 & C & Diagnostic \\
38 & 60.0005 & 0.026000 & 0.019000 & $60\,\mu{\rm m}$ & 1 & S32 & -- & C & Low-S/N dust constraint\\
39 & 90.0000 & $<0.021$ & 0.007000 & $90\,\mu{\rm m}$ & 1 & S32 & -- & C & $3\sigma$ UL \\

\hline
\end{tabular}

\vspace{0.15cm}

\raggedright
\scriptsize
\textit{Source identifiers:}
S1=\texttt{II/358/smss};
S2=\texttt{II/379/smssdr4};
S3=\texttt{J/ApJ/925/164/catalog};
S4=\texttt{I/353/gsc242};
S5=\texttt{I/350/tyc2tdsc};
S6=\texttt{II/346/jsdc\_v2};
S7=\texttt{I/305/out};
S8=\texttt{I/320/spm4};
S9=\texttt{II/336/apass9};
S10=\texttt{IV/38/tic};
S11=\texttt{J/MNRAS/463/4210/ucac4rpm};
S12=\texttt{J/MNRAS/471/770/table1};
S13=\texttt{I/350/gaiaedr3};
S14=\texttt{I/345/gaia2};
S15=\texttt{J/PASP/120/1128/catalog};
S16=\texttt{II/366/catv2021};
S17=\texttt{J/A+A/653/A98/aspic1\_1};
S18=\texttt{J/ApJS/203/32/table4};
S19=\texttt{I/339/hsoy};
S20=\texttt{J/A+A/664/A105/main};
S21=\texttt{J/ApJ/867/105/refcat2};
S22=\texttt{B/denis/denis};
S23=\texttt{I/280B/ascc};
S24=\texttt{II/246/out};
S25=\texttt{I/319/xpm};
S26=\texttt{I/317/sample};
S27=\texttt{II/311/wise};
S28=\texttt{II/328/allwise};
S29=\texttt{II/365/catwise};
S30=\texttt{II/369/vexassm};
S31=\texttt{IV/39/tic82};
S32=\texttt{J/ApJ/555/932/targets}.

\par\medskip
\textit{Note.}
The source identifiers refer to the catalogue tables associated with
the distinct photometric records retained after duplicate removal.
The complete VizieR provenance, including catalogue entries carrying
duplicate copies of the same photometric measurement, was retained
during the cleaning procedure but is not reproduced here.
Catalogue-specific native quality flags are not propagated as a common
field by the homogenized VizieR SED response and therefore are not
represented by a single generic quality column in this table.

\end{table*}

\subsection{Time-Series Photometry}
\subsubsection{TESS Observations}

CPD--82 291 was observed by \textit{TESS} in eight sectors (11, 13, 38, 39, 65, 66, 93, and 94), spanning different observing epochs and cadences. The observations were obtained with exposure times of 1800\,s (Sectors 11 and 13), 600\,s (Sectors 38 and 39), and 200\,s (Sectors 65, 66, 93, and 94).

\begin{figure*}[htbp!]
    \centering 
    \includegraphics[width=0.75\linewidth]{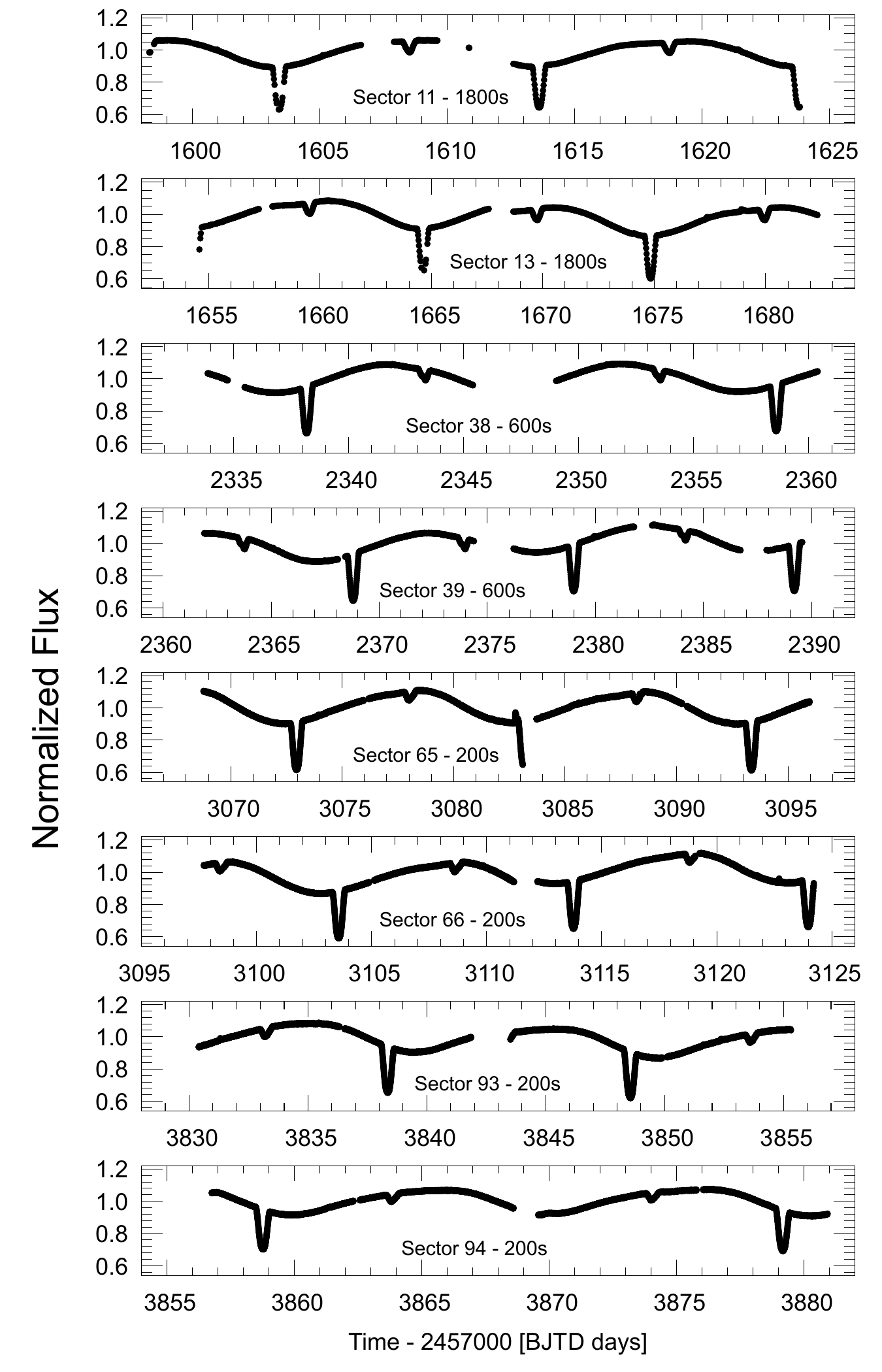}
    \caption{\textit{TESS} light curves of CPD--82 291 obtained in eight different sectors with varying cadences.}
    \label{fig:lc_tess}
\end{figure*}

We used the Quick-Look Pipeline \citep[QLP,][]{Huang2020} light curves for all sectors with \texttt{lightkurve} \citep{lk}while using the quality flag setting of ``hard'', without applying any externally detrending, or offset correction. A total of 49\,427 photometric measurements were collected from these sectors, with none rejected. The light curves from all sectors were combined to construct a homogeneous data set, which is shown in Figure~\ref{fig:lc_tess}.

The multi-sector data reveal a stable overall light-curve morphology, while significant variations are observed in the out-of-eclipse regions. In particular, large-amplitude wave-like modulations are present at the light-curve maxima, and the phase and shape of these modulations change from sector to sector. This behavior indicates the presence of evolving surface inhomogeneities, most likely related to stellar activity.

Furthermore, asymmetries in the secondary minimum are consistently observed in all sectors, whereas the primary eclipse remains largely unaffected. The combination of these features suggests that the system exhibits strong and time-dependent activity, which primarily influences the out-of-eclipse flux levels and the morphology of the secondary eclipse.

\subsubsection{ASAS-SN Observations}
\label{phot:asas-sn}

In the ASAS-SN database \citep{Shappee2014,Jayasinghe2019}, CPD--82~291 is listed with the identifier ASASSN-V J085305.31$-$824400.0. 
The same astrophysical source was previously discussed by \citet{Christy2022} as ASASSN-V J085305.34$-$824360.0; those authors also noted the earlier $V$-band catalogue identifier ASASSN-V J085305.74$-$824401.0. The small differences among these coordinate-based identifiers reflect the catalogue/reference positions adopted in different ASAS-SN data products and do not correspond to distinct nearby sources.

A total of 1063 photometric measurements were obtained between May 2014 and July 2021, consisting of 241 observations in the $V$ band and 822 in the $g$ band. 
The data set used here was retrieved from the ASAS-SN database and overlaps the observations considered by \citet{Christy2022}, although we do not assume a point-by-point identity with the light-curve extraction used in their Citizen ASAS-SN analysis. In particular, \citet{Christy2022} primarily analysed the newer $g$-band data while referring separately to the earlier $V$-band classification.

\begin{figure}[htbp!]
    \centering 
    \includegraphics[width=0.5\linewidth]{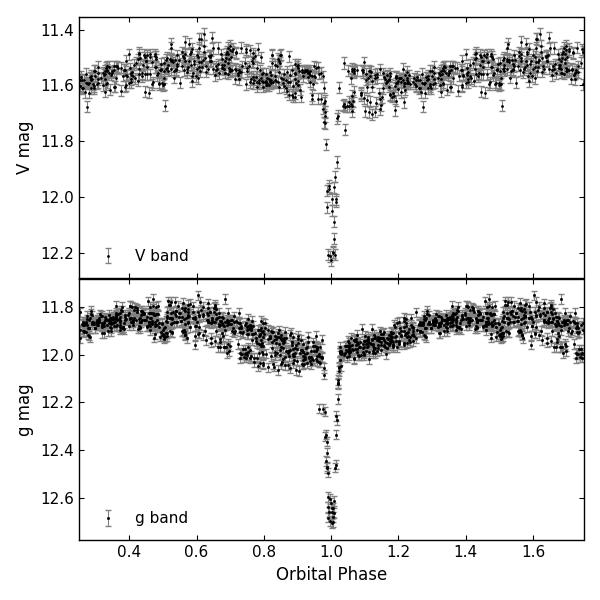}
    \caption{Phase-folded ASAS-SN light curves of CPD-82 291 in the $V$ (top) and $g$ (bottom) bands.}
    \label{fig:lc_asassn}
\end{figure}

Each observation was acquired on a separate night; therefore, the data set does not include continuous coverage during eclipse phases, making the determination of precise times of minima challenging. The $V$-band data span the period 2014--2018, whereas the $g$-band observations extend from 2017 to 2021. The phase-folded light curves in both passbands are shown in Figure~\ref{fig:lc_asassn}, constructed using the ephemeris given in Eq.~\ref{eq:ephem}.

\section{Analysis}
\subsection{Eclipse-Timing and Orbital-Period Analysis}

From the \textit{TESS} observations, a total of
18 primary-eclipse times were determined using \cite{Kwee1956}. The individual timings and their uncertainties are listed in Table~\ref{tab:minima_times}. The secondary eclipse is relatively shallow and is strongly affected by activity-related light-curve distortions; therefore, the secondary minima were not included in the eclipse-timing analysis.

\begin{table}[!ht]
    \centering
    \caption{Calculated times of minima for CPD-82 291.}
    \begin{tabular}{cc}
    \hline
    Observed ToM & Error \\
    HJD & (days) \\
    \hline
        2458603.40291 & 0.00087 \\ 
        2458613.60774 & 0.00051 \\ 
        2458664.63403 & 0.00127 \\ 
        2458674.83977 & 0.00043 \\ 
        2459338.17604 & 0.00017 \\ 
        2459358.58730 & 0.00011 \\ 
        2459368.79215 & 0.00016 \\ 
        2459378.99722 & 0.00019 \\ 
        2459389.20361 & 0.00012 \\ 
        2460072.95482 & 0.00009 \\ 
        2460093.36529 & 0.00010 \\ 
        2460103.57034 & 0.00008 \\ 
        2460113.77488 & 0.00010 \\ 
        2460123.98078 & 0.00015 \\ 
        2460838.35533 & 0.00009 \\ 
        2460848.56124 & 0.00010 \\ 
        2460858.76658 & 0.00013 \\ 
        2460879.17644 & 0.00017 \\ 
        \hline
    \end{tabular}
    \label{tab:minima_times}
\end{table}

\begin{figure}[htbp!]
    \centering 
    \includegraphics[width=0.5\linewidth]{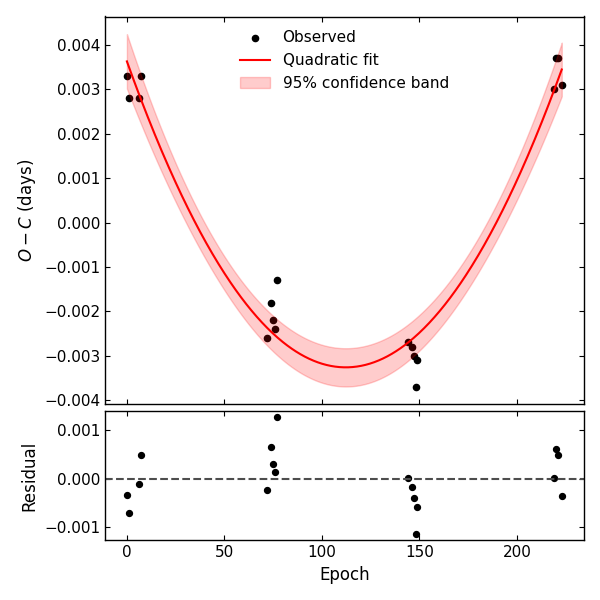}
    \caption{O--C diagram constructed from the \textit{TESS} primary-eclipse timings. The solid curve represents the best-fitting quadratic model, while the shaded region denotes the 95\% confidence interval of the fit.}
    \label{fig:oc_tess}
\end{figure}

A quadratic fit to the O--C data yields

\begin{equation}
\begin{aligned}
O-C =\;&\,(3.63 \pm 0.31)\times10^{-3} \\
     &- (1.23 \pm 0.07)\times10^{-4} E \\
     &+ (5.5 \pm 0.3)\times10^{-7} E^{2}
\end{aligned}
\label{eq:oc}
\end{equation}

The corresponding updated quadratic ephemeris is given by

\begin{equation}
\begin{aligned}
T (\mathrm{HJD}) =\;& 2458603.4033 \\
                  &+ 10.20514\,E \\
                  &+ 5.5\times10^{-7} E^2.
\end{aligned}
\label{eq:ephem}
\end{equation}

The positive quadratic term provides a phenomenological
description of the curvature seen in the current O--C measurements,
but it should not be interpreted as secure evidence for a secular
increase in the orbital period. The timing measurements are confined
to a limited and temporally clustered \textit{TESS} baseline, and the strong
surface activity of CPD--82~291 may introduce additional systematic
shifts in the measured eclipse times through distortions of the eclipse
profiles. Consequently, the observed curvature could represent only
part of a longer-term variation, and interpretations involving secular
period evolution, magnetic activity, or the light-travel-time effect
of a tertiary component cannot presently be distinguished. A longer
timing baseline and quantitative comparison of alternative timing
models are required to establish the physical origin of the observed
trend.

\subsection{Light-curve Modelling}
\label{sec:lc_analysis}
We first modelled the Sector~94 light curve to establish a reference binary solution. Sector~94 was selected because it provides the highest-quality light curve and well-defined eclipse profiles. The remaining \textit{TESS} sectors exhibit different out-of-eclipse
modulations and were subsequently analysed with the reference binary
parameters held fixed, as described in Sect.~\ref{sec:spot_evolution}.

Although additional photometric data are available from the ASAS-SN survey in the $V$ and $g$ bands (see Sect.~2.2.2 and Figure~\ref{fig:lc_asassn}), these data were not included in the light-curve modeling. Compared to the \textit{TESS} observations, the ASAS-SN light curves have significantly lower photometric precision and sparse temporal sampling, limiting their ability to constrain the eclipse morphology and surface-brightness variations. We therefore used only the \textit{TESS} photometry for the binary light-curve solution.

The Sector~94 light curve was modelled with the PHysics Of Eclipsing BinariEs code
\citep[\texttt{PHOEBE},][]{phoebe1,phoebe2,phoebe3,phoebe4,phoebe5}, version 2.4, adopting a detached binary configuration. The adjustable parameters included the volume-equivalent fractional radii,
photometric mass ratio, orbital inclination, component temperature
ratio, and luminosity contribution in the \textit{TESS} bandpass.

The third-light contribution was initially included as a free parameter to test for unresolved contaminating flux. Trial solutions converged to negligible values of $l_3\simeq0.001$--$0.002$, without a meaningful improvement in the fit. We therefore fixed $l_3=0$ in the final solution. This test constrains only additional flux in the photometric aperture and does not by itself test for the dynamical presence of a tertiary component.

The primary effective temperature adopted in the
light-curve calculation does not represent an independent
temperature measurement from the \textit{TESS} data. It sets
the reference temperature scale used by \texttt{PHOEBE} for
the stellar-atmosphere, passband-intensity, and limb-darkening
calculations, whereas the principal temperature constraint
provided by the light curve is the component temperature ratio.
The absolute effective temperatures of both components are subsequently determined from the SED analysis while preserving
the temperature ratio obtained from the light-curve solution
(Sect.~\ref{sec:sed_analysis}).

Although the model was sampled using
$T_{\rm eff,1}/T_{\rm eff,2}$, its reciprocal,
$T_{\rm eff,2}/T_{\rm eff,1}$, is reported in
Table~\ref{tab:lc_pars} for consistency with the notation used
in the SED analysis. The parameters allowed to vary during the optimization, together with the fixed model inputs and their
adopted values or sources, are listed separately in
Table~\ref{tab:lc_pars}.

\begin{figure}[htbp!]
    \centering
    \includegraphics[width=0.6\linewidth]{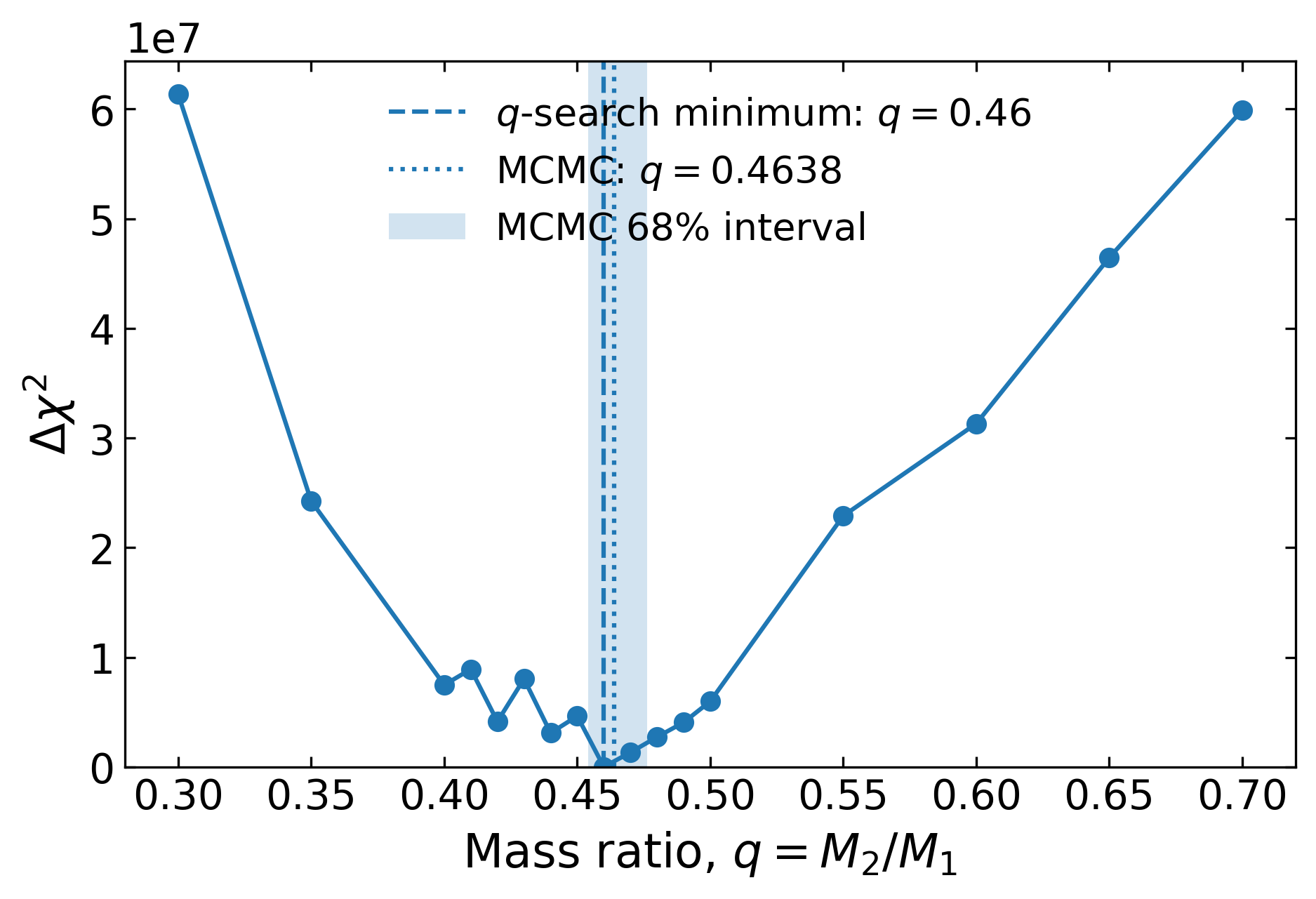}
    \caption{Photometric mass-ratio search for CPD--82 291. The ordinate shows $\Delta\chi^{2}=\chi^{2}-\chi^{2}_{\rm min}$ obtained by fixing the mass ratio at successive trial values and re-optimizing the remaining adjustable light-curve parameters. The minimum occurs near $q=0.46$. The vertical dotted line and shaded interval indicate the median and 68\% credible interval of the subsequent MCMC solution, $q_{\rm phot}=0.4638^{+0.0126}_{-0.0097}$.}
    \label{fig:q_search}
\end{figure}

Before performing the Markov chain Monte Carlo \citep[MCMC,][]{mcmc} analysis to obtain fundamental parameters and uncertainties, we carried out a preliminary photometric mass-ratio search to identify the region of parameter space providing the best light-curve solution. The mass ratio was fixed at successive values over the interval $0.30\leq q\leq0.70$, while the remaining adjustable light-curve parameters were re-optimized at each trial value. The goodness of fit was evaluated using the corresponding $\chi^{2}$ value. As shown in Figure~\ref{fig:q_search}, the $\chi^{2}$ distribution reaches its minimum near $q=0.46$, while the quality of the fit deteriorates toward both smaller and larger mass ratios.

The value obtained from this preliminary search was used only as the initial estimate for the subsequent MCMC analysis. In the MCMC calculation, $q$ was allowed to vary freely together with the other fitted binary parameters. The resulting posterior solution, $q_{\rm phot}=0.4638^{+0.0126}_{-0.0097}$, is consistent with the minimum of the independent $q$-search. Because the system is detached and its out-of-eclipse light curve is strongly affected by surface-brightness inhomogeneities, this value is treated as a photometric and model-dependent mass ratio rather than a direct dynamical measurement. An independent spectroscopic orbit is required to determine the dynamical mass ratio and to assess possible systematic uncertainties associated with the adopted surface-brightness model.

Parameter uncertainties were estimated through an MCMC analysis with 128 walkers and 1000 iterations. The parameters identified as fitted in
Table~\ref{tab:lc_pars}, including the photometric mass ratio, were allowed to vary. The posterior medians and corresponding 16th and 84th percentiles were adopted as the parameter estimates and their 68\% credible intervals. The quoted intervals describe the statistical uncertainties within the adopted light-curve and surface-brightness model and do not include all possible systematic uncertainties associated with the fixed assumptions.

\begin{figure}[htbp!]
    \centering
    \includegraphics[width=0.75\linewidth]{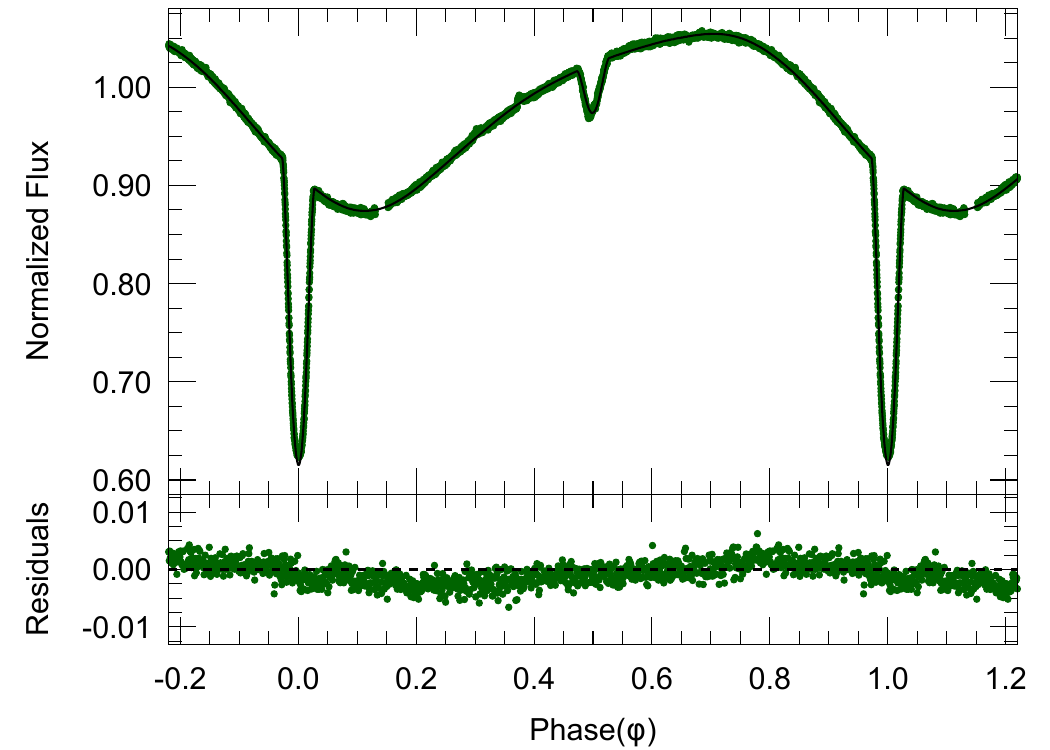}
    \caption{\textit{TESS} Sector~94 light curve and the corresponding median posterior model obtained with \texttt{PHOEBE} v2.4. The lower panel shows the residuals, $O-C$.}
    \label{fig:TESS_lc_model}
\end{figure}

The median posterior model for Sector~94 is shown in
Figure~\ref{fig:TESS_lc_model}, and the corresponding parameter values are summarized in Table~\ref{tab:lc_pars}. The posterior distributions and parameter correlations are shown in the corner plot in Appendix~B, Figure~\ref{fig:mesh2}.

\begin{table*}[htbp!]
\centering
\begin{tabular}{lcl}
\hline
Parameter & Value & Source/Assumption \\
\hline

\multicolumn{3}{c}{\textit{Fitted parameters}}\\
\hline
Primary fractional radius, $r_1$ &
$0.0568^{+0.0050}_{-0.0050}$ & MCMC \\

Secondary fractional radius, $r_2$ &
$0.1687^{+0.0050}_{-0.0050}$ & MCMC \\

Photometric mass ratio, $q=M_2/M_1$ &
$0.4638^{+0.0126}_{-0.0097}$ & MCMC \\

Orbital inclination, $i$ ($^\circ$) &
$82.1^{+0.057}_{-0.069}$ & MCMC \\

Temperature ratio, $T_{\rm eff,2}/T_{\rm eff,1}$ &
$0.6413^{+0.0031}_{-0.0026}$ & MCMC \\

Primary \textit{TESS} light fraction, $l_{1,\mathrm{TESS}}$ &
$0.342^{+0.004}_{-0.003}$ & MCMC \\

\hline
\multicolumn{3}{c}{\textit{Fixed parameters and assumptions}}\\
\hline

Orbital period, $P$ (d) &
$10.20514$ & Eclipse-timing analysis \\

Reference epoch, $T_0$ (BJD$_{\rm TDB}$) &
$2458613.642535$ & Eclipse-timing analysis \\

Eccentricity, $e$ &
$0$ & Circular orbit adopted \\

Adopted primary effective temperature,
$T_{\rm eff,1}$ (K)
& $6900$
& Ref. temp. for atm. and
LD calculations \\

Bolometric albedos, $A_{1,2}$ &
$0.500$ & Convective envelopes \\

Gravity-darkening exponents, $g_{1,2}$ &
$0.320$ & Convective envelopes \\

Third light, $l_3$ &
$0$ & Adopted \\

Synchronicity parameters, $F_{1,2}$ &
$1.0$ & Synchronous rotation adopted \\

\textit{TESS} LD coefficient, $x_1$ &
$0.640$ & Adopted LD tables \\

\textit{TESS} LD coefficient, $y_1$ &
$0.361$ & Adopted LD tables \\

\textit{TESS} LD coefficient, $x_2$ &
$0.738$ & Adopted LD tables \\

\textit{TESS} LD coefficient, $y_2$ &
$0.257$ & Adopted LD tables \\

Bolometric LD coefficient, $x_{\rm Bol,1}$ &
$0.681$ & Adopted LD tables \\

Bolometric LD coefficient, $y_{\rm Bol,1}$ &
$0.184$ & Adopted LD tables \\

Bolometric LD coefficient, $x_{\rm Bol,2}$ &
$0.714$ & Adopted LD tables \\

Bolometric LD coefficient, $y_{\rm Bol,2}$ &
$0.045$ & Adopted LD tables \\

\hline
\multicolumn{3}{c}{\textit{Derived parameters}}\\
\hline

Secondary \textit{TESS} light fraction, $l_{2,\mathrm{TESS}}$ &
$0.658^{+0.003}_{-0.004}$ & $1-l_{1,\mathrm{TESS}}$, assuming $l_3=0$ \\

\hline
\end{tabular}

\caption{Light-curve solution parameters derived from the
\textit{TESS} Sector~94 data using \texttt{PHOEBE} v2.4.
Quoted uncertainties represent the 16th and 84th percentiles of
the posterior distributions. The mass ratio is photometric and
model dependent.}
\label{tab:lc_pars}
\end{table*}

\subsection{SED Modelling}
\label{sec:sed_analysis}

The SED of the system was first modelled using the stellar
photospheres alone and was subsequently compared with a model including an additional circumstellar dust component. The total observed flux was
represented as

\begin{equation}
F_{\nu}^{\mathrm{obs}} =
\left(
F_{\nu,1}+F_{\nu,2}+F_{\nu,\mathrm{disk}}
\right)
10^{-0.4A_{\lambda}},
\end{equation}

where $F_{\nu,1}$ and $F_{\nu,2}$ are the contributions from the
primary and secondary components, respectively, and
$F_{\nu,\mathrm{disk}}$ is the thermal emission from the adopted disk model. For the stellar-only solution, $F_{\nu,\mathrm{disk}}$ was set to zero. The disk component was included only to investigate whether the tentative far-infrared emission could be reproduced by a cool circumbinary-dust model. The extinction was parameterized through the colour excess $E(B-V)$ using the G23 extinction law \citep{Gordon2023}, with $R_V=3.1$.

The stellar fluxes were represented by blackbody spectra,

\begin{equation}
F_{\nu,\star} =
\left(\frac{R_{\star}}{d}\right)^2
\pi B_{\nu}(T_{\star}),
\end{equation}

where $R_{\star}$ and $T_{\star}$ are the stellar radius and effective temperature, respectively, and $d$ is the distance to the system. The
distance was fixed at $d=486.9\pm3.2$~pc, obtained by direct inversion
of the source-dependent zero-point-corrected \textit{Gaia} DR3
parallax, $\varpi_{\rm corr}=2.0538\pm0.0134$~mas. The component temperature and
radius ratios were fixed to the posterior medians obtained from the light-curve analysis,

\begin{equation}
\frac{T_{\star,2}}{T_{\star,1}}=0.6413,
\qquad
\frac{R_{\star,2}}{R_{\star,1}}=2.971.
\end{equation}

Fixing the light-curve temperature and radius ratios reduces
the stellar part of the SED model from four independent parameters
($T_{\star,1}$, $T_{\star,2}$, $R_{\star,1}$, and $R_{\star,2}$)
to two ($T_{\star,1}$ and $R_{\star,1}$), while the Gaia distance
provides an external scale constraint. These constraints restrict the
allowed stellar parameter space but do not eliminate the remaining
covariances among stellar temperature, radius, and reddening. They
also do not independently establish the presence or geometry of a
circumbinary disk.

The disk emission was calculated by integrating the thermal emission from concentric annuli,

\begin{equation}
F_{\nu,\mathrm{disk}} =
\frac{\cos i_{\rm disk}}{d^2}
\int_{R_{\mathrm{in}}}^{R_{\mathrm{out}}}
2\pi r B_\nu[T(r)]
\left[
1-\exp\left(
-\frac{\tau_\nu(r)}{\cos i_{\rm disk}}
\right)
\right],dr .
\end{equation}

The radial structure follows the parametrized model presented by
\citet{Bakis2026}, incorporating a power-law surface-density
distribution and a passive irradiation temperature profile. The model was extended to include the combined irradiation from both stellar components.

Because the disk is spatially unresolved, its inclination cannot be
determined independently from the SED. We therefore adopted
$i_{\rm disk}=i_{\rm binary}=82.1^\circ$ as a fiducial coplanar
configuration. To examine the sensitivity to this assumption, the
deterministic optimization was repeated for fixed inclinations of
$0^\circ$, $30^\circ$, $60^\circ$, and $82.1^\circ$.
These tests used the same revised likelihood as the fiducial
analysis, including the $60\,\mu$m flux measurement and the one-sided
$90\,\mu$m upper-limit constraint. The corresponding total $\chi^2$
values were 6.778, 6.778, 6.778, and 6.779, respectively.

The stellar parameters and inner disk radius remained broadly similar,
whereas the surface-density normalization decreased from
approximately $0.30$~g\,cm$^{-2}$ at $0^\circ$ to
$0.050$~g\,cm$^{-2}$ at $82.1^\circ$, a change by a factor of about
six. In these deterministic sensitivity tests, the optimizer
converged to outer radii close to the adopted lower boundary
($R_{\rm out}\simeq105$--107~AU) for all four inclinations.
This behaviour should not be interpreted as a posterior constraint on
$R_{\rm out}$: in the fiducial $i_{\rm disk}=82.1^\circ$ MCMC
analysis, the posterior is broad and asymmetric, with
$R_{\rm out}=188_{-52}^{+150}$~AU. The difference between the
deterministic optimum and the posterior distribution further
illustrates the weak constraint on the outer disk radius. These tests show that the unresolved
photometry does not independently constrain the disk inclination and
that the inferred surface density, dust mass, and radial extent remain
model-dependent.

We used the cleaned and band-combined photometric measurements
described in Sect.~\ref{sec:sed_data}. To account for cross-survey calibration differences, non-simultaneous observations, source variability, and limitations of the adopted photospheric and disk prescriptions, a fractional systematic uncertainty equal to 20\% of the measured flux
was added in quadrature to the adopted uncertainty of each fitted
measurement:

\begin{equation}
\sigma_{\nu,\mathrm{eff}}=
\sqrt{
\sigma_{\nu,\mathrm{phot}}^2+
\left(0.20F_{\nu}\right)^2
}.
\label{eq:sed_sigma_eff}
\end{equation}

Here, $\sigma_{\nu,\mathrm{phot}}$ denotes either the combined
catalogue uncertainty described in Sect.~\ref{sec:sed_data} or the adopted uncertainty for measurements lacking a valid catalogue error. This conservative prescription prevents measurements with very small formal errors from dominating the broadband fit.

\begin{table}[H]
\centering
\caption{Parameters of the stellar SED and conditional
circumbinary-dust model. Uncertainties are the 16th and 84th
percentiles of the MCMC posterior distributions. Uniform priors are
listed for fitted parameters; disk quantities are conditional and
model-dependent.}
\label{tab:sed_pars}

\small
\renewcommand{\arraystretch}{0.95}
\setlength{\tabcolsep}{5pt}

\begin{tabular}{lcc}
\hline
Parameter & Value & Prior \\
\hline

\multicolumn{3}{c}{\textit{Fitted parameters}} \\
\hline

$T_{\star,1}$ (K) &
\textbf{$6898_{-480}^{+670}$} &
$\mathcal{U}(5020,8020)$ \\

$R_{\star,1}$ ($R_\odot$) &
\textbf{$1.245_{-0.087}^{+0.094}$} &
$\mathcal{U}(0.666,1.997)$ \\

$\Sigma_0$ (g cm$^{-2}$) &
\textbf{$0.0256_{-0.022}^{+1.0}$} &
$\mathcal{U}(10^{-5},2)$ \\

$R_{\rm in}$ (AU) &
\textbf{$147_{-48}^{+37}$} &
$\mathcal{U}(10,200)$ \\

$R_{\rm out}$ (AU) &
\textbf{$188_{-52}^{+150}$} &
$\mathcal{U}(100,500)$ \\

$E(B-V)$ &
\textbf{$0.130_{-0.089}^{+0.110}$} &
$\mathcal{U}(0,1)$ \\

\hline
\multicolumn{3}{c}{\textit{Fixed parameters and model inputs}} \\
\hline

$T_{\star,2}/T_{\star,1}$ &
$0.6413$ &
-- \\

$R_{\star,2}/R_{\star,1}$ &
\textbf{$2.970$} &
-- \\

Gaia DR3 parallax, $\varpi$ (mas) &
$2.0365\pm0.0134$; corr. $2.0538\pm0.0134$ &
-- \\

Distance, $d$ (pc) &
$486.9\pm3.2$ &
-- \\

Disk inclination, $i_{\rm disk}$ ($^\circ$) &
$82.1$ &
-- \\

Stellar photosphere model &
Blackbody &
-- \\

Extinction law; $R_V$ &
G23 \citep{Gordon2023}; $3.1$ &
-- \\

Systematic photometric uncertainty &
$0.20\,F_\nu$ &
-- \\

Disk scale height, $H_0$ (AU) &
$0.03$ &
-- \\

Surface-density exponent, $p_{\Sigma}$ &
$1.0$ &
-- \\

Disk flaring exponent, $\beta_H$ &
$1.1$ &
-- \\

Dust opacity $(\kappa_0,\lambda_0,\beta_\kappa)$ &
$(0.5,\ 100\,\mu{\rm m},\ 1.25)$ &
-- \\

\hline
\multicolumn{3}{c}{\textit{Derived parameters}} \\
\hline

$T_{\star,2}$ (K) &
\textbf{$4423_{-308}^{+430}$} &
-- \\

$R_{\star,2}$ ($R_\odot$) &
\textbf{$3.697_{-0.258}^{+0.279}$} &
-- \\

$A_V$ (mag)\tablenotemark{a} &
\textbf{$0.403_{-0.276}^{+0.341}$} &
-- \\

$M_{\rm dust}$ ($M_\odot$) &
\textbf{$\left(8.1_{-5.7}^{+16.0}\right)\times10^{-5}$} &
-- \\

\hline
\end{tabular}

\tablecomments{
The temperature and radius ratios are adopted from the
light-curve solution. The Gaia distance is based on the
zero-point-corrected parallax ($Z_{\varpi}=-0.0173$~mas) and was held
fixed in the SED fit. The disk inclination was fixed at
$82.1^\circ$ assuming coplanarity.}

\tablenotetext{a}{$A_V=R_VE(B-V)$ with $R_V=3.1$.}

\end{table}

Special treatment was applied to the far-infrared measurements
reported by \citet{Spangler2001}. The $60\,\mu$m value,
$F_{60}=0.026\pm0.019$~Jy, was included in the conditional disk fit as a flux measurement with its published uncertainty, but its significance of only $1.37\sigma$ was not regarded as a secure detection. The $90\,\mu$m non-detection was represented by a
$3\sigma$ upper limit of $0.021$~Jy, corresponding to $\sigma_{\nu,\mathrm{ul}}=0.007$~Jy. Its contribution to the
likelihood was implemented through the one-sided term

\begin{equation}
\chi^2_{\rm ul} =
\left[
\max\left(
0,
\frac{
F_{\nu,\mathrm{model}}-
F_{\nu,\mathrm{lim}}
}{
\sigma_{\nu,\mathrm{ul}}
}
\right)
\right]^2,
\end{equation}

so that no penalty was applied when the model remained below the upper limit.

The best-fitting parameters were initially determined through
non-linear least-squares optimization. The stellar-only model included $T_{\star,1}$, $R_{\star,1}$, and $E(B-V)$ as free parameters. For the conditional disk solution, $\Sigma_0$, $R_{\rm in}$, and $R_{\rm out}$ were additionally varied. All other quantities were held fixed at the values listed in Table~\ref{tab:sed_pars}.

The best-fitting stellar-plus-disk solution was then used to initialize
an MCMC analysis. The MCMC sampling was used to characterize the
posterior distributions and parameter correlations within the adopted
conditional disk model. Uniform priors were adopted within the parameter boundaries
listed in Table~\ref{tab:sed_pars}, with the additional physical
requirement $R_{\rm out}-R_{\rm in}\geq1$~AU. No Gaussian or other
informative prior was imposed on $E(B-V)$.

The posterior sampling was performed with the
\texttt{emcee} affine-invariant ensemble sampler using 32 walkers.
The walkers were initialized in the vicinity of the deterministic
least-squares solution with a dispersion corresponding to 2\% of the allowed range of each fitted parameter; all initial walker positions were required to satisfy the adopted priors. The sampler was allowed to run for a maximum of 70000 steps per walker.

Convergence was assessed from the integrated autocorrelation
time, $\tau$, estimated separately for each fitted parameter at intervals of 5000 steps. Sampling was considered practically converged when the total chain length exceeded $50\tau$ for every parameter and the maximum fractional change in the estimated autocorrelation times between successive checks was less than 5\%. These criteria were satisfied after 50000 steps. The final integrated autocorrelation times ranged from 393.7 to 970.7 steps, corresponding to a minimum
chain-length-to-autocorrelation-time ratio of 51.5. The maximum
fractional change in $\tau$ at the final convergence check was 2.5\%.

After convergence was established, the first 4854 steps were discarded as burn-in, corresponding to approximately five times the largest integrated autocorrelation time. The retained chains were thinned by a factor of 196, approximately one half of the shortest autocorrelation time, yielding 7360 posterior samples used for parameter estimation.

The mean acceptance fraction of the converged run was 0.192. This quantity is reported as a sampling diagnostic and was not used as evidence of convergence. The posterior medians and the corresponding 16th and 84th percentiles were adopted as the parameter
estimates and 68\% credible intervals. These intervals describe the allowed parameter ranges within the conditional disk model and do not
constitute evidence for the existence of the disk.

As an additional assessment of the adequacy of the fitted model, posterior-predictive realizations were generated from samples
drawn from the converged posterior distribution. For each posterior draw, the model SED was evaluated at the observed wavelengths and synthetic measurements were generated using the same effective uncertainty prescription adopted in the likelihood.

Using the revised photometric uncertainties and the additional
20\% systematic term defined above, the residuals were evaluated
quantitatively using the same effective uncertainties adopted in the
SED likelihood. Relative to the stellar-only photospheric model, the
W3 and W4 measurements differ from the model by approximately
$-0.39\sigma$ and $-0.37\sigma$, respectively, while the
$60\,\mu$m measurement lies only $+1.30\sigma$ above the stellar
prediction. Thus, the photometry through WISE W4 is consistent with
the stellar photospheres, and the $60\,\mu$m point represents only a
low-significance positive far-infrared deviation.

As a separate diagnostic test of whether the shorter-wavelength
SED itself requires an additional dust component, the W4 and
$60\,\mu$m measurements were excluded from both models, leaving
30 photometric points in the comparison data set. This diagnostic
comparison is distinct from the conditional disk fit described above,
in which W4 and the $60\,\mu$m measurement were retained and the
$90\,\mu$m upper limit was applied. The stellar-only model, with three free parameters, yielded
$\chi^2=5.830$ for 27 degrees of freedom
($\chi^2_\nu\simeq0.22$). The low value of $\chi^2_\nu$ is expected
in part from the deliberately conservative 20\% systematic uncertainty
included in the likelihood and should not be interpreted as evidence
for an exceptionally precise model.

The stellar-plus-disk model yielded essentially the same
$\chi^2$ despite introducing three additional free parameters.
The stellar-only solution gives
${\rm AIC_c}=12.75$ and ${\rm BIC}=16.03$, compared with
${\rm AIC_c}=21.48$ and ${\rm BIC}=24$ for the
stellar-plus-disk model. Thus,
$\Delta{\rm AIC_c}=8.73$ and $\Delta{\rm BIC}=10.20$ favour the
stellar-only model, while the disk parameters become poorly
constrained when the tentative long-wavelength measurements are
removed. The optical-to-W3 photometry therefore does not require an
additional circumstellar-dust component.

\begin{figure}[htbp!]
    \centering \includegraphics[width=0.75\columnwidth]{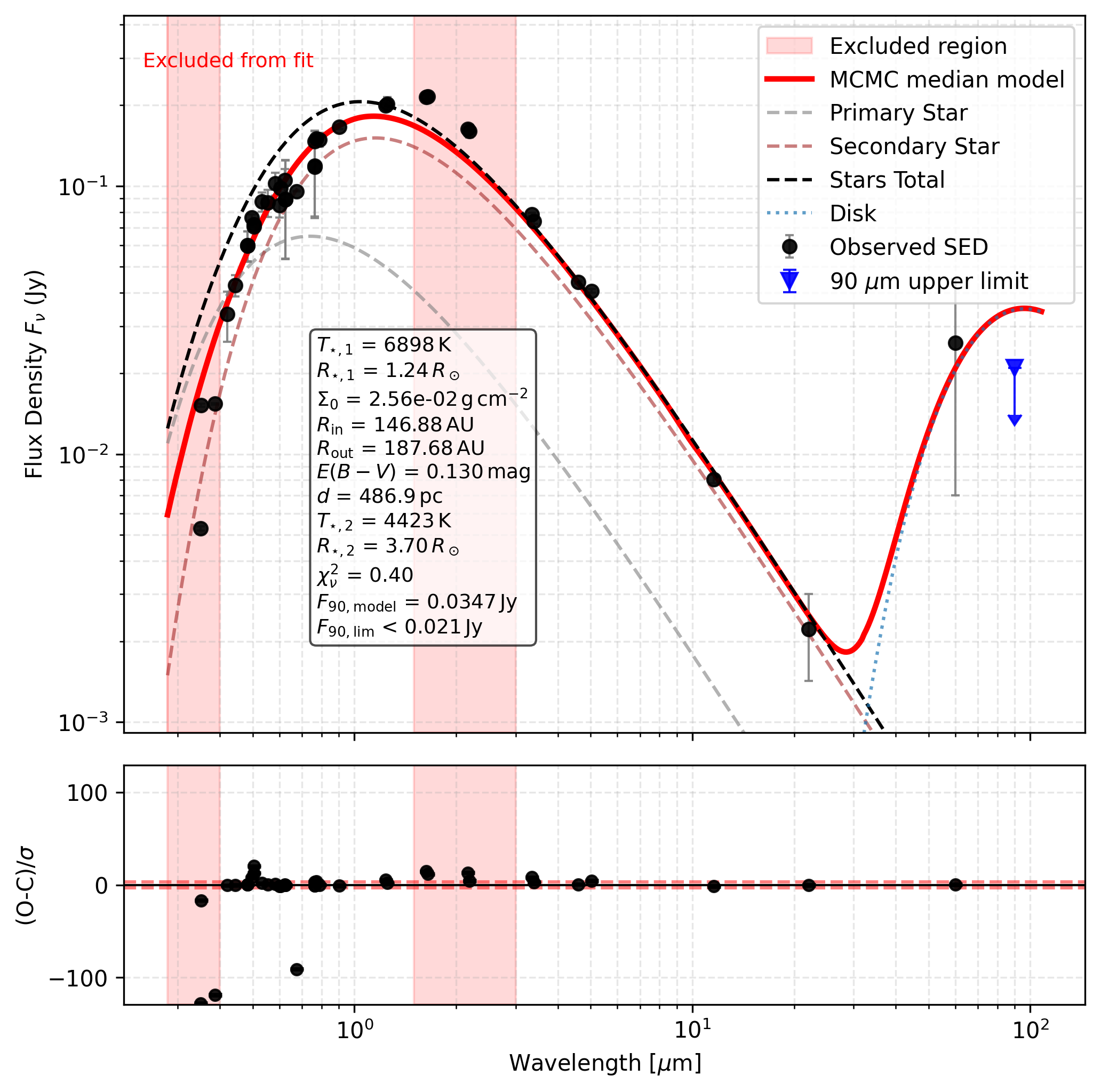}
    \caption{Observed SED and model comparison.
\textit{Top panel:} The photometric measurements are shown together with the stellar-only model (black dashed line) and the conditional
stellar-plus-disk model (red solid line), as identified in the legend. The stellar photospheres reproduce the SED through the WISE W4 band, and neither W3 nor W4 shows a significant infrared excess. The additional disk component illustrates one possible interpretation of the
low-significance $60\,\mu$m measurement. The blue downward arrow denotes the $90\,\mu$m upper limit and does not represent a flux detection. Shaded regions indicate wavelength intervals excluded from the
main SED fitting procedure for the reasons described in the text. \textit{Bottom panel:} Normalized residuals $(O-C)/\sigma$
computed using the revised effective uncertainties adopted in the SED
likelihood. The W3 and W4 measurements remain consistent with the
stellar photospheric contribution, while the $60\,\mu$m point represents
only a low-significance positive deviation. The disk model is
conditional and should not be interpreted as a unique determination of
the disk properties.}
    \label{fig:sed_model}
\end{figure}

The observed SED and the two model representations are shown in
Figure~\ref{fig:sed_model}. The MCMC parameter estimates are listed in Table~\ref{tab:sed_pars}, and their posterior distributions are
presented in Appendix~B, Figure~\ref{fig:corner_sed}. The physical interpretation and limitations of the conditional disk solution are discussed in Sects.~\ref{sec:dust_emission} and
\ref{sec:model_limitations}.

\subsection{\textbf{Sector-by-Sector Surface-Brightness Modelling}}
\label{sec:spot_evolution}

To characterize the changing out-of-eclipse modulations, we modelled the light curves from Sectors~11, 13, 38, 39, 65, 66, 93, and 94 separately. The orbital and stellar parameters were fixed to the reference solution derived from Sector~94
(Sect.~\ref{sec:lc_analysis}), because these parameters are not expected to vary over the time span of the observations. Only the parameters describing the surface-brightness inhomogeneities were adjusted between the individual data segments.

The same parametrized configuration was adopted for all segments, consisting of one circular surface-brightness region on the secondary
component and two regions on the primary. For each region, the spot-to-local-surface temperature ratio, colatitude, longitude, and angular radius were allowed to vary. The suffixes "--1" and "--2"
in Table~\ref{tab:spot_pars} denote the temporal subdivisions of the corresponding sectors. The observation baseline of \textit{TESS} is approximately 27 days per sector. Given the orbital period of CPD-82 291 of $\sim10$ days, each sector covers approximately two orbital cycles. We therefore analyzed the individual cycles within each sector independently.

The spot angular radius, $\alpha_{\rm spot}$, denotes the
angular radius of the circular cap used to represent an individual
surface-brightness region on the stellar surface; it is therefore an
angular size rather than a physical radius in units of $R_\odot$.
For an individual circular cap, the fraction of the total stellar
surface covered by the region is
$f_{\rm spot}=(1-\cos\alpha_{\rm spot})/2$. Thus, angular radii of
$65^\circ$--$82^\circ$ correspond to individual surface-covering
fractions of approximately 29--43\%. These fractions should not be
interpreted as a unique total spotted area when multiple regions are
present. The large fitted regions should instead be regarded as a
parametrization of the asymmetric surface-brightness distribution
rather than as uniquely determined individual physical spots.

The circular regions used in the model should be regarded as a parametrized representation of large-scale surface-brightness
inhomogeneities rather than as unique maps of individual physical
spots. Their fitted sizes, temperature contrasts, and positions are mutually degenerate and may represent combinations of multiple unresolved active regions.

The best-fitting parameters obtained for the individual segments are listed in Table~\ref{tab:spot_pars}, and their sector-to-sector variation is shown in Figure~\ref{fig:spot_evolution}. Representative model surfaces at four orbital phases are presented in Appendix~B, Figures~\ref{fig:mesh1}--\ref{fig:mesh2}. These surfaces illustrate the adopted parametrized solutions and should not be interpreted as uniquely reconstructed stellar maps.

The sector-specific solutions require different surface-brightness configurations to reproduce the observed out-of-eclipse asymmetries. The physical interpretation and limitations of these changes are discussed in Sect.~\ref{sec:surface_activity}.

\begin{table*}
\centering
\caption{Fitted spot parameters obtained from the sector-by-sector
light-curve solutions, illustrating the temporal changes in the
parametrized surface-brightness distributions. Mid-times are given as
$\mathrm{BJD}-2450000$. The tabulated quantities
are $T_{\rm rel}$, the spot-to-local-surface temperature ratio; Colat., the fitted spot colatitude; Long., the fitted spot longitude; and Rad., the angular radius of the circular cap adopted in the spot model. These parameters describe a non-unique representation of the observed light-curve asymmetries and should not be interpreted as direct measurements of the size, location, or evolution of individual physical
spots.}

\label{tab:spot_pars}
\scriptsize
\setlength{\tabcolsep}{3.5pt}
\renewcommand{\arraystretch}{1.15}

\begin{tabular}{l c cccc cccc cccc}
\hline
\hline
Sector & Mid-BJTD &
\multicolumn{4}{c}{Secondary spot} &
\multicolumn{4}{c}{Primary spot 1} &
\multicolumn{4}{c}{Primary spot 2} \\
\cline{3-6}\cline{7-10}\cline{11-14}
& &
$T_{\rm rel}$ & Colat. & Long. & Rad. &
$T_{\rm rel}$ & Colat. & Long. & Rad. & 
$T_{\rm rel}$ & Colat. & Long. & Rad. \\
& &
 & ($^\circ$) & ($^\circ$) & ($^\circ$) &
 & ($^\circ$) & ($^\circ$) &  ($^\circ$) & 
 & ($^\circ$) & ($^\circ$) & ($^\circ$)\\
\hline

Sector11-2 & 8613.61 & 0.9602 & 92.4601 & 195.1268 & 66.4772 & 0.9413 & 89.3431 & 189.4455 & 18.7151 & 1.3785 & 92.0224 & 174.5958 & 18.2822 \\ 
        Sector13-1 & 8664.63 & 0.9570 & 94.5761 & 190.9004 & 66.3762 & 0.9416 & 89.2880 & 189.4982 & 18.7085 & 1.3799 & 91.9399 & 173.5865 & 18.4206 \\ 
        Sector13-2 & 8674.84 & 0.9516 & 92.2384 & 187.3188 & 66.3807 & 0.9417 & 89.2708 & 189.4340 & 18.7021 & 1.3801 & 91.9368 & 173.6960 & 18.4200 \\ 
        Sector38-1 & 9338.18 & 0.9544 & 92.2557 & 238.7179 & 66.3370 & 0.9411 & 89.4494 & 249.5484 & 18.7081 & 1.3636 & 91.8946 & 233.6335 & 18.4076 \\ 
        Sector38-2 & 9358.59 & 0.9546 & 92.5193 & 245.6491 & 66.3677 & 0.9424 & 89.4187 & 249.4313 & 18.7073 & 1.3656 & 91.8978 & 235.4834 & 18.4081 \\ 
        Sector39-1 & 9368.79 & 0.9483 & 92.2652 & 246.4660 & 65.7037 & 0.9212 & 89.4030 & 251.1126 & 18.7621 & 1.3902 & 91.8030 & 235.7709 & 18.4009 \\ 
        Sector39-2 & 9379.00 & 0.9526 & 91.9191 & 246.4698 & 65.6879 & 0.9188 & 89.3107 & 249.0609 & 18.7070 & 1.3283 & 91.6347 & 232.7646 & 18.4475 \\ 
        Sector65-1 & 10072.95 & 0.9518 & 91.8628 & 203.2935 & 82.6160 & 0.9451 & 88.6403 & 260.0475 & 17.3238 & 1.4289 & 89.8332 & 158.5490 & 18.4915 \\ 
        Sector65-2 & 10093.37 & 0.9509 & 91.8863 & 212.3859 & 81.8096 & 0.9204 & 88.9056 & 259.7706 & 17.2947 & 1.4266 & 88.7743 & 157.5119 & 17.8060 \\ 
        Sector66-1 & 10103.57 & 0.9466 & 93.3694 & 207.1445 & 79.7251 & 0.9422 & 89.1131 & 258.0376 & 17.2638 & 1.3221 & 88.7337 & 157.4687 & 17.8828 \\ 
        Sector66-2 & 10113.77 & 0.9541 & 91.8508 & 212.5278 & 81.6854 & 0.9201 & 88.9371 & 259.7188 & 17.2761 & 1.3594 & 88.6679 & 157.6116 & 17.7987 \\ 
        Sector93-1 & 10838.36 & 0.9290 & 90.0129 & 137.9898 & 66.8479 & 0.9399 & 89.0158 & 212.0375 & 47.7732 & 0.8250 & 90.0095 & 77.2543 & 18.2667 \\ 
        Sector93-2 & 10848.56 & 0.9271 & 90.8100 & 137.8200 & 66.8600 & 0.9400 & 88.9300 & 212.5000 & 46.5800 & 0.8249 & 90.2300 & 77.3000 & 18.1500 \\ 
        Sector94-1 & 10858.77 & 0.9413 & 90.0122 & 138.0094 & 66.9009 & 0.9400 & 89.0113 & 212.0173 & 46.5984 & 0.8250 & 90.0013 & 77.0279 & 18.1516 \\ 
        Sector94-2 & 10879.18 & 0.9374 & 90.0356 & 137.9923 & 66.9174 & 0.9403 & 89.0078 & 212.0714 & 46.5678 & 0.8249 & 90.0100 & 77.0275 & 18.1542 \\ \hline

\hline
\end{tabular}
\end{table*}

\begin{figure*}
\centering
\includegraphics[width=0.95\textwidth]{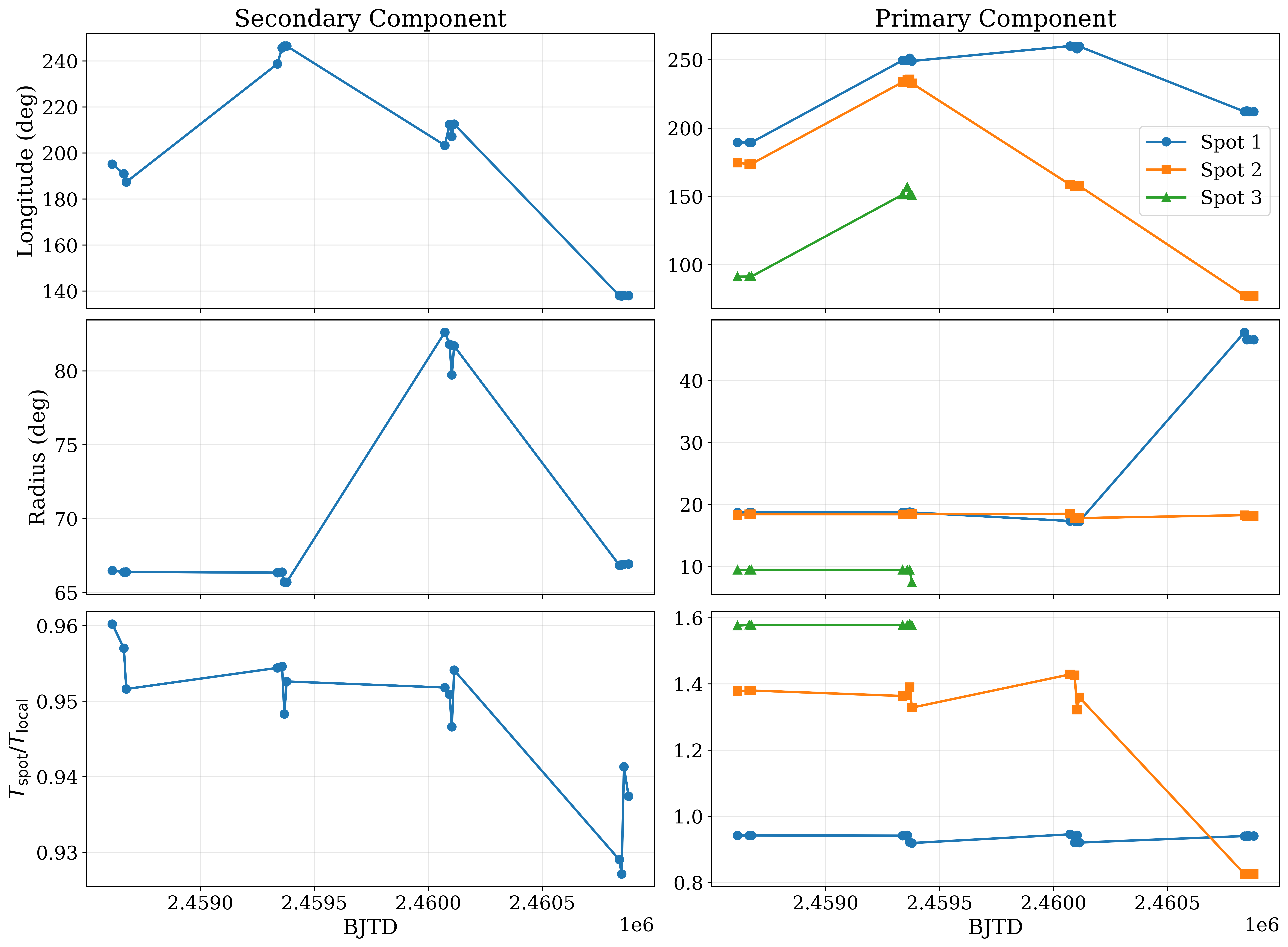}
\caption{Temporal evolution of the spot parameters obtained from the sector-by-sector light-curve solutions. The left and right columns show the spot parameters for the secondary and primary components, respectively. From top to bottom, the panels show the spot longitude, angular radius, and relative temperature ($T_{\rm spot}/T_{\rm local}$). The variations indicate that the surface activity pattern changes between different \textit{TESS} observing epochs.}
\label{fig:spot_evolution}
\end{figure*}


\subsection{Flare Identification and Measurement}

The flare events were identified and analyzed using an interactive procedure applied to the \textit{TESS} light curves. First, a baseline model was constructed by selecting flare-free regions and fitting a low-order polynomial to remove long-term trends. For each candidate event, the selected flare-free regions surrounding the event were used to represent the local quiescent baseline, $F_q$. The detrended light curve was then expressed in terms of the relative excess flux,

\begin{equation}
    \frac{F-F_q}{F_q} = \frac{F}{F_q}-1,
\end{equation}

\noindent thereby removing the local brightness variation while preserving the fractional flare amplitude.

\begin{figure*}[htbp!]
    \centering
    \includegraphics[width=\textwidth]{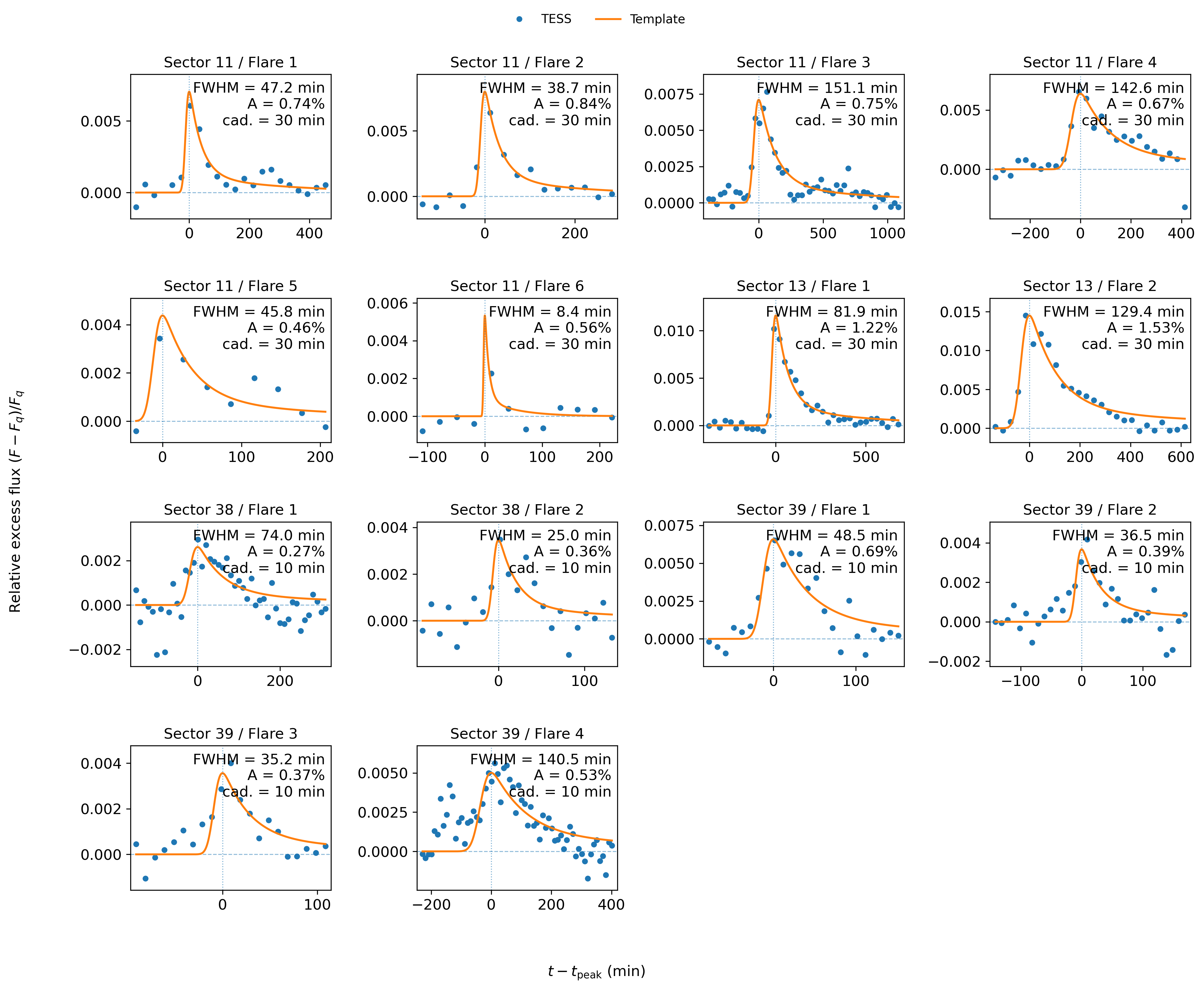}
    \caption{Relative excess-flux profiles of the first 14 flare candidates
    identified in the \textit{TESS} observations of CPD--82 291. Points show the
    detrended \textit{TESS} measurements, and the solid curves show the best-fitting
    continuous analytic flare templates. Time is expressed relative to the
    fitted template peak. The different sampling densities reflect the
    cadence of the corresponding \textit{TESS} observations.}
    \label{fig:flare_fits1}
\end{figure*}

\begin{figure*}[htbp!]
    \centering
    \includegraphics[width=\textwidth]{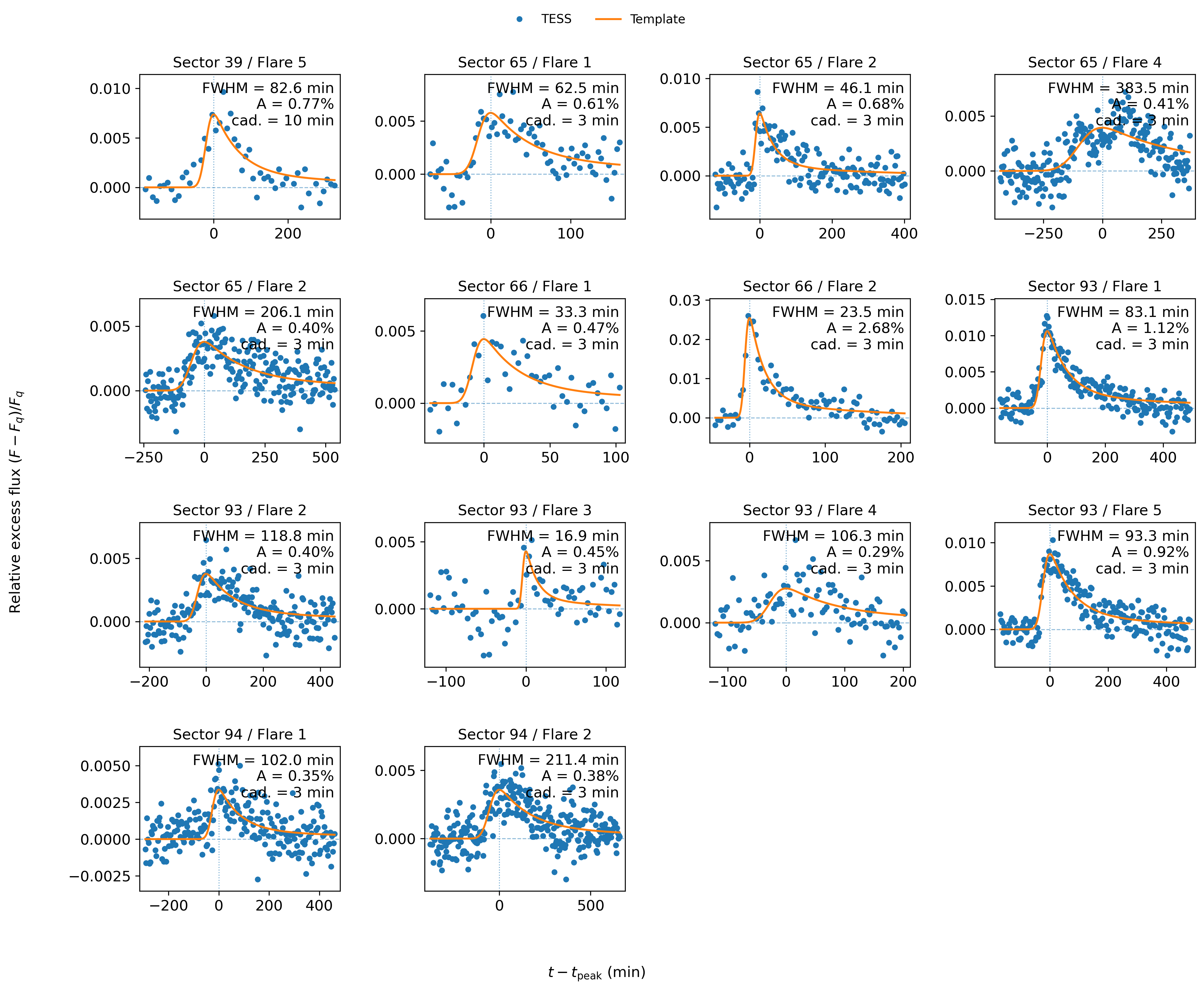}
    \caption{Same as Figure~\ref{fig:flare_fits1}, but for the remaining
    14 flare candidates identified in the \textit{TESS} observations of CPD--82 291.}
    \label{fig:flare_fits2}
\end{figure*}

Flare candidates were selected manually in the detrended data, ensuring that both the rise and decay phases were included. The selected windows were sufficiently wide to include the rise, peak, and decay portions of each event. The start and end times of each flare were determined using a threshold-based approach, with the local noise level, $\sigma$, estimated from the relative residual flux in the selected flare-free baseline regions. The first and last points exceeding $2.5\sigma$ within the candidate window were adopted as the flare start and end times, respectively. We therefore refer to the corresponding interval as the threshold-defined duration, rather than as an intrinsic physical flare duration.

The flare peak was identified as the maximum residual flux within the selected region. The observed flare amplitude was defined as the maximum relative excess flux. The flare duration was defined as the time interval between the start and end points, while the equivalent duration was calculated by integrating the flare excess flux above the baseline. Specifically, the equivalent duration was calculated directly from the observed relative excess flux as

\begin{equation}
    {\rm ED} =
    \int \frac{F-F_q}{F_q}\,dt ,
\end{equation}

\noindent where the integration was performed over the manually selected flare interval.

Rather than approximating the flare profiles with symmetric Gaussian functions, we fitted the continuous analytic white-light flare template of \cite{TovarMendoza2022}. The model describes the characteristic rapid heating and asymmetric cooling morphology of classical stellar flares and is parameterized by the flare peak time, amplitude, and full width at half maximum (FWHM). To account for the finite integration time of the \textit{TESS} observations, the analytic profile was evaluated at a higher temporal resolution and subsequently averaged to the cadence of the corresponding light curve before comparison with the observations. This treatment is particularly relevant for the sectors obtained at relatively long cadence.

The equivalent duration obtained by integrating the best-fitting analytic template was also retained as a model-dependent diagnostic. For events whose morphology departs from that of a single classical flare, the directly integrated equivalent duration provides the less model-dependent measurement.

Figures~\ref{fig:flare_fits1} and~\ref{fig:flare_fits2} show the relative excess-flux profiles and best-fitting analytic templates for all 28 flare events identified in the \textit{TESS} observations.

A total of 28 flare events were retained in the final sample. All measured parameters, including start time, peak time, threshold-defined duration, amplitude, equivalent duration, and the peak time, amplitude, and FWHM obtained from the analytic flare-template fit, were compiled into a flare catalog for further statistical analysis.

\section{Results and Discussion}
\subsection{Binary Configuration and Model-Inferred Absolute Parameters}
\label{sec:binary_configuration}

The reference light-curve solution derived from the
\textit{TESS} Sector~94 photometry constrains the orbital inclination, volume-equivalent fractional radii, component temperature ratio, luminosity ratio, and photometric mass ratio. These quantities describe the relative geometry and surface-brightness distribution of the system, but the light curve alone does not determine its absolute
physical scale.

The absolute radius scale was obtained by combining the light-curve constraints with the primary radius derived from the SED analysis using
the adopted \textit{Gaia} distance. Because no radial-velocity orbit is available, the resulting component masses and orbital dimensions are not direct dynamical measurements. They should instead be regarded as
model-inferred quantities that depend on the SED radius scale, the light-curve geometry, and the photometric mass ratio.

The light-curve solution provides the volume-equivalent fractional
radii,

\begin{equation}
r_1=\frac{R_1}{a},
\qquad
r_2=\frac{R_2}{a},
\end{equation}

where $a$ is the orbital semi-major axis. The semi-major axis was therefore calculated from the primary radius as

\begin{equation}
a=\frac{R_1}{r_1}.
\label{eq12}
\end{equation}

The secondary radius was independently obtained from the light-curve radius ratio,

\begin{equation}
R_2=R_1\frac{r_2}{r_1}.
\label{eq13}
\end{equation}

The total system mass was then inferred from Kepler's third law,

\begin{equation}
M_{\rm tot}=M_1+M_2
=\frac{4\pi^2a^3}{GP^2},
\label{eq14}
\end{equation}

and the individual component masses were calculated using the
photometric mass ratio,
$q_{\rm phot}=M_2/M_1$:

\begin{equation}
M_1=\frac{M_{\rm tot}}{1+q_{\rm phot}},
\qquad
M_2=\frac{q_{\rm phot}M_{\rm tot}}{1+q_{\rm phot}}.
\label{eq15}
\end{equation}

The corresponding surface gravities were calculated from

\begin{equation}
g_j=\frac{GM_j}{R_j^2},
\qquad j=1,2.
\label{eq16}
\end{equation}

Combining Eqs.~(\ref{eq12})--(\ref{eq15}), the dependence of
the inferred component masses on the directly adopted quantities can
be written explicitly as

\begin{equation}
M_1=
\frac{4\pi^2}{GP^2}
\left(\frac{R_1}{r_1}\right)^3
\frac{1}{1+q_{\rm phot}},
\label{eq17}
\end{equation}

\begin{equation}
M_2=
\frac{4\pi^2}{GP^2}
\left(\frac{R_1}{r_1}\right)^3
\frac{q_{\rm phot}}{1+q_{\rm phot}}.
\label{eq18}
\end{equation}

Uncertainties in the absolute radii, semi-major axis, component
masses, and surface gravities were propagated through Monte Carlo
sampling of the posterior distributions of $R_1$, $r_1$, $r_2$, and
$q_{\rm phot}$, together with the orbital-period uncertainty. For each
Monte Carlo realization, $a$, $R_2$, $M_{\rm tot}$, $M_1$, $M_2$,
$g_1$, and $g_2$ were recalculated sequentially using
Eqs.~(\ref{eq12})--(\ref{eq16}). Full parameter vectors were drawn
from the corresponding SED and light-curve posterior samples so that
correlations among parameters within each analysis were retained.
Because the light-curve and SED analyses were performed separately,
no additional covariance between the two posterior samples was
assumed. The reported values and uncertainties were obtained from the
median and the 16th and 84th percentiles of the resulting derived
distributions.

Using the current posterior solution, the component radii are
$R_1=1.245_{-0.087}^{+0.094}\,R_{\odot}$ and
$R_2=3.70_{-0.27}^{+0.29}\,R_{\odot}$, and the inferred orbital
semi-major axis is

\begin{equation}
a=21.95_{-2.32}^{+2.69}\,R_{\odot}
\simeq0.1021_{-0.0108}^{+0.0125}~{\rm AU}.
\end{equation}

The corresponding model-inferred total mass is

\begin{equation}
M_{\rm tot}=1.36_{-0.39}^{+0.57}\,M_{\odot},
\end{equation}

and the component masses are

\begin{equation}
M_1=0.93_{-0.27}^{+0.39}\,M_{\odot},
\qquad
M_2=0.43_{-0.12}^{+0.18}\,M_{\odot},
\end{equation}

with surface gravities

\begin{equation}
\log g_1=4.21_{-0.11}^{+0.12},
\qquad
\log g_2=2.94_{-0.12}^{+0.12},
\end{equation}

in cgs units.

The uncertainties in the masses are substantially larger than those in
the fractional light-curve parameters because the total mass scales as
$M_{\rm tot}\propto a^3$. Consequently, uncertainties or systematic
offsets in the SED-derived radius scale are amplified when propagated
to the component masses. The mass estimates also depend directly on
$q_{\rm phot}$, which is model-dependent in this detached and strongly
active system.

The derived radii, masses, and surface gravities are therefore useful
for assessing the possible evolutionary state of the components, but
they should not be treated as precise mass--radius benchmarks. A
double-lined spectroscopic orbit is required to determine the dynamical
mass ratio and orbital scale and to test the reliability of the
model-inferred absolute parameters. The evolutionary implications of
the current solution are considered in
Sect.~\ref{sec:comparison_systems}.

\subsection{Evolutionary status and comparison with similar systems}
\label{sec:comparison_systems}

\begin{figure*}[htbp!]
    \centering \includegraphics[width=0.75\linewidth]{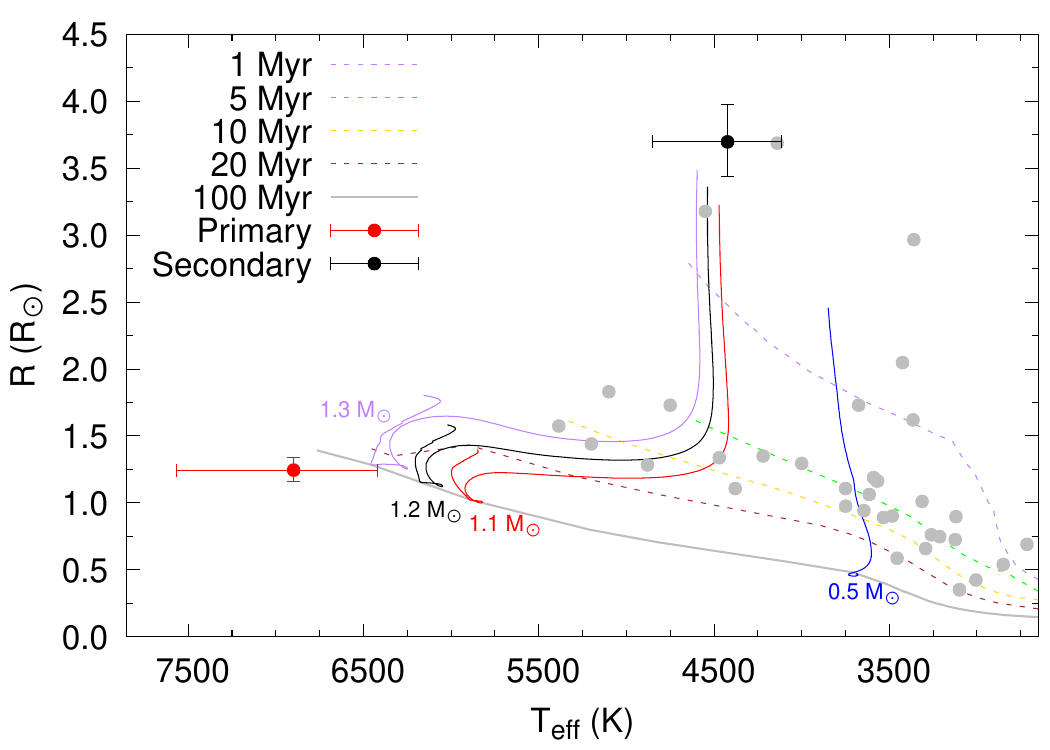}
\caption{
Positions of the components of CPD--82~291 in the
$T_{\rm eff}$--$R$ plane, together with the pre-main-sequence evolutionary tracks and isochrones of \citet{Baraffe2015}. The primary
and secondary components are shown with black and red symbols,
respectively. Components of the comparison eclipsing binaries listed in Table~\ref{tab:pms_eb_comparison} are shown as grey circles. The
primary lies close to the ZAMS region, whereas the secondary is located well above the main-sequence locus. Because the absolute parameters of CPD--82~291 are model-inferred and are not yet supported by a
spectroscopic orbit, the diagram illustrates possible evolutionary
interpretations rather than establishing a unique pre-main-sequence age or coeval solution.
}\label{fig:evolution}
\end{figure*}

The model-inferred stellar parameters derived in
Sect.~\ref{sec:binary_configuration} were compared with the
pre-main-sequence evolutionary tracks and isochrones of
\citet{Baraffe2015} in the $T_{\rm eff}$--$R$ plane
(Figure~\ref{fig:evolution}). Because the absolute dimensions of
CPD--82~291 are not yet supported by a spectroscopic orbital solution,
this comparison is intended to explore possible evolutionary
interpretations rather than to establish a unique age or evolutionary
state.

The primary component lies close to the zero-age main-sequence region. Within the current uncertainties, its position may be consistent either with a star approaching the main sequence or with one that has already reached the ZAMS. Its location therefore does not independently demonstrate a
pre-main-sequence evolutionary state.

The secondary component occupies a markedly different region of the
diagram. Its comparatively low effective temperature, large inferred radius, and low surface gravity place it well above the main-sequence locus. If the adopted absolute radius scale is approximately correct,
this position could be consistent with a low-mass star at an early
evolutionary stage. However, the two components do not follow a single well-defined coeval isochrone in the adopted evolutionary models.

Such an apparent discrepancy could be affected by magnetic activity, extensive surface-brightness inhomogeneities, and limitations of standard non-magnetic evolutionary tracks. Starspots may alter the
observed effective temperatures and luminosity distribution and can contribute to apparently enlarged radii in active low-mass stars. Nevertheless, activity alone cannot be assumed to explain the position
of the secondary. Alternative possibilities include systematic errors in the model-inferred absolute scale, a more evolved secondary, or a previous episode of binary interaction. The available photometry does
not distinguish among these interpretations.

\begin{table*}
\centering
\caption{Comparison of the inferred parameters of CPD--82 291 with selected pre-main-sequence eclipsing-binary systems from the literature.}
\label{tab:pms_eb_comparison}

\scriptsize
\setlength{\tabcolsep}{3.5pt}
\renewcommand{\arraystretch}{1.15}

\begin{tabular}{lccccccc}
\hline
System & $P$ (d) & $M_1$ & $M_2$ & $R_1$ & $R_2$ & RV basis & Ref. \\
       &         & ($M_\odot$) & ($M_\odot$) &
       ($R_\odot$) & ($R_\odot$) & & \\
\hline

CPD--82 291 &
10.205 &
$1.13\pm0.36$ &
$0.53\pm0.17$ &
$1.31\pm0.08$ &
$3.88\pm0.25$ &
No RV orbit & This work \\

\hline

2MASS J05352184-0546085 &
9.780 &
$0.0572\pm0.0033$ &
$0.0366\pm0.0022$ &
$0.690\pm0.011$ &
$0.540\pm0.009$ & SB2 &
(1) \\

JW 380 &
5.299 &
$0.262\pm0.025$ &
$0.151\pm0.013$ &
$1.189\pm0.175$ &
$0.897\pm0.170$ & SB2 &
(1,2) \\

Par 1802 &
4.674 &
$0.391\pm0.032$ &
$0.385\pm0.032$ &
$1.73\pm0.02$ &
$1.62\pm0.02$ & SB2 &
(1,2) \\

CoRoT 223992193 &
3.875 &
$0.668\pm0.012$ &
$0.4953\pm0.0073$ &
$1.295\pm0.040$ &
$1.107\pm0.050$ & SB2 &
(1,2) \\

V1174 Ori &
2.635 &
$1.006\pm0.013$ &
$0.7271\pm0.0096$ &
$1.338\pm0.011$ &
$1.063\pm0.011$ & SB2 &
(1,2) \\

RX J0529.4+0041A &
3.038 &
$1.27\pm0.01$ &
$0.93\pm0.01$ &
$1.44\pm0.10$ &
$1.35\pm0.10$ & SB2 &
(1,2) \\

ASAS J0528+03 &
3.873 &
$1.375\pm0.028$ &
$1.329\pm0.020$ &
$1.83\pm0.07$ &
$1.73\pm0.07$ & SB2 &
(1,2) \\

MML 53 &
2.098 &
$1.0400\pm0.0067$ &
$0.8907\pm0.0058$ &
$1.283\pm0.043$ &
$1.107\pm0.049$ & SB2 &
(3) \\

Mon--735 &
1.975 &
$0.2918\pm0.0099$ &
$0.2661\pm0.0095$ &
$0.762\pm0.022$ &
$0.748\pm0.023$ & SB2 &
(4) \\

THOR--42 &
0.859 &
$0.497\pm0.005$ &
$0.205\pm0.002$ &
$0.659\pm0.002$ &
$0.424\pm0.002$ & Two-component RV$^\dagger$ &
(5) \\

USco 48 &
2.874 &
$0.738\pm0.035$ &
$0.709\pm0.035$ &
$1.164\pm0.051$ &
$1.164\pm0.051$ & SB2 &
(6) \\

TOI--450 &
10.715 &
$0.1768\pm0.0004$ &
$0.1767\pm0.0003$ &
$0.351\pm0.003$ &
$0.351\pm0.004$ & SB2 &
(7) \\

2M1222--57 &
3.072 &
$0.7354\pm0.0057$ &
$0.6680\pm0.0044$ &
$0.976\pm0.013$ &
$0.942\pm0.015$ & SB2 &
(8) \\

2M05--06 &
1.704 &
$0.424\pm0.023$ &
$0.522\pm0.028$ &
$0.902\pm0.023$ &
$0.891\pm0.024$ & SB2 &
(9) \\

2M05--00 &
1.817 &
$0.348\pm0.021$ &
$0.339\pm0.019$ &
$1.011\pm0.093$ &
$0.725\pm0.077$ & SB2 &
(9) \\

MML 48 &
2.017 &
$1.2\pm0.07$ &
$0.2509\pm0.0078$ &
$1.574\pm0.054$ &
$0.587\pm0.051$ & SB1 + model &
(10) \\

\hline
\end{tabular}

\vspace{0.15cm}

\raggedright
\scriptsize
References:
(1) \citet{Stassun2014};
(2) \citet{Gillen2017};
(3) \citet{GomezMaqueoChew2019};
(4) \citet{Gillen2020};
(5) \citet{Murphy2020};
(6) \citet{David2019};
(7) \citet{Tofflemire2023};
(8) \citet{Stassun2022};
(9) \citet{KounkelStassun2024};
(10) \citet{MML48_2025}.

\par\medskip
\textit{Note.} The ``RV basis'' column identifies the
spectroscopic information entering each solution. ``SB2'' denotes a
double-lined spectroscopic eclipsing binary for which radial velocities
of both eclipsing components are available. For THOR--42
($^\dagger$), the primary velocity was measured from photospheric
absorption lines whereas the secondary velocity was obtained from
resolved H$\alpha$ emission. MML~48 is an SB1 eclipsing binary, and its
primary mass includes stellar-model information. CPD--82~291 has no
spectroscopic orbital solution; its absolute dimensions are
model-inferred from the light-curve and SED/Gaia scaling, together
with the photometric mass ratio. Accordingly, the systems in this
table are shown for qualitative evolutionary comparison and should
not be interpreted as a methodologically homogeneous dynamical
mass--radius sample.

\end{table*}

This comparison is not methodologically homogeneous.
The spectroscopic basis of each literature solution is therefore
identified explicitly in Table~\ref{tab:pms_eb_comparison}. Most of
the comparison systems are double-lined spectroscopic eclipsing
binaries (SB2 EBs), for which both component radial-velocity curves,
together with the eclipse light curves, provide dynamical masses.
Exceptions are indicated separately in the table; in particular,
THOR--42 has a two-component velocity solution in which the secondary
velocities were obtained from H$\alpha$ emission, while MML~48 is a
single-lined spectroscopic eclipsing binary whose primary mass was
estimated using stellar models. In contrast, no spectroscopic orbit is
currently available for CPD--82~291. Its absolute radius scale is
inferred from the light-curve and SED solutions using the Gaia
distance, and its masses are subsequently estimated from the inferred
orbital scale and the photometric mass ratio. The comparison should
therefore be regarded as qualitative evolutionary context rather than
as a homogeneous dynamical mass--radius benchmark.

The primary component lies within the range occupied by several young eclipsing-binary components of comparable mass and radius. Its
properties are compatible with a star approaching the main sequence, but this agreement does not independently establish its youth. The secondary is considerably more unusual. Young systems such as Par~1802, JW~380, and CoRoT~223992193 demonstrate that low-mass pre-main-sequence stars can possess radii substantially larger than
their main-sequence counterparts. However, the inferred secondary
radius of CPD--82~291 is larger than those of most comparison objects, and unlike the literature values, it has not been validated by a dynamical orbital solution.

The strong and changing out-of-eclipse modulation supports the
presence of substantial surface activity. Such activity may contribute to the disagreement between the inferred component properties and a single coeval non-magnetic isochrone. However, the displacement of the
secondary should not be interpreted uniquely as evidence of pre-main-sequence radius inflation because the absolute scale remains model-dependent.

CPD--82~291 also has a relatively long orbital period of
$P=10.20514$~d compared with many known pre-main-sequence eclipsing
binaries. If its youth were independently confirmed, it would represent an interesting long-period system for testing early stellar evolution and magnetic activity. At present, however, its Gaia astrometry is
inconsistent with physical membership in Chamaeleon~I, and no suitable spectroscopy is available to test lithium absorption, H$\alpha$
emission, surface-gravity indicators, or the systemic radial velocity. CPD--82~291 should therefore not yet be regarded as a secure member of
the pre-main-sequence eclipsing-binary population.


\subsubsection{Possible far-infrared dust emission}
\label{sec:dust_emission}

The stellar-only model reproduces the observed SED from the optical
region through the WISE W4 band. In particular, the W3 and W4
measurements differ from the stellar-only predictions by only
$-0.39\sigma$ and $-0.37\sigma$, respectively. The available
optical-to-mid-infrared photometry therefore provides no evidence for a
statistically significant infrared excess at wavelengths up to
approximately \textbf{$22\,\mu$m}.

The only positive long-wavelength deviation is the $60\,\mu$m
measurement reported by \citet{Spangler2001},
$F_{60}=0.026\pm0.019$~Jy, corresponding to a raw significance
of only $1.37\sigma$ and lying $1.30\sigma$ above the stellar-only
prediction when the adopted systematic uncertainty is included.
This value should therefore not be regarded as a secure far-infrared
detection. The corresponding $90\,\mu$m result is a non-detection and
was incorporated only as a one-sided upper limit. Moreover, the
\textit{Spitzer}/FEPS observations showed no significant excess at
16, 24, or $70\,\mu$m. Thus, the possible ISO $60\,\mu$m signal has
not been independently confirmed by later mid- or far-infrared
measurements.

Under the conditional assumption that the low-significance
$60\,\mu$m emission is physically associated with CPD--82~291, it can
be reproduced by a cool circumbinary-dust model. In the adopted
parametrization, the converged MCMC posterior gives
characteristic inner and outer radii of
$R_{\rm in}=147_{-48}^{+37}$~AU and
$R_{\rm out}=188_{-52}^{+150}$~AU.
The corresponding dust mass is
$\left(8.1_{-5.7}^{+16.0}\right)\times10^{-5}\,M_\odot$.
This quantity refers specifically to the dust component; no
gas-to-dust ratio has been assumed and no total gas+dust disk mass is
inferred from the present data. The inferred radii and dust mass are
conditional, model-dependent quantities rather than direct
measurements of the circumstellar material.

The inferred radial extent and dust mass are strongly model-dependent.
The disk is spatially unresolved; its inclination was fixed to the
binary inclination under an assumed coplanar configuration, and the inferred surface-density normalization depends strongly
on the assumed inclination. The outer radius is also poorly
constrained by the limited far-infrared information. Consequently,
comparisons with resolved disk radii or dynamically characterized
Class~II disks are not methodologically equivalent, and the fitted
mass and radial extent should not be used to classify the system as
possessing a typical Class~II disk.

The large fitted inner radius should also be considered in
relation to the binary orbit. The model-inferred binary semimajor axis
is approximately $0.102$~AU, whereas the conditional disk solution
places the inner edge at $R_{\rm in}\simeq147$~AU, giving

\begin{equation}
\frac{R_{\rm in}}{a_{\rm binary}}
\simeq 1.4\times10^{3}.
\end{equation}

This scale is far larger than the cavity sizes of a few binary
semimajor axes normally associated with tidal truncation by a central
binary \citep{Artymowicz1994}. We therefore do not interpret the fitted
inner radius as evidence for a cavity dynamically cleared by the
present binary. Given the weak far-infrared constraint and the model
dependence of the inferred disk parameters, the present data cannot
establish the physical origin of such a large characteristic inner
radius.

The radial-emission diagnostics shown in Figure~\ref{fig:sed_summary}
illustrate the temperature profile and wavelength-dependent emitting
regions predicted by the conditional disk model. They are entirely
model-derived and do not represent spatially resolved observations.
In particular, they should not be used as independent evidence for the
inferred disk geometry or radial extent.

The spectral index measured between 2 and $24\,\mu$m is
$\alpha=-2.68\pm0.05$. According to the classification boundaries
adopted by \citet{Grossschedl2019}, this value lies in the
Class~III/no-IR-excess regime. In the present case, however, the
evolutionary status of the source is not independently established and
the WISE and \textit{Spitzer} photometry is consistent with
photospheric emission. We therefore use $\alpha$ only as a descriptive
measure of the optical-to-mid-infrared SED slope. It supports the
absence of a strong warm inner-disk contribution but does not establish
the presence of an evolved circumbinary disk.

Taken together, the available observations do not securely demonstrate
that CPD--82~291 hosts a gravitationally bound circumbinary disk. The
adopted cool-dust model provides one possible representation of the
tentative far-infrared signal if that signal is physically associated
with the binary. Confirmation requires new far-infrared or millimetre
observations with substantially improved sensitivity and angular
resolution.

\subsection{Surface and flare activity}
\label{sec:surface_activity}

The sector-specific models show that the out-of-eclipse modulation
cannot be represented by a single static surface-brightness
configuration. Broad cool regions assigned to the secondary provide an effective description of the dominant modulation, while different cool and hot regions are required on the primary at different epochs. This
supports the general conclusion that CPD--82~291 exhibits substantial surface activity.

The fitted parameters are generally more similar in temporally adjacent
sectors than between epochs separated by several years, which is consistent with short-term persistence and longer-term changes in the overall activity pattern. However, the large fitted angular radii of
approximately $65^\circ$--$82^\circ$ should not be interpreted as the
sizes of single physical spots. Their fitted dimensions and positions are degenerate with the adopted temperature contrast, viewing geometry, luminosity ratio, and number of surface-brightness regions,
and may represent combinations of multiple unresolved active
structures.

Likewise, changes in the fitted longitudes do not demonstrate the
migration of the same physical spot, and the fitted colatitudes near $90^\circ$ do not establish an equatorial concentration of activity. The robust result is therefore the changing large-scale
surface-brightness distribution rather than the detailed geometry,
latitude, lifetime, or migration of individual fitted regions.

The changing surface-brightness distribution inferred from the
light-curve modelling is accompanied by recurrent flare activity in the
\textit{TESS} photometry. A total of 28 flare events were retained in the
final sample. Because the temporal sampling differs substantially among the
observed sectors, ranging from approximately 30 min to 3.33 min, the flare
timescales were interpreted with explicit reference to the cadence of the
corresponding data. One event in Sector~11 was detected above the adopted
threshold in only a single cadence and therefore did not provide a resolved
threshold-defined duration or FWHM. Consequently, the duration and FWHM
statistics below refer to the remaining 27 temporally resolved events, whereas
the amplitude and equivalent-duration statistics include all 28 flares.

The threshold-defined durations span approximately
$10$--$620$ min, with a median of $150.0$ min and a mean of
$218.6$ min. The corresponding analytic-template FWHM values range from
$16.9$ to $383.5$ min, with a median of $81.9$ min and a mean of
$95.0$ min. These quantities should not be interpreted as exact intrinsic
flare timescales, particularly for events sampled at the longer \textit{TESS}
cadences, but they provide a consistent characterization of the observed
flare profiles across the available sectors.

\begin{figure*}[htbp!]
    \centering
    \includegraphics[width=\linewidth]{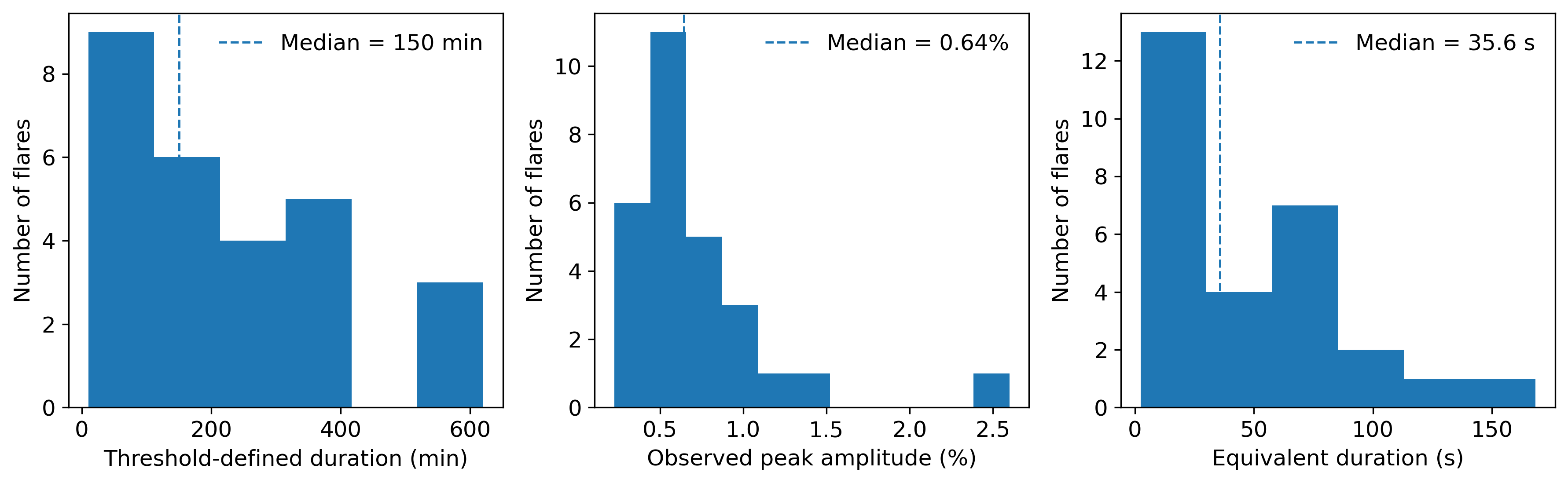}
    \caption{Statistical properties of the 28 flare events retained
    in the final sample. Left: distribution of the threshold-defined
    durations for the 27 events with resolved start and end times.
    Middle: distribution of the observed peak amplitudes for all 28 events.
    Right: distribution of the directly integrated equivalent durations.
    Dashed vertical lines indicate the corresponding median values.}
    \label{fig:flare_stats}
\end{figure*}

The observed peak amplitudes are generally modest. The median
relative flux enhancement is $0.643\%$, while the largest event reaches
$2.60\%$. The directly integrated equivalent durations range from
$2.39$ to $168.37$ s, with a median of $35.65$ s and a mean of
$47.92$ s. The distributions of threshold-defined duration, observed peak
amplitude, and directly integrated equivalent duration are shown in
Figure~\ref{fig:flare_stats}.

Within the detected flare sample, the equivalent duration is positively correlated with both the observed flare timescale and the
peak amplitude. For the directly integrated equivalent duration and threshold-defined duration, the Spearman rank coefficient is $\rho_{\rm S}=0.718$ with a 95\% bootstrap confidence interval of
$[0.433,0.865]$ ($p=2.5\times10^{-5}$; $N=27$).
A similar correlation is obtained between equivalent duration and the template FWHM, with $\rho_{\rm S}=0.689$ and a 95\% confidence interval of $[0.410,0.871]$
($p=7.0\times10^{-5}$; $N=27$). The correlation between equivalent duration and observed peak amplitude is comparably strong,
$\rho_{\rm S}=0.720$, with a 95\% confidence interval of
$[0.450,0.854]$ ($p=1.5\times10^{-5}$; $N=28$).
In contrast, the relation between peak amplitude and threshold-defined
duration is weaker and is not statistically significant,
with $\rho_{\rm S}=0.322$ and a 95\% confidence interval of
$[-0.089,0.634]$ ($p=0.101$; $N=27$). Thus, the present sample does not indicate that event duration is the dominant factor associated with
flare equivalent duration; instead, both flare amplitude and temporal extent show comparably strong correlations with the observed equivalent
duration. These relations are descriptive of the detected events and should not be interpreted as completeness-corrected population
relations.

\begin{figure*}[htbp!]
    \centering
    \includegraphics[width=\linewidth]{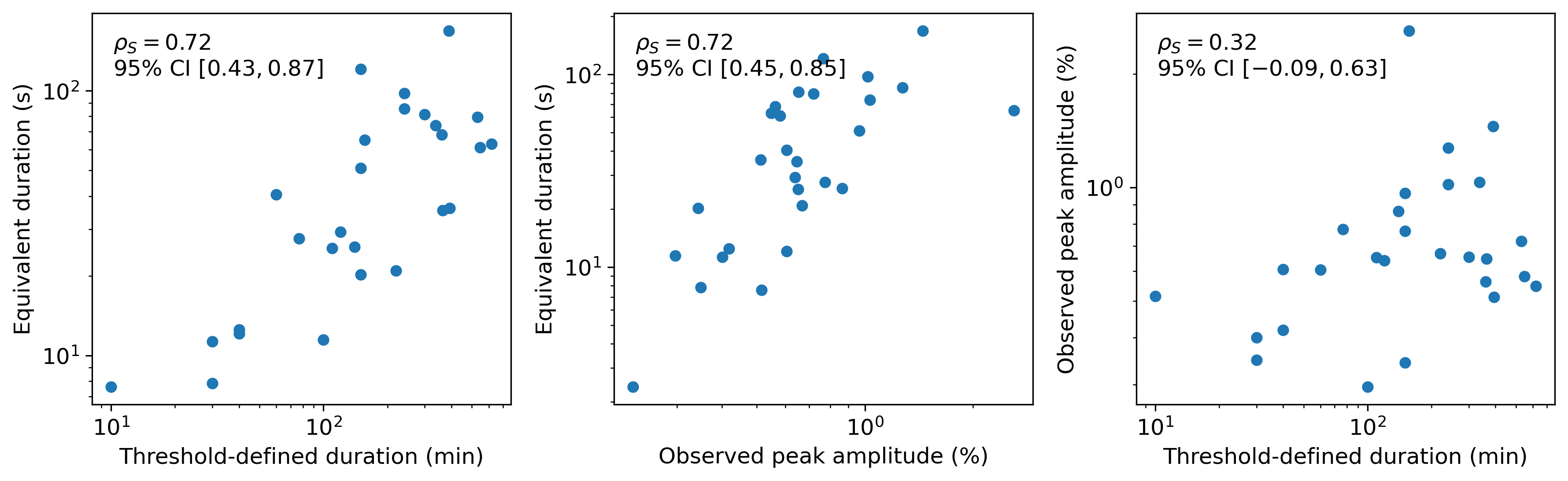}
    \caption{Relations among the measured flare properties.
    Left: directly integrated equivalent duration versus threshold-defined
    duration. Middle: equivalent duration versus observed peak amplitude.
    Right: observed peak amplitude versus threshold-defined duration.
    The Spearman rank coefficient, $\rho_{\rm S}$, is indicated in each
    panel.}
    \label{fig:flare_correlations}
\end{figure*}

The equivalent durations obtained from the analytic flare templates
closely track those measured directly from the detrended data. The two
quantities show a very strong rank correlation
($\rho_{\rm S}=0.976$, $p=7.6\times10^{-19}$), demonstrating that the
adopted analytic profiles reproduce the overall flare-strength ordering of
the sample. The template-derived equivalent durations are nevertheless
systematically larger, with a median ratio
$ED_{\rm model}/ED_{\rm data}=1.44$. This difference is expected because
the analytic profiles include extended decay tails beyond the portion of the
event directly sampled within the manually selected integration interval.
For this reason, the directly integrated equivalent duration is adopted as
the primary observational measure in the statistical analysis, while the
template-derived value is retained as a model-dependent diagnostic.

The number of identified events varies among the individual
\textit{TESS} sectors, from two to six flares per sector. These raw counts
should not, however, be interpreted directly as changes in the intrinsic
flare rate. The sectors differ both in temporal sampling and in usable
observing duration, and the detectability of low-amplitude or short-duration
events is correspondingly cadence dependent. A quantitative comparison of
flare occurrence rates between sectors would therefore require an explicit
completeness analysis and normalization by the effective observing time. We
accordingly restrict the present discussion to the properties of the detected
flare sample rather than interpreting the sector-to-sector counts as evidence
for long-term changes in the magnetic activity level.

Taken together, the evolving large-scale surface-brightness
distribution and the recurrent flare detections establish CPD--82~291 as an
active eclipsing system. However, neither the parametrized surface-brightness
regions nor the flare measurements uniquely identify the active component,
the physical sizes of individual magnetic structures, or the presence of
persistent active longitudes or magnetic cycles. The robust observational
result is therefore the coexistence of evolving photospheric modulation and
recurrent flare activity over multiple \textit{TESS} observing epochs.

\subsection{Model limitations and future observational tests}
\label{sec:model_limitations}

The posterior distributions of the SED parameters reveal
significant degeneracies within the adopted model. The strongest
correlations are found between the stellar effective temperature and
stellar radius ($r=-0.82$), and between the stellar effective
temperature and reddening ($r=+0.93$). Such behaviour is
expected in broadband SED modelling because an increase in effective
temperature can be partly compensated by either a smaller stellar
radius or a larger reddening, producing a similar optical continuum
shape. A weaker anti-correlation is present between the outer disk
radius and the surface-density normalization
($r=-0.39$), reflecting the partial degeneracy between the
radial extent and dust content of the conditional disk model.

The stellar parameters are better constrained than the
circumstellar parameters because the optical-to-mid-infrared photometry
provides substantially greater leverage on the stellar contribution,
while the light-curve ratios and \textit{Gaia} distance further reduce
the allowed stellar parameter space. Significant covariance nevertheless
remains, particularly among $T_{\star,1}$, $R_{\star,1}$, and
$E(B-V)$. The stellar parameters obtained from the converged
SED posterior,
$T_{\star,1}=6898_{-480}^{+670}$~K and
$R_{\star,1}=1.245_{-0.087}^{+0.094}\,R_{\odot}$,
are consistent with the observed multi-wavelength photometry under the
adopted light-curve constraints on the temperature and radius ratios.
As an external comparison, the StarHorse estimate \citep{Anders2022} gives
$A_V=0.55^{+0.01}_{-0.07}$~mag
[$E(B-V)=0.177^{+0.004}_{-0.024}$ for $R_V=3.1$],
which is consistent, within the broad posterior uncertainties, with our
SED result of $E(B-V)=0.130_{-0.089}^{+0.110}$. We use the StarHorse
estimate only as an external consistency check and do not interpret the
comparison as evidence for local or circumbinary extinction.

The circumstellar parameters are substantially more weakly
constrained than the stellar parameters, with broad and asymmetric
posterior distributions. In particular, the outer radius is
$R_{\rm out}=188_{-52}^{+150}$~AU for the present conditional model.
This weak constraint results from the very limited far-infrared
information, which consists of a low-significance $60\,\mu$m
measurement and a $90\,\mu$m upper limit.
The photometry through the WISE W4 band is consistent with the
stellar photospheres alone. Consequently, the circumstellar posterior
quantifies the range of parameters permitted under the conditional
assumption that the tentative far-infrared signal is physically
associated with the system, rather than demonstrating that such a
component is required by the data. The inferred surface density, dust
mass, and disk radii remain model dependent and also depend on
the assumed inclination, opacity law, temperature prescription, and
radial density profile.

A further major limitation is the absence of a spectroscopic
orbital solution. The light-curve analysis constrains the fractional
radii, orbital inclination, temperature ratio, and a model-dependent
photometric mass ratio, while the primary radius is inferred
from the SED using the \textit{Gaia} distance and the secondary radius
follows from the light-curve radius ratio. The orbital scale is then
obtained from the combination of the SED-based radius and the
corresponding fractional radius. Consequently, the component masses,
absolute radii, and orbital scale are model inferred rather than
dynamically determined. Double-lined radial-velocity observations
covering the orbit are required to determine the dynamical
mass ratio and orbital scale and to test the reliability of the
model-inferred absolute parameters.

The evolutionary status of CPD--82~291 also remains uncertain. Its
\textit{Gaia} astrometry is incompatible with physical membership in
Chamaeleon~I, and no suitable spectra are presently available from
which lithium absorption, H$\alpha$ emission, surface-gravity
indicators, or a systemic radial velocity can be measured. The PMS
interpretation must therefore remain tentative. Spectroscopy covering
the Li~I $6708$~\AA\ line, H$\alpha$, and suitable photospheric
absorption lines would provide direct tests of its youth, activity, and
orbital properties.

The sector-by-sector light-curve solutions demonstrate a changing
large-scale surface-brightness distribution, but the fitted circular
regions are non-unique representations of the observed asymmetries.
Their sizes, temperature contrasts, and locations are mutually
degenerate and should not be interpreted as uniquely reconstructed
physical spots. Multiband photometry and time-resolved spectroscopy
would help separate temperature-contrast effects from geometric
surface-brightness variations.

Finally, confirmation of the possible far-infrared dust
component requires new observations at far-infrared or millimetre
wavelengths with substantially improved sensitivity and angular
resolution. Such observations would determine whether the tentative
long-wavelength emission is associated with CPD--82~291 and would
provide meaningful constraints on the temperature, mass, and radial
extent of any surrounding material.



\section{Conclusions}

In this study, we present a multi-sector \textit{TESS}
photometric and spectral energy distribution analysis of the eclipsing
binary CPD--82~291. Although the system has historically been classified
as a young star in the direction of Chamaeleon~I, the \textit{Gaia} DR3
astrometry rules out physical membership in that association, and no
spectroscopic youth diagnostics or radial-velocity orbit are currently
available. The evolutionary status of the system therefore remains
uncertain. Our main results can be summarized as follows:

\begin{itemize}

\item The multi-sector \textit{TESS} light curves show pronounced
out-of-eclipse variability whose morphology changes between observing
epochs. Sector-by-sector surface-brightness modelling requires different
configurations at different epochs, supporting evolving large-scale
photospheric inhomogeneities. The fitted regions should, however, be
regarded as non-unique parametrizations of the surface-brightness
distribution rather than literal maps of individual starspots. The
configurations are generally more similar in temporally adjacent sectors,
but the changing sizes and longitudes do not uniquely establish
long-lived individual spots, spot migration, active longitudes, or
magnetic cycles.

\item A total of 28 flare events were retained in the final
\textit{TESS} sample. Their observed peak amplitudes have a median value
of $0.64\%$, while the directly integrated equivalent durations have a
median of $35.6$~s. For the 27 events with resolved threshold-defined
durations, the median duration is approximately $150$~min. Equivalent
duration is significantly correlated with both flare timescale and peak
amplitude, whereas the correlation between peak amplitude and
duration is weak and not statistically significant. The flare
detections, together with the evolving photospheric modulation, establish
CPD--82~291 as an active eclipsing system.

\item The reference light-curve solution is consistent with a
detached binary viewed at an orbital inclination of
$i\simeq82.1^\circ$. The photometric modelling yields
$q_{\rm phot}\simeq0.464$; in the absence of a spectroscopic orbit, this
mass ratio is model dependent and should not be interpreted as a
dynamical measurement.

\item The stellar photospheric model reproduces the observed SED
through the WISE W4 band without requiring an infrared-excess component.
The reported $60\,\mu$m flux is only a low-significance measurement
($0.026\pm0.019$~Jy), while the $90\,\mu$m result is a
non-detection and is included only as a one-sided upper limit.
Consequently, the available photometry does not provide secure evidence
for a bound circumbinary dust disk.

\item Under the conditional assumption that the marginal
far-infrared signal is physically associated with the system, a cool-dust
model can reproduce it, with posterior median radii of
$R_{\rm in}\simeq147$~AU and $R_{\rm out}\simeq188$~AU. These quantities
are strongly model dependent and non-unique. The inferred inner radius is
approximately $1.4\times10^{3}$ times the model-inferred binary separation
and therefore should not be interpreted as a conventional tidally
truncated circumbinary cavity.

\item In the absence of radial velocities, the absolute scale is
model-inferred by combining the SED-derived primary radius at
the adopted \textit{Gaia} distance with the fractional light-curve
parameters. The current solution gives approximately
$R_1\simeq1.25\,R_\odot$,
$R_2\simeq3.70\,R_\odot$,
$M_1\simeq0.93\,M_\odot$, and
$M_2\simeq0.43\,M_\odot$.
These values are not dynamical measurements, and the mass
uncertainties are amplified by the cubic dependence of the total mass
on the inferred orbital scale, while the absolute radii depend on the
SED-derived radius scale and light-curve radius ratio.

\item Comparison with stellar evolutionary calculations places
the primary close to the ZAMS, whereas the secondary lies substantially
above the main-sequence locus if the inferred absolute scale is correct.
The two components do not define an unambiguous coeval PMS solution. The
secondary's position may reflect activity-related structural effects,
limitations of the inferred absolute scale, or limitations of
the evolutionary models; it cannot by itself be taken as evidence for
youth or PMS radius inflation

\item CPD--82~291 should therefore presently be regarded as an
active eclipsing binary of uncertain evolutionary status, with marginal
far-infrared emission but no secure evidence for a circumbinary disk.
Time-series double-lined spectroscopy is the most important next step,
as it would provide the dynamical orbital scale and component masses,
systemic velocity, and independent atmospheric and youth diagnostics.
More sensitive far-infrared or millimetre observations with improved
angular resolution are likewise required to determine whether any cold
circumstellar or circumbinary material is physically associated with the
system.

\end{itemize}

\begin{acknowledgments}

We thank the anonymous referee for their insightful and constructive suggestions that significantly improved the paper. We thank T\"{U}B\.{I}TAK for funding this research under project number 126F230.

The numerical calculations reported in this paper were partially performed at the T\"{U}B\.{I}TAK ULAKB\.{I}M High Performance and Grid Computing Center (TRUBA resources).

We acknowledge the use of TESS High Level Science Products (HLSP) produced by the Quick-Look Pipeline (QLP) at the TESS Science Office at MIT, which are publicly available from the Mikulski Archive for Space Telescopes (MAST). Funding for the TESS mission is provided by NASA's Science Mission directorate. The data described here may be obtained from the MAST archive via \cite{Huang2020}. This research made use of Lightkurve, a Python package for Kepler and TESS data analysis \citep{lk}.

This work has made use of data from the European Space Agency (ESA) mission \textit{Gaia} (\url{https://www.cosmos.esa.int/gaia}), processed by the \textit{Gaia} Data Processing and Analysis Consortium (DPAC).

This research has made use of the SIMBAD database and the VizieR catalog access tool, operated at CDS, Strasbourg, France.

This publication makes use of data products from the Wide-field Infrared Survey Explorer (\textit{WISE}), which is a joint project of the University of California, Los Angeles, and the Jet Propulsion Laboratory/California Institute of Technology, funded by NASA.

\end{acknowledgments}

\begin{contribution}

All editors contribute equally to the operation of PASP.


\end{contribution}

%
\facilities{HST(STIS), Swift(XRT and UVOT), AAVSO, CTIO:1.3m, CTIO:1.5m, CXO}

\software{
\texttt{PHOEBE} \citep{phoebe1,phoebe2,phoebe3,phoebe4,phoebe5}, 
\texttt{Astropy} \citep{astro1,astro2,astro3}, 
\texttt{corner} \citep{corner}, 
\texttt{Lightkurve} \citep{lk},
\texttt{Matplotlib} \citep{matplotlib}, 
\texttt{NumPy} \citep{numpy},
}


\appendix
\section{Additional MCMC and Model Figures}

This appendix presents the posterior probability distributions obtained
from the MCMC analyses performed for both the light-curve and SED
modelling. The corner plots provide a visual representation of the
parameter uncertainties, correlations, and possible degeneracies within
the adopted models. Median parameter values and confidence intervals
reported in the main text were derived from these posterior
distributions.

\subsection{Light-Curve MCMC and Surface Geometry}

This subsection presents additional figures related to the light-curve
modelling of CPD--82 291. The appendix includes the corner plot obtained
from the MCMC analysis and the component surface mesh plots corresponding
to the adopted binary configuration.

The corner plot illustrates the posterior distributions and parameter
correlations derived from the MCMC analysis of the light-curve solution.
The distributions indicate that the principal orbital and stellar
parameters are generally well constrained by the high-precision
\textit{TESS} photometry, although moderate correlations remain between
inclination, surface potentials, temperature ratios, and spot-related
parameters.

The component surface mesh plots provide a model-based visualization of
the Roche geometry and adopted surface-brightness distributions of the
binary components, including the spot configurations used to reproduce
the observed out-of-eclipse asymmetries. These surface representations
should not be interpreted as unique maps of individual starspots.

\begin{figure*}[htbp!]
    \centering
    \includegraphics[width=0.24\linewidth]{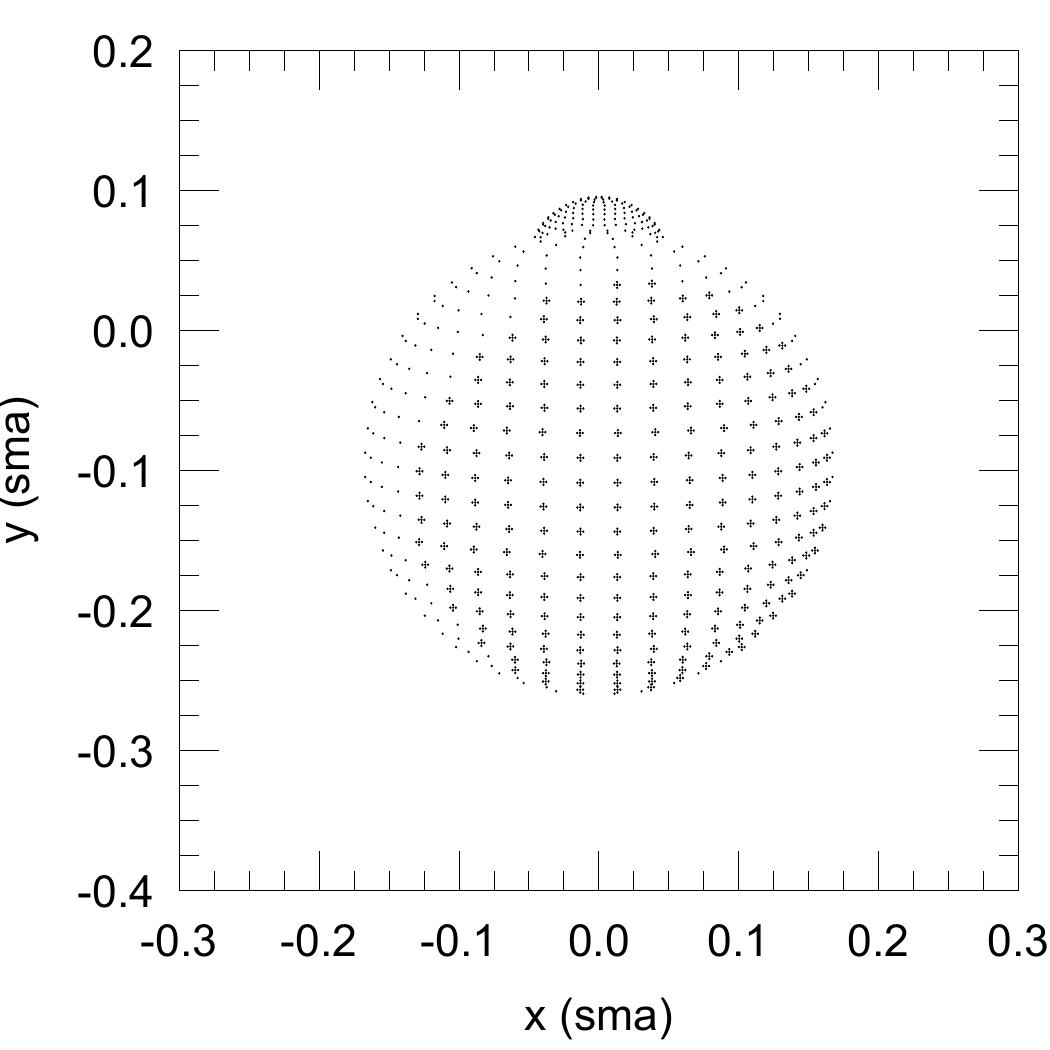}
    \includegraphics[width=0.24\linewidth]{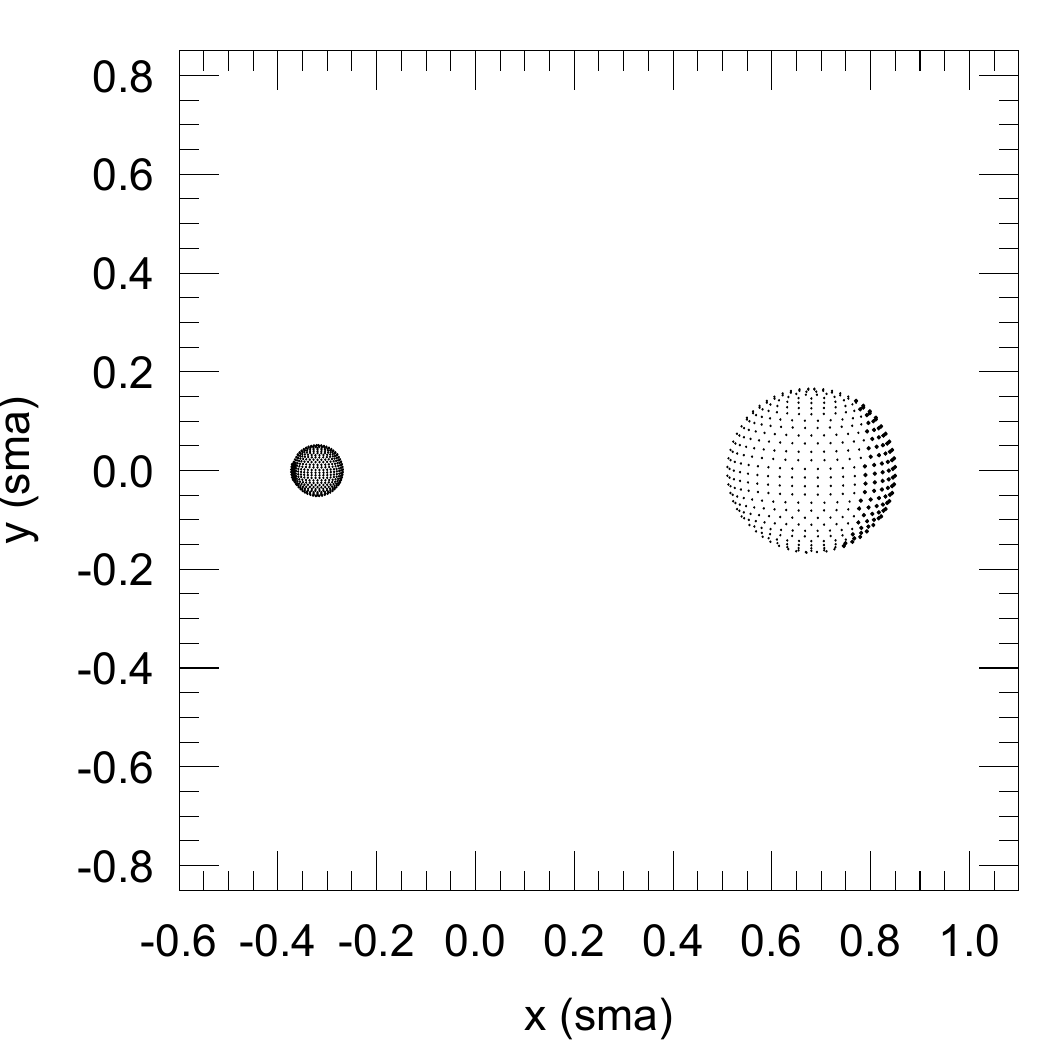}
    \includegraphics[width=0.24\linewidth]{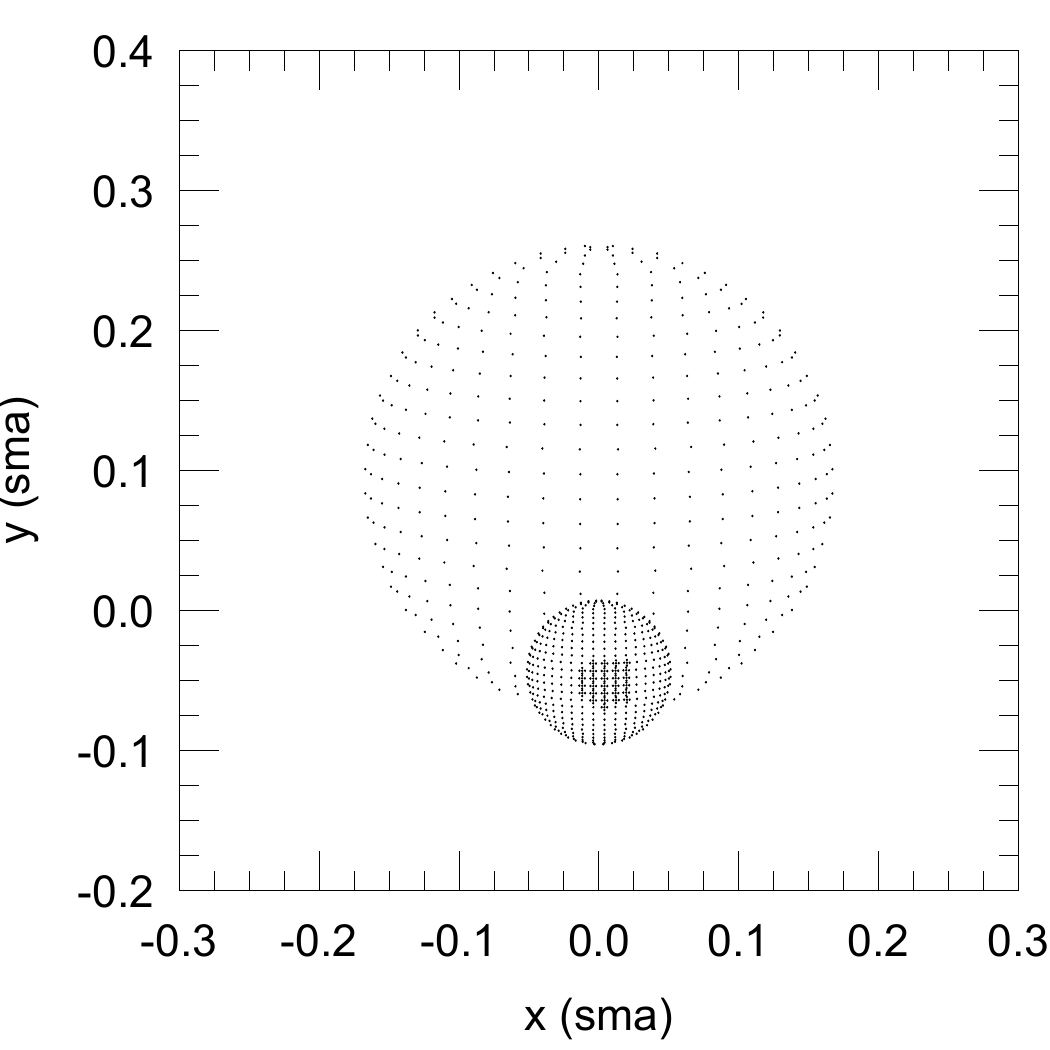}
    \includegraphics[width=0.24\linewidth]{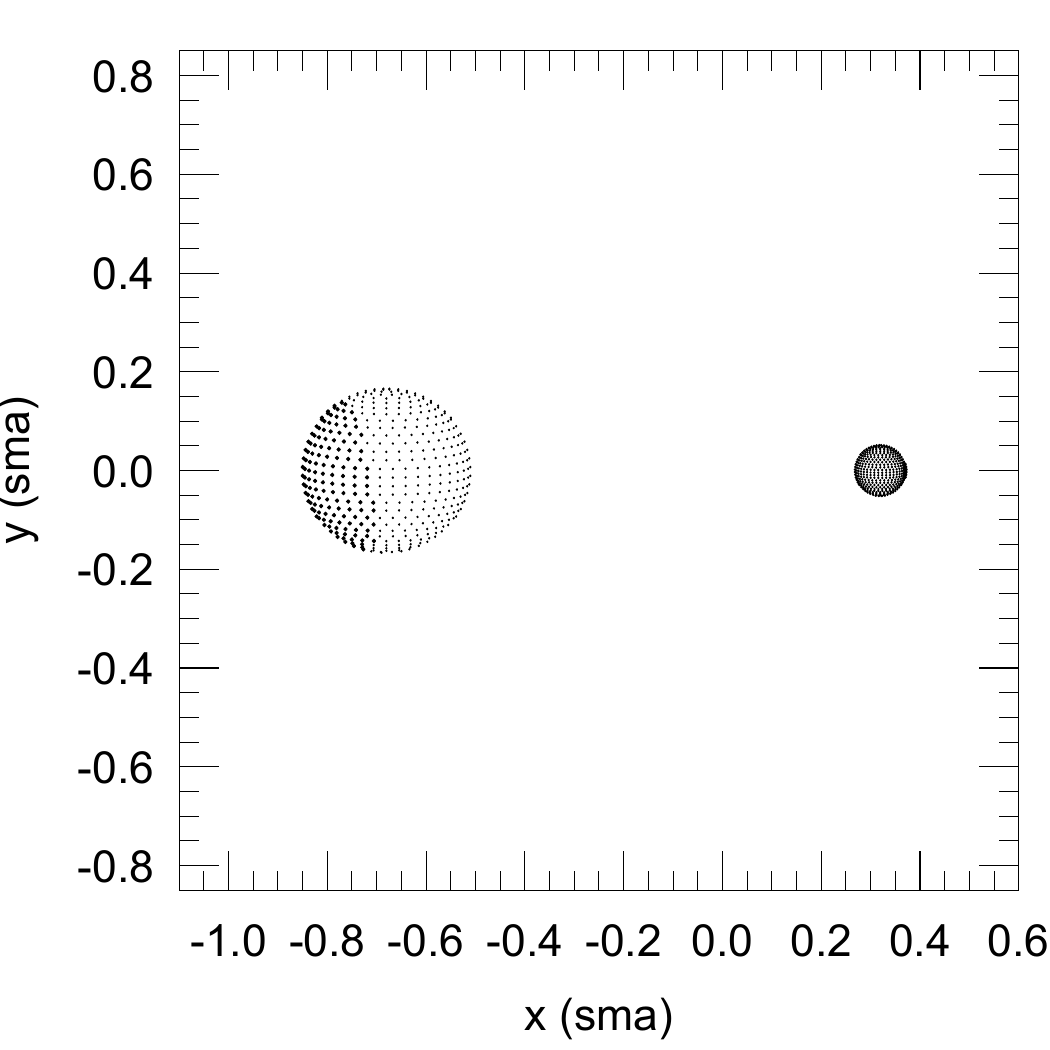}
    
    
    \includegraphics[width=0.24\linewidth]{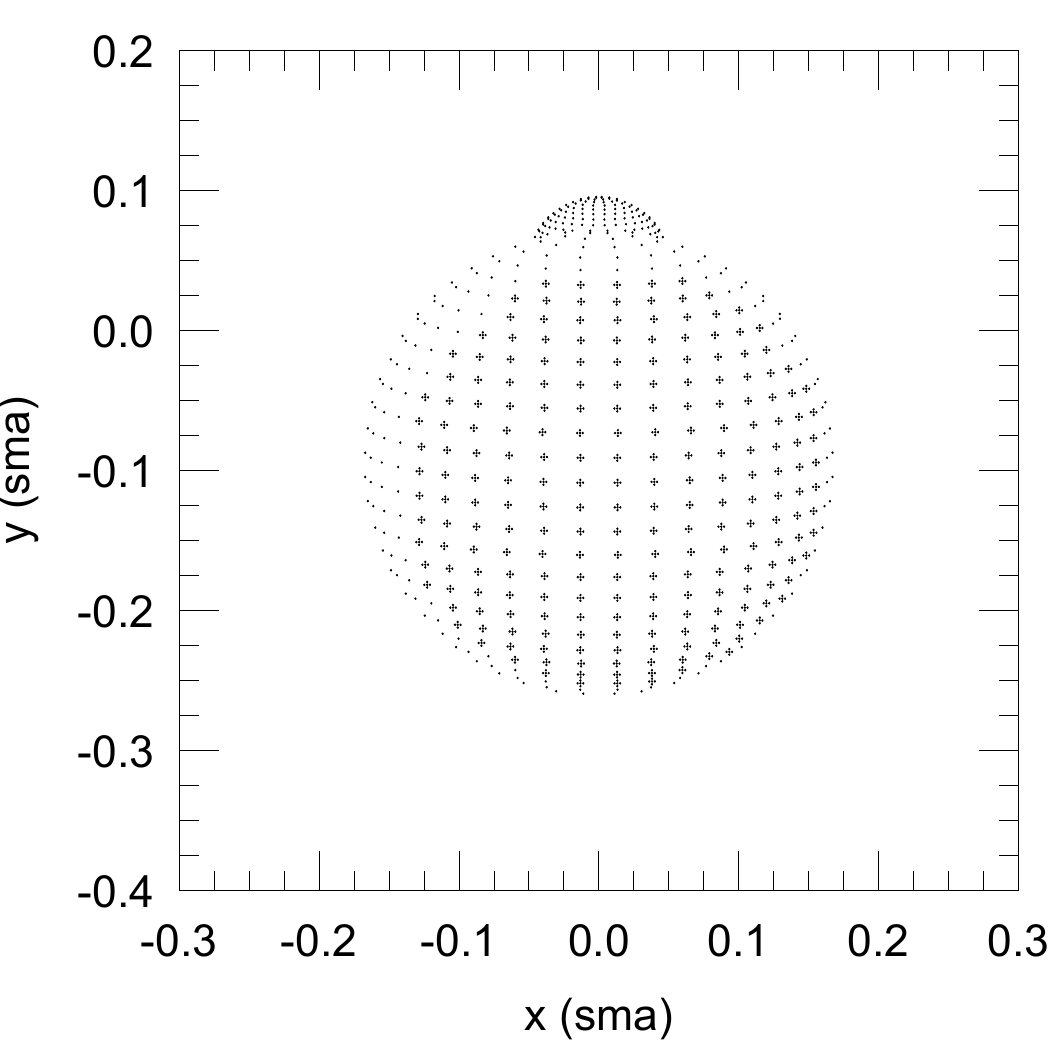}
    \includegraphics[width=0.24\linewidth]{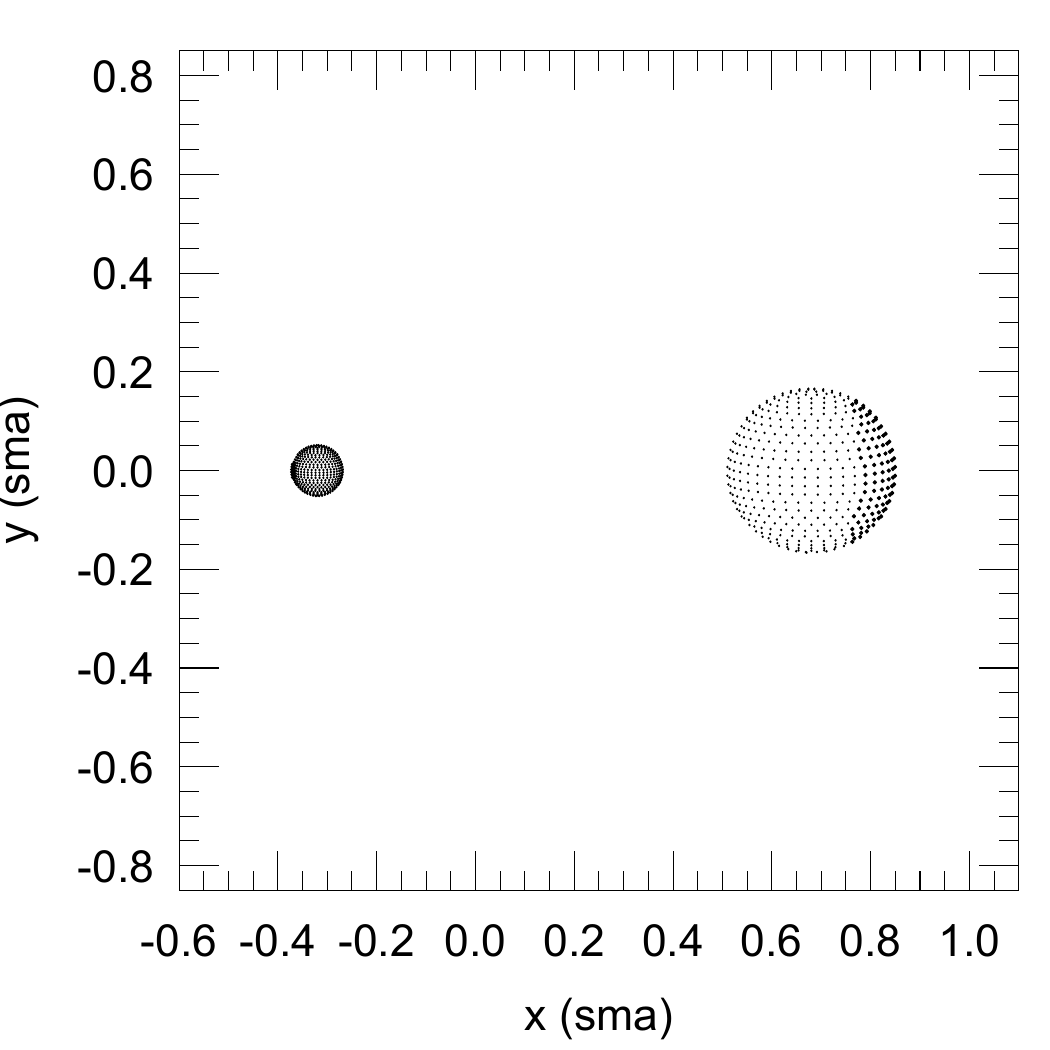}
    \includegraphics[width=0.24\linewidth]{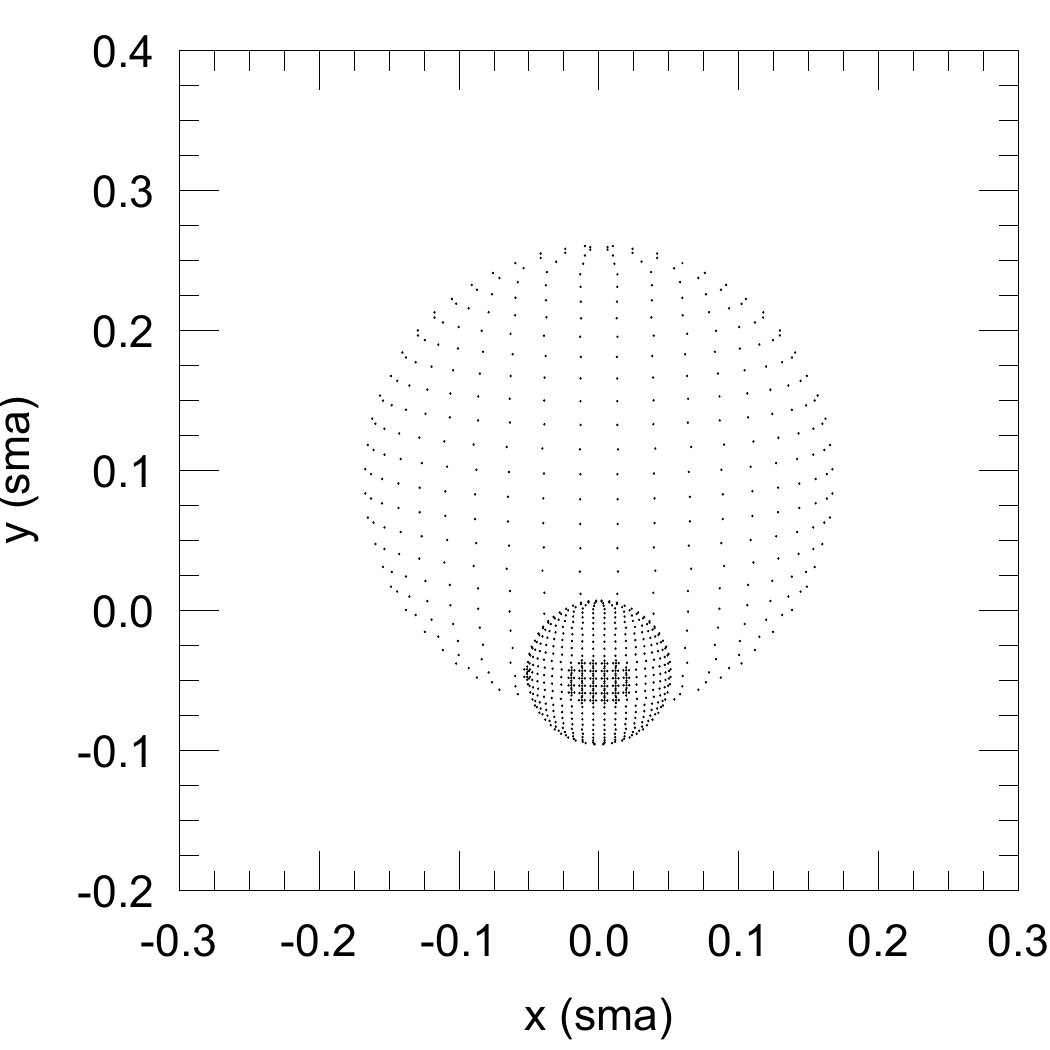}
    \includegraphics[width=0.24\linewidth]{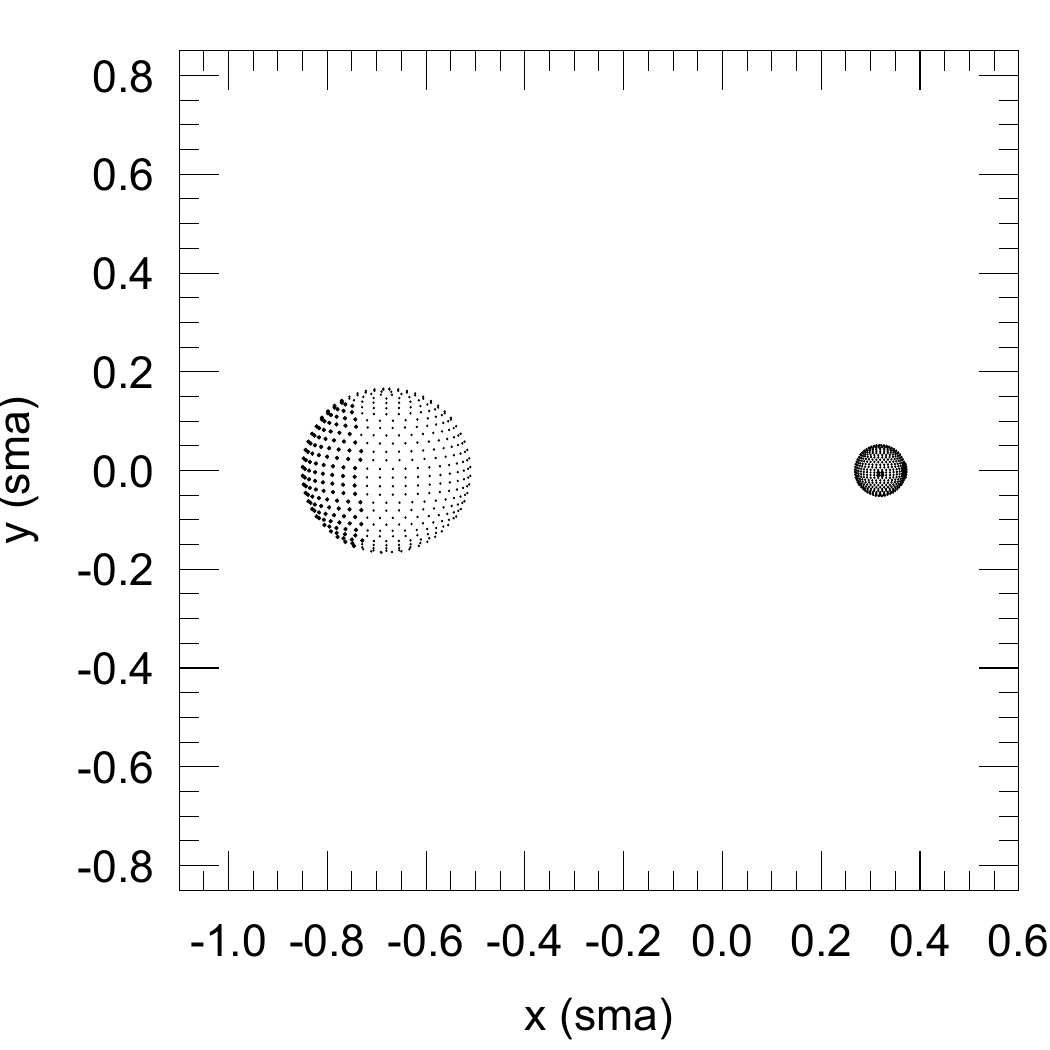}
    
    
    \includegraphics[width=0.24\linewidth]{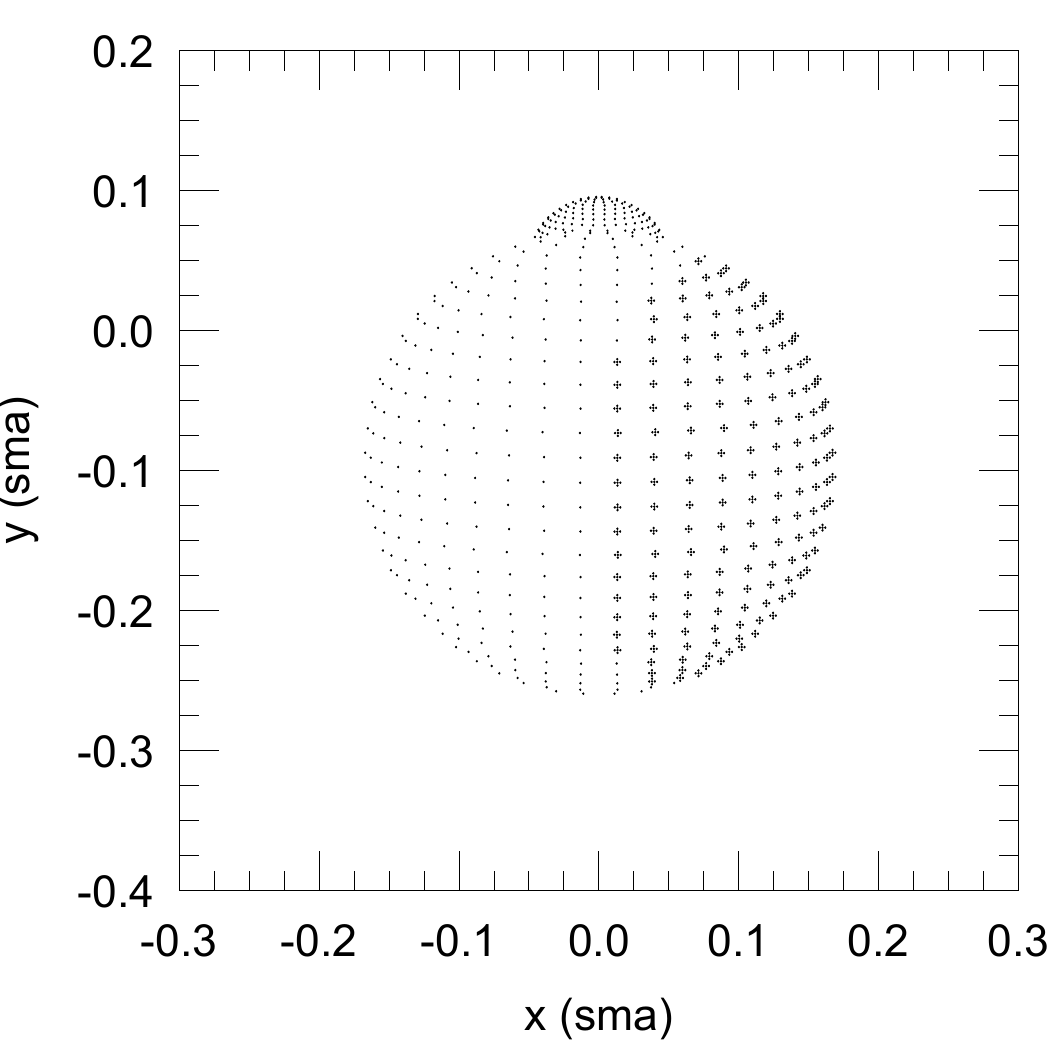}
    \includegraphics[width=0.24\linewidth]{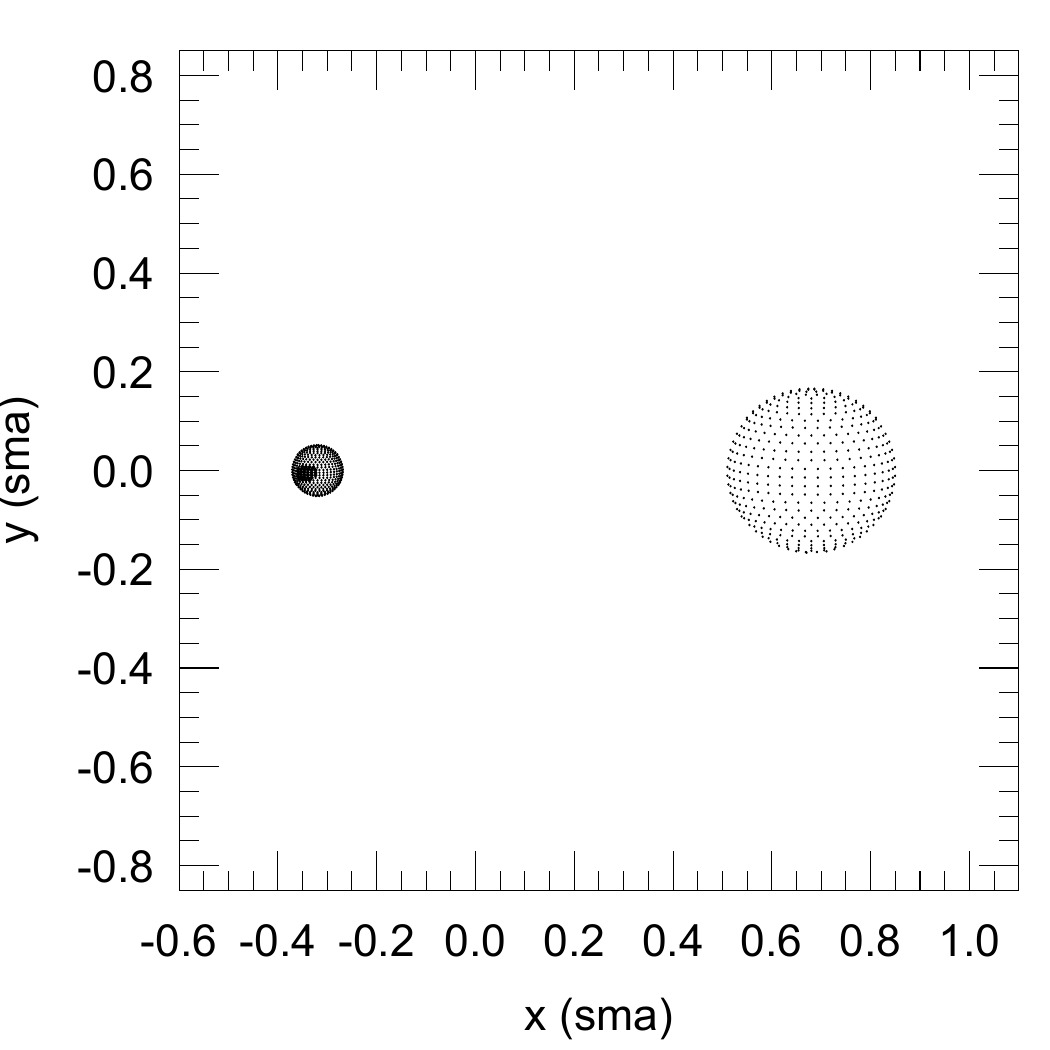}
    \includegraphics[width=0.24\linewidth]{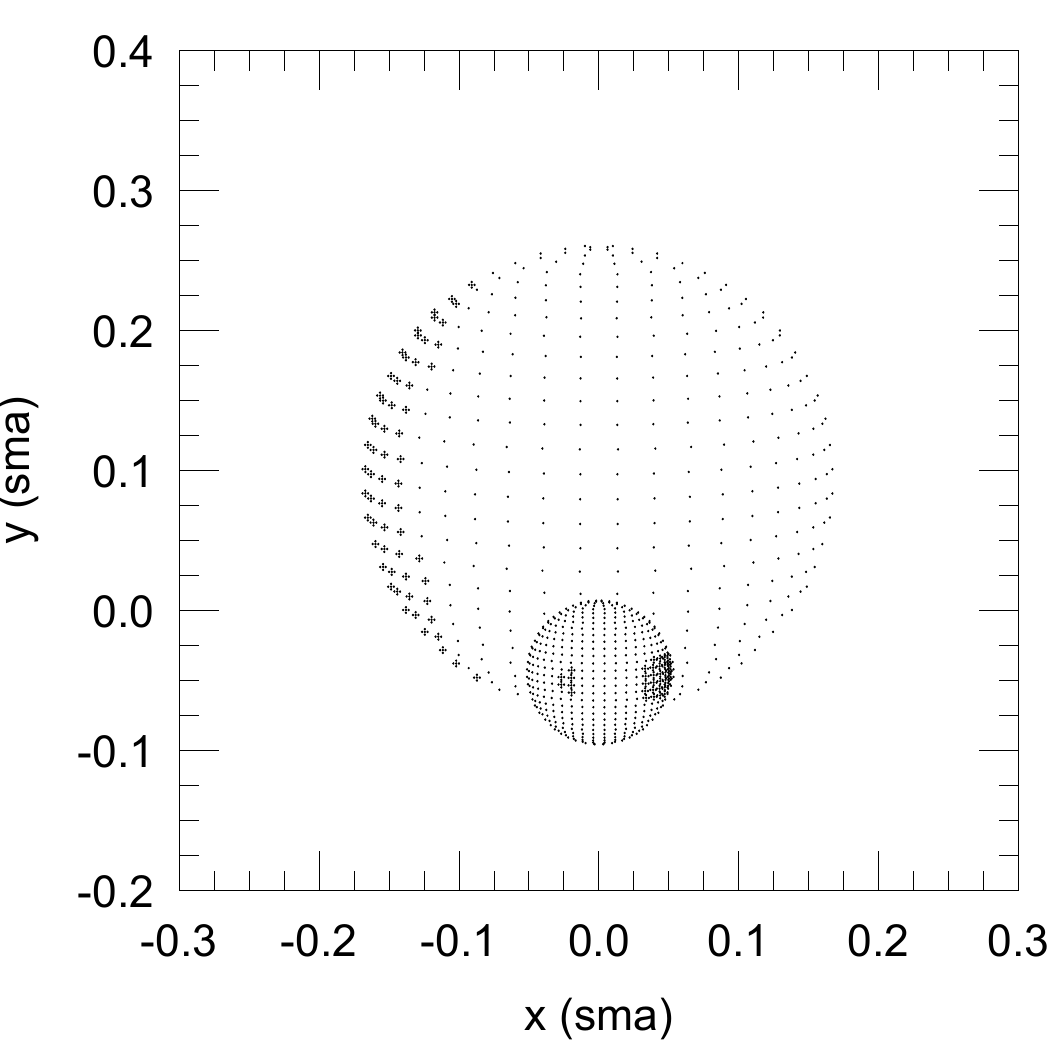}
    \includegraphics[width=0.24\linewidth]{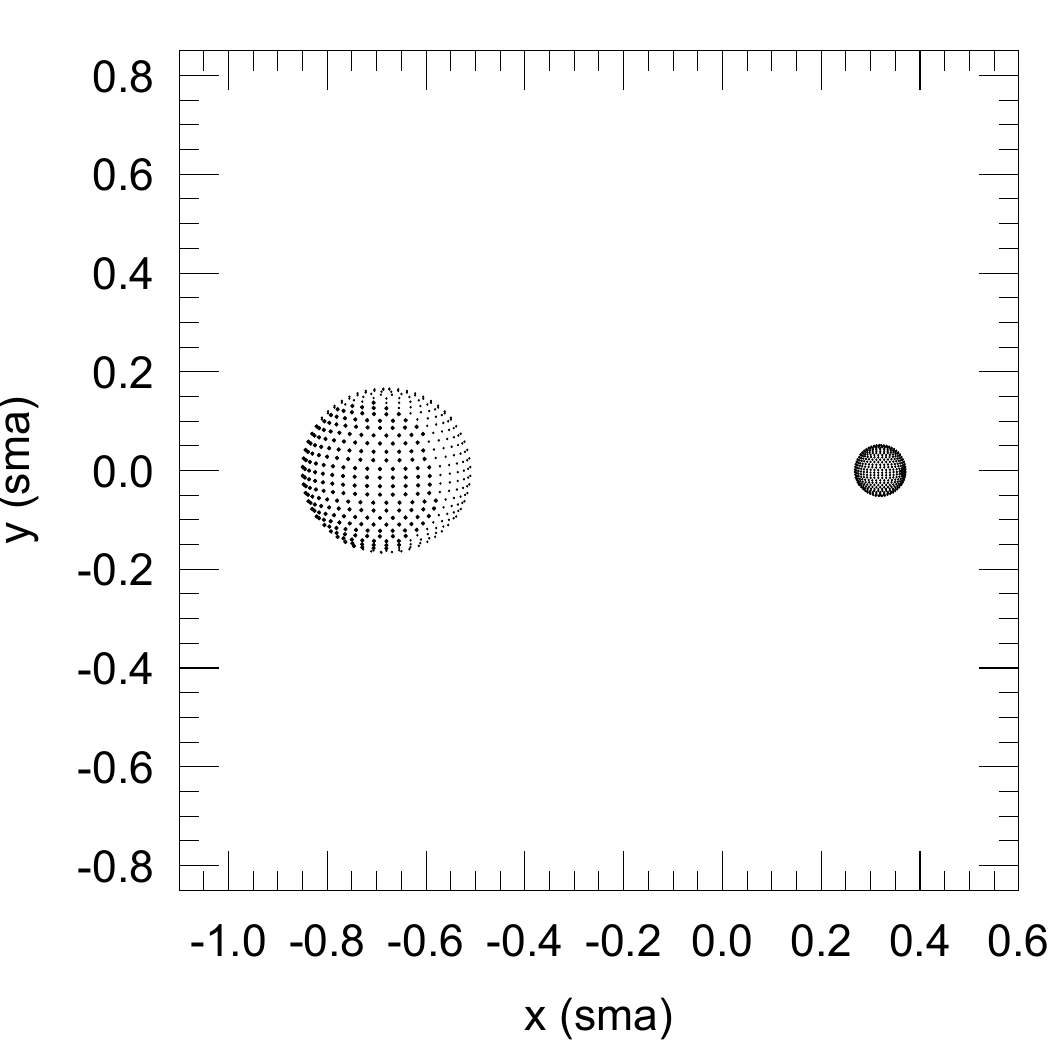}
    
    
    \includegraphics[width=0.24\linewidth]{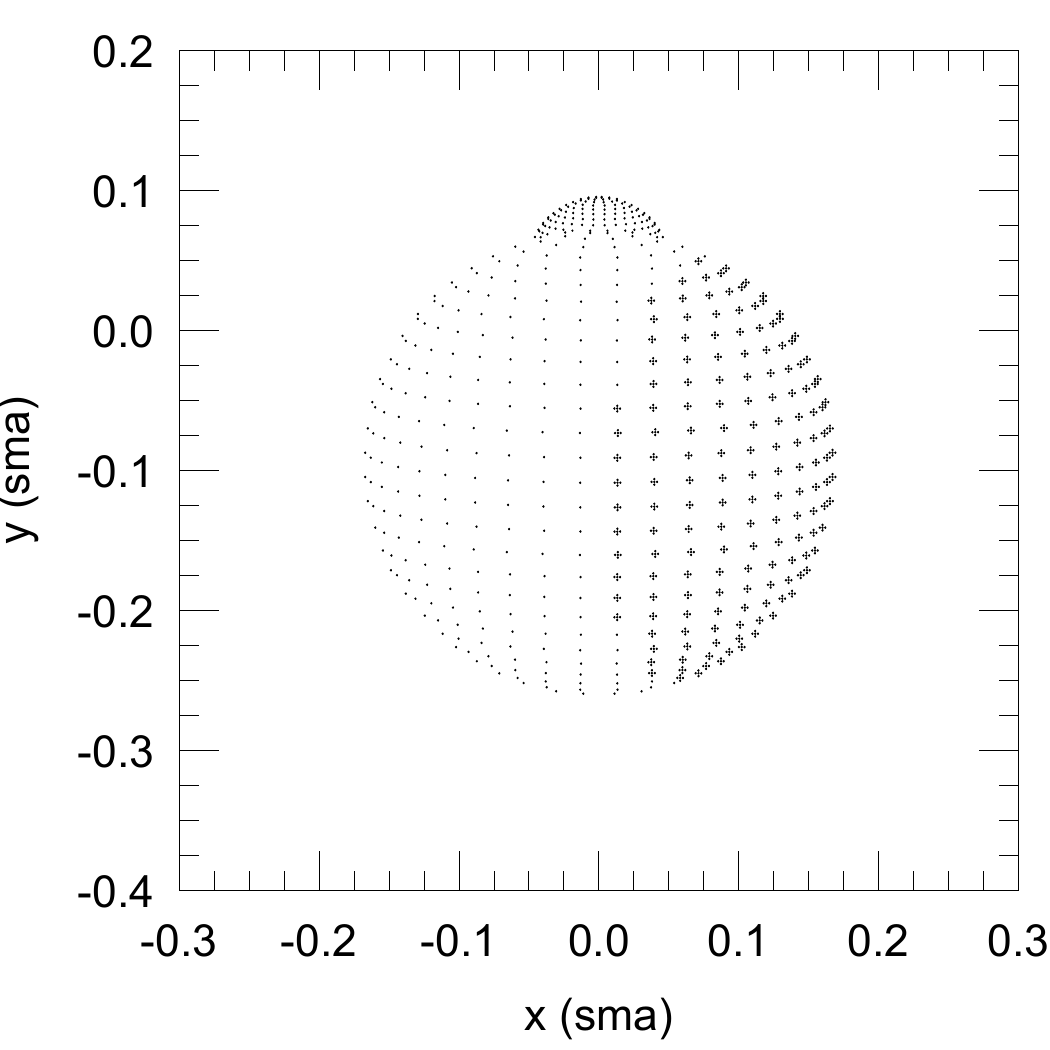}
    \includegraphics[width=0.24\linewidth]{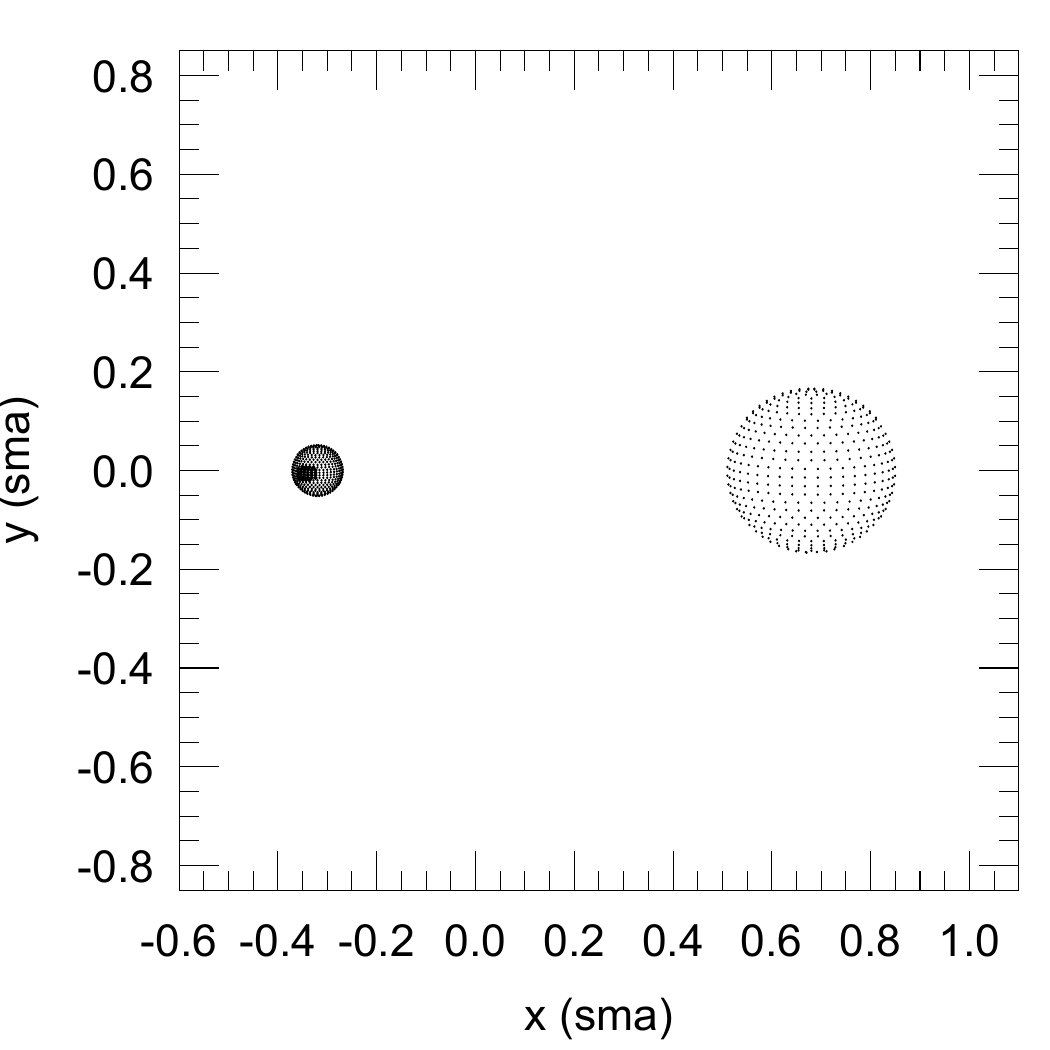}
    \includegraphics[width=0.24\linewidth]{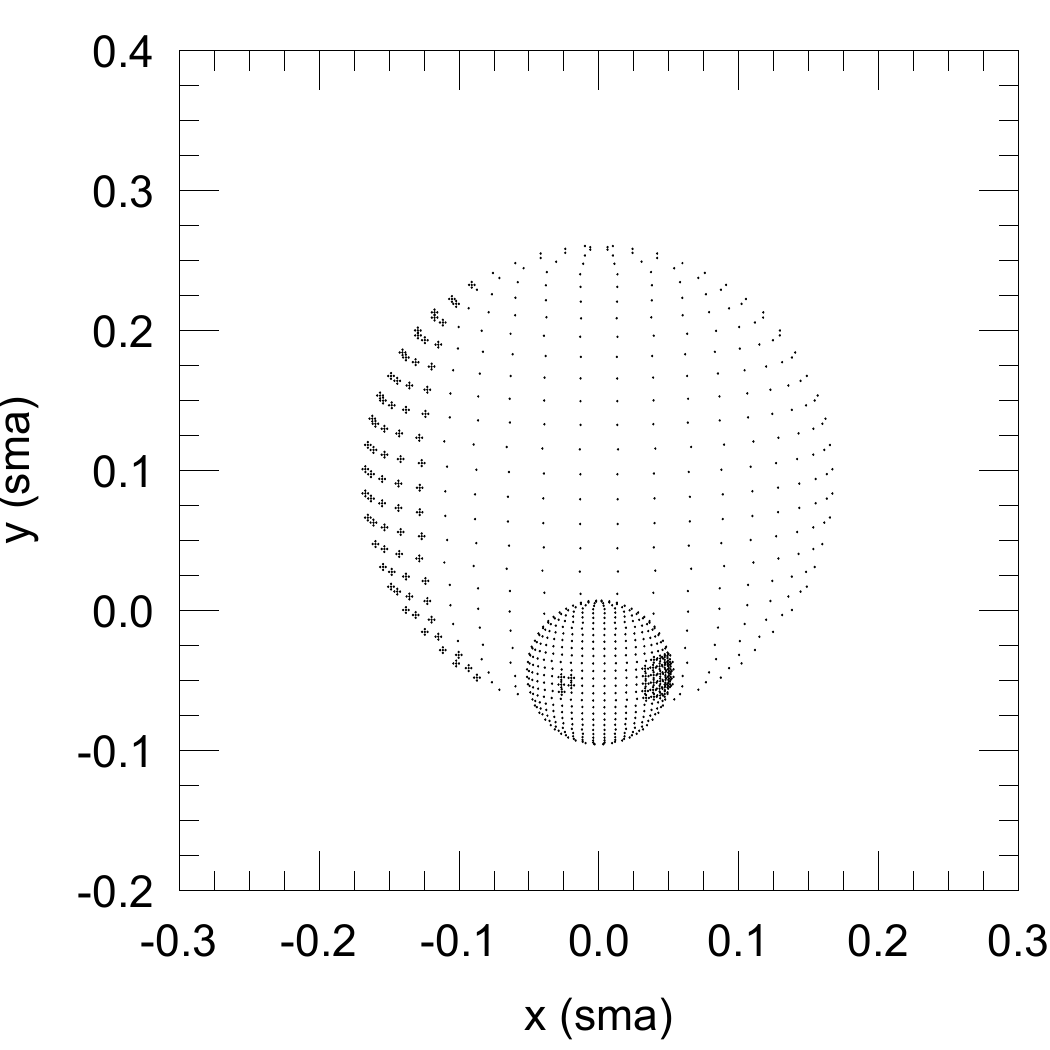}
    \includegraphics[width=0.24\linewidth]{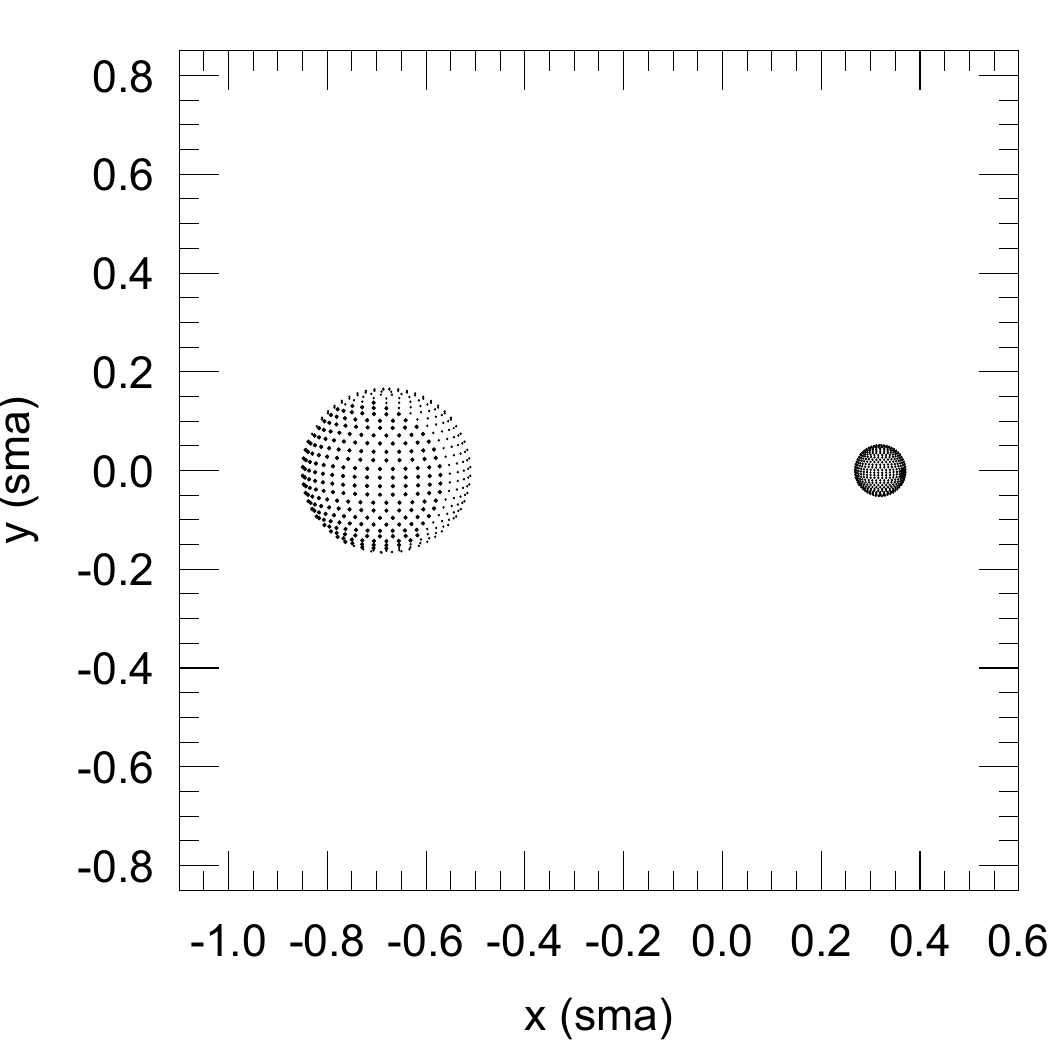}
    
    \caption{Surface mesh representations of the binary components at four different orbital phases, from left to right, 0, 0.25, 0.50, and 0.75, respectively, for the light-curve solutions obtained from \textit{TESS} Sectors 11, 13, 38, and 39 (from top to bottom, respectively). The models illustrate the evolution of the spot configurations and the corresponding changes in the longitudinal distribution of the active regions between different observing epochs. The adopted spot parameters are listed in Table~\ref{tab:spot_pars}.}
    \label{fig:mesh1}
\end{figure*}

\begin{figure*}[htbp!]
    \centering
    \includegraphics[width=0.24\linewidth]{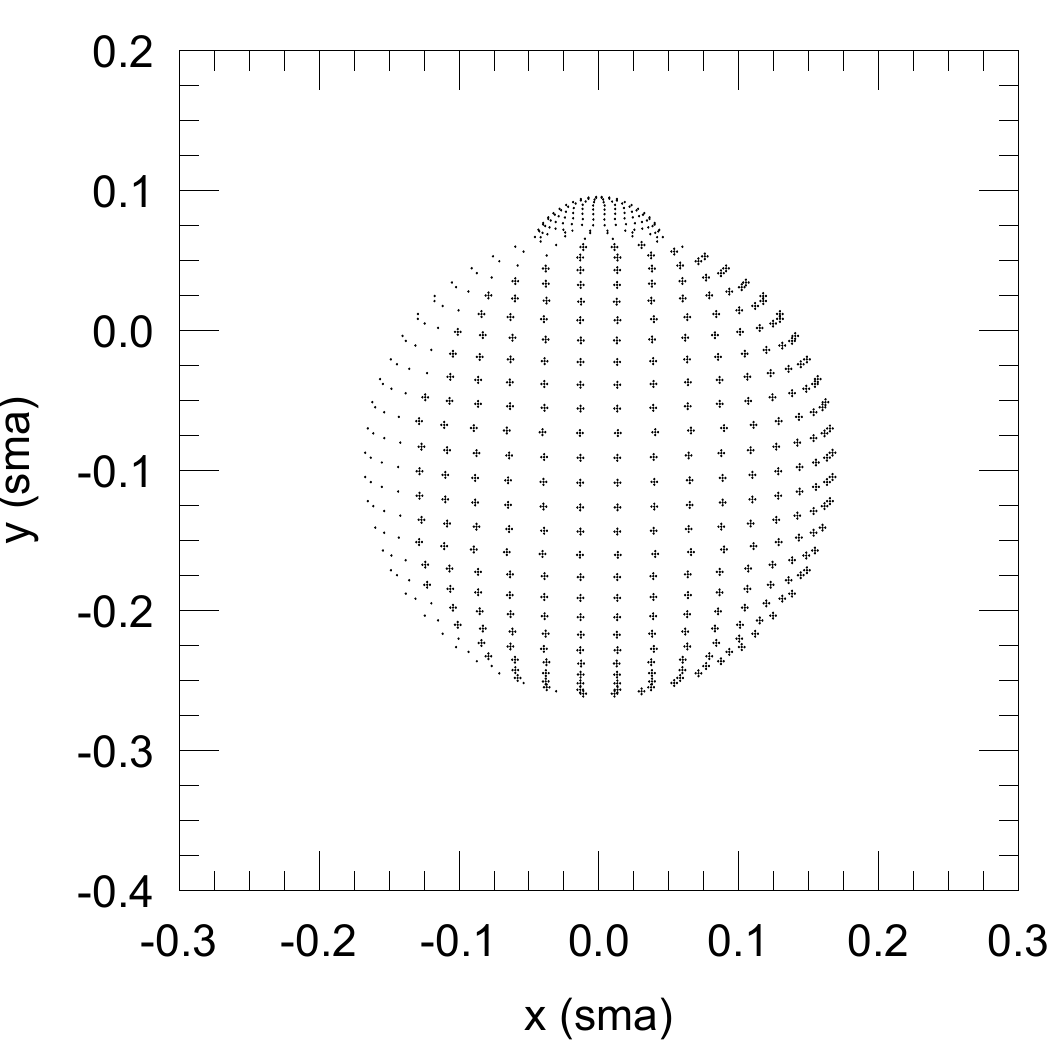}
    \includegraphics[width=0.24\linewidth]{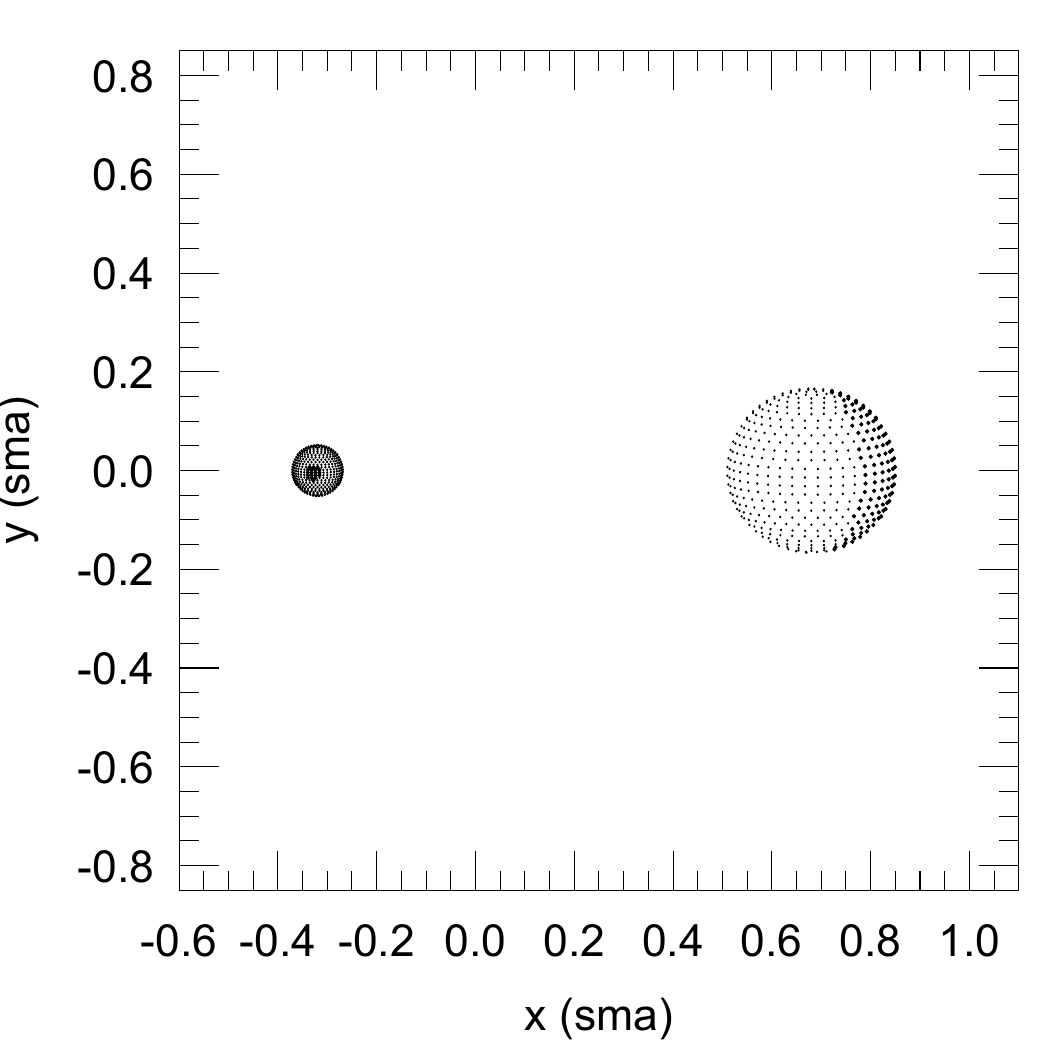}
    \includegraphics[width=0.24\linewidth]{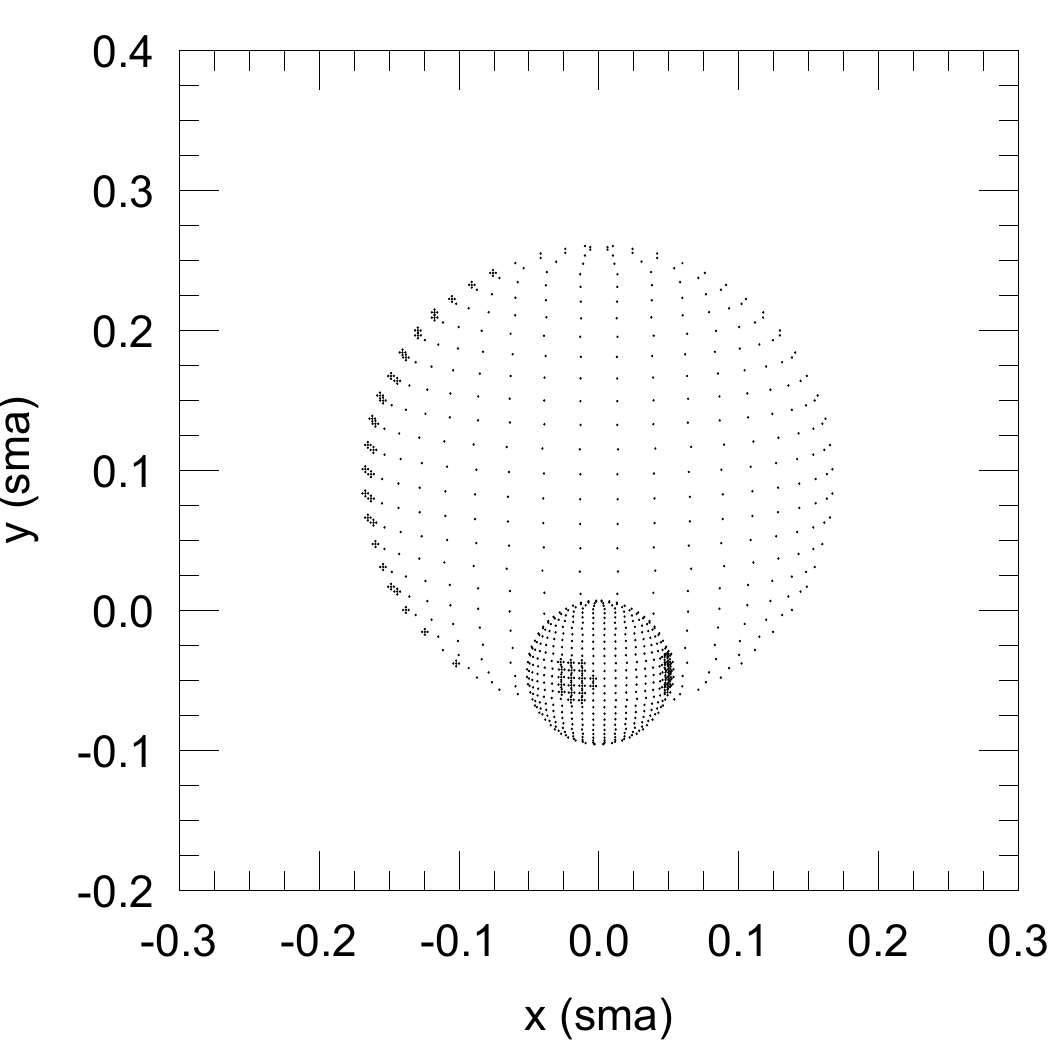}
    \includegraphics[width=0.24\linewidth]{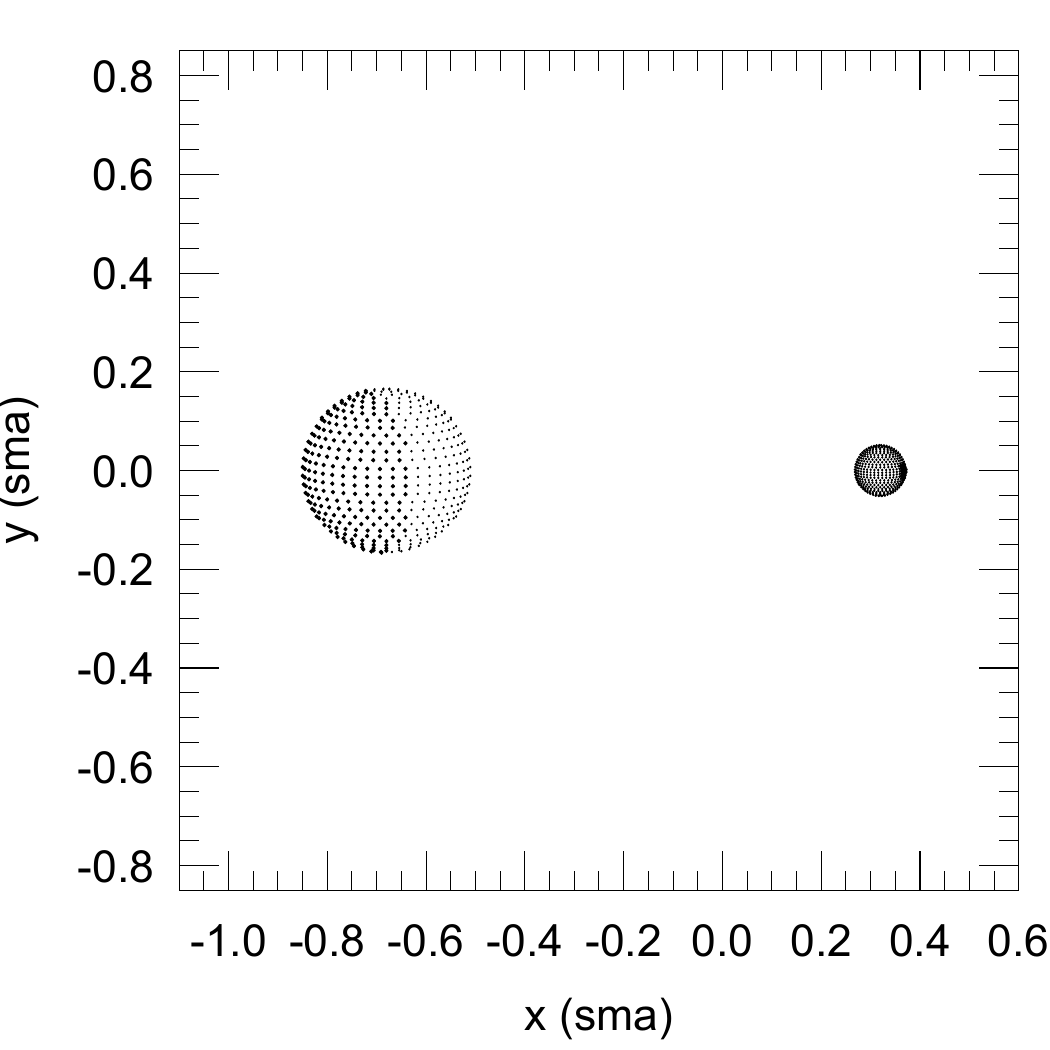}
    
    
    \includegraphics[width=0.24\linewidth]{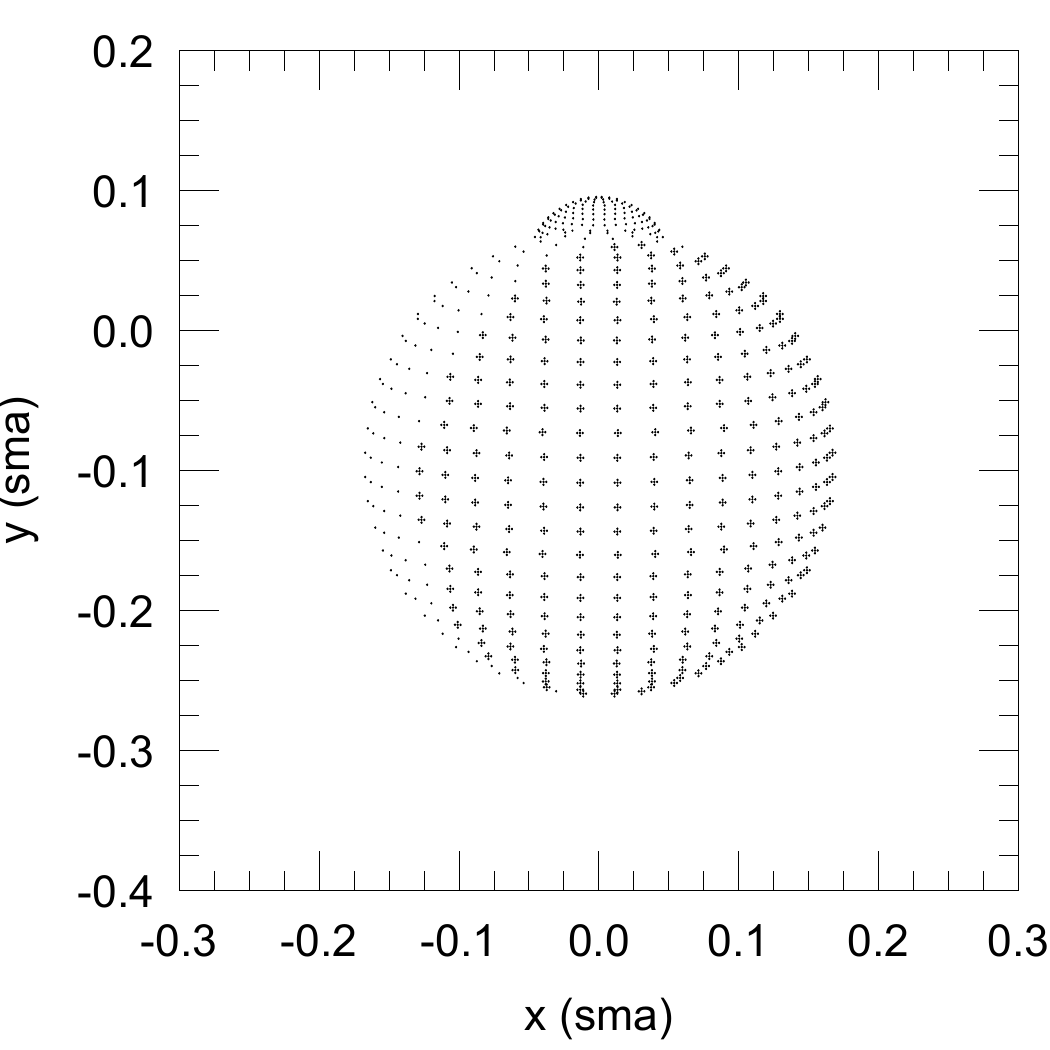}
    \includegraphics[width=0.24\linewidth]{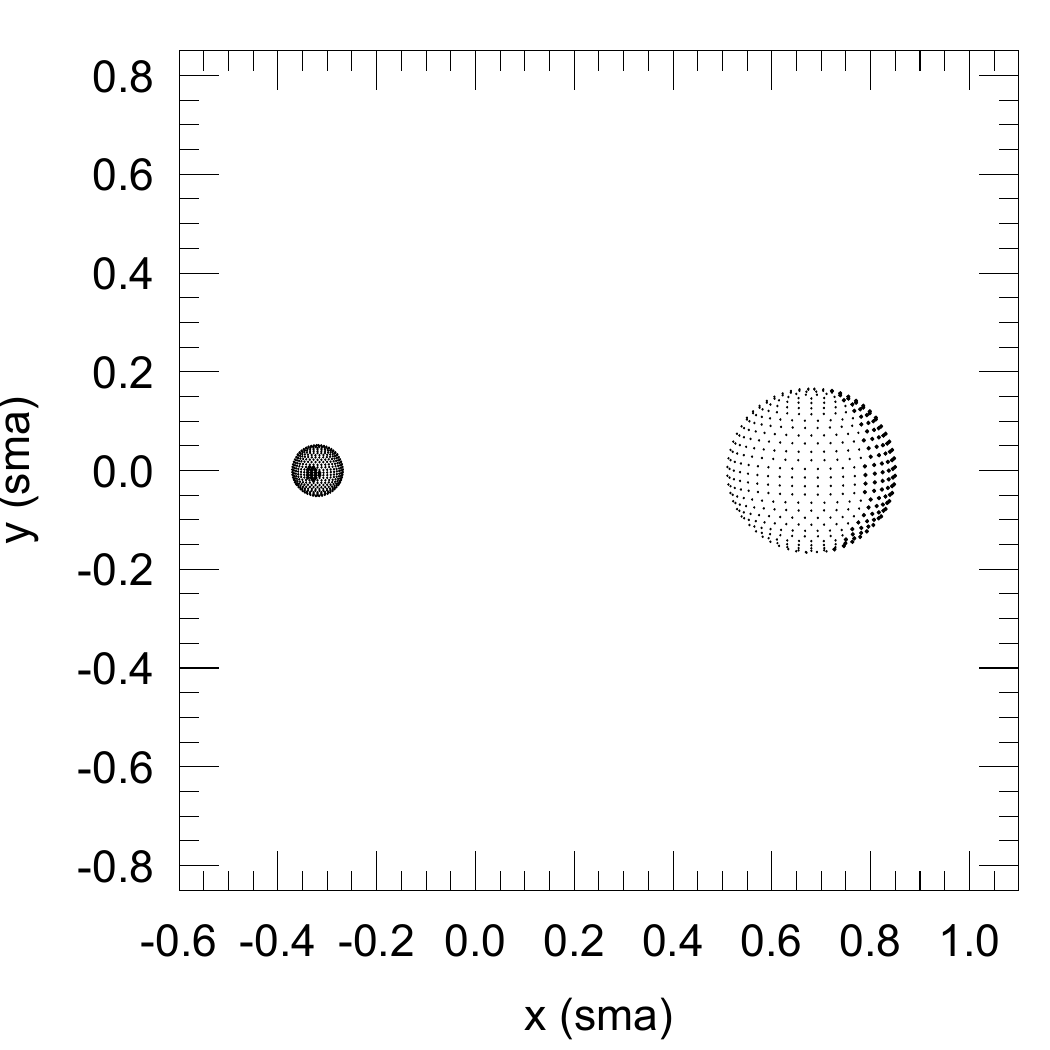}
    \includegraphics[width=0.24\linewidth]{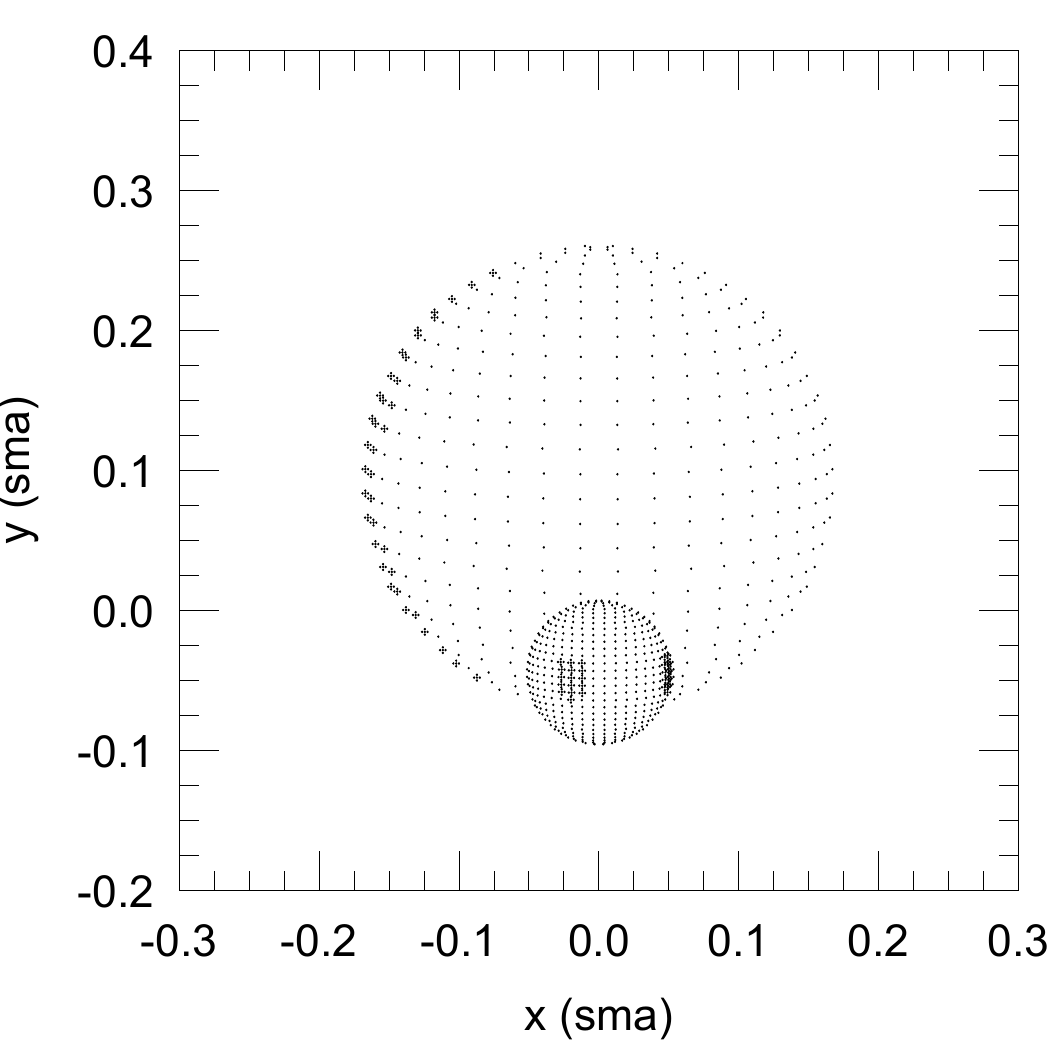}
    \includegraphics[width=0.24\linewidth]{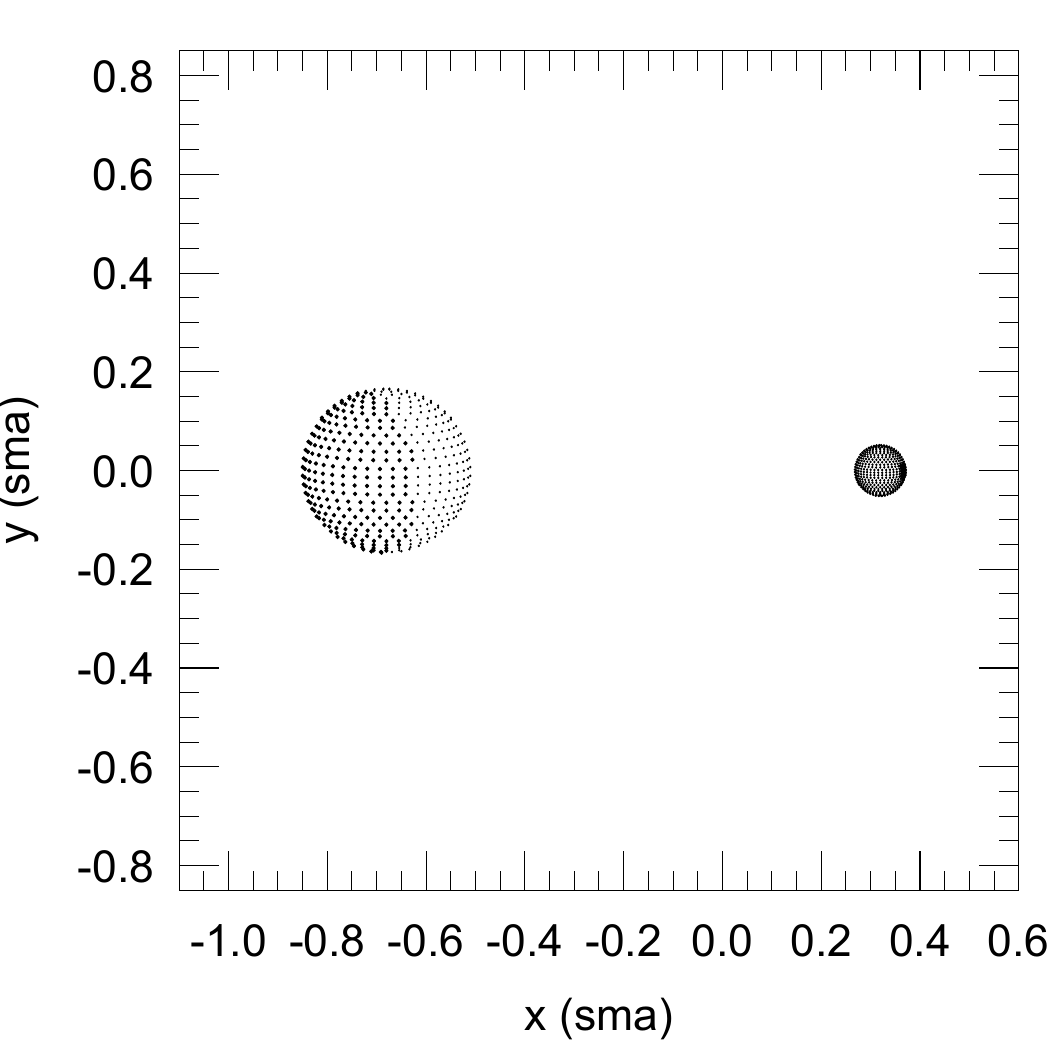}
    
    
    \includegraphics[width=0.24\linewidth]{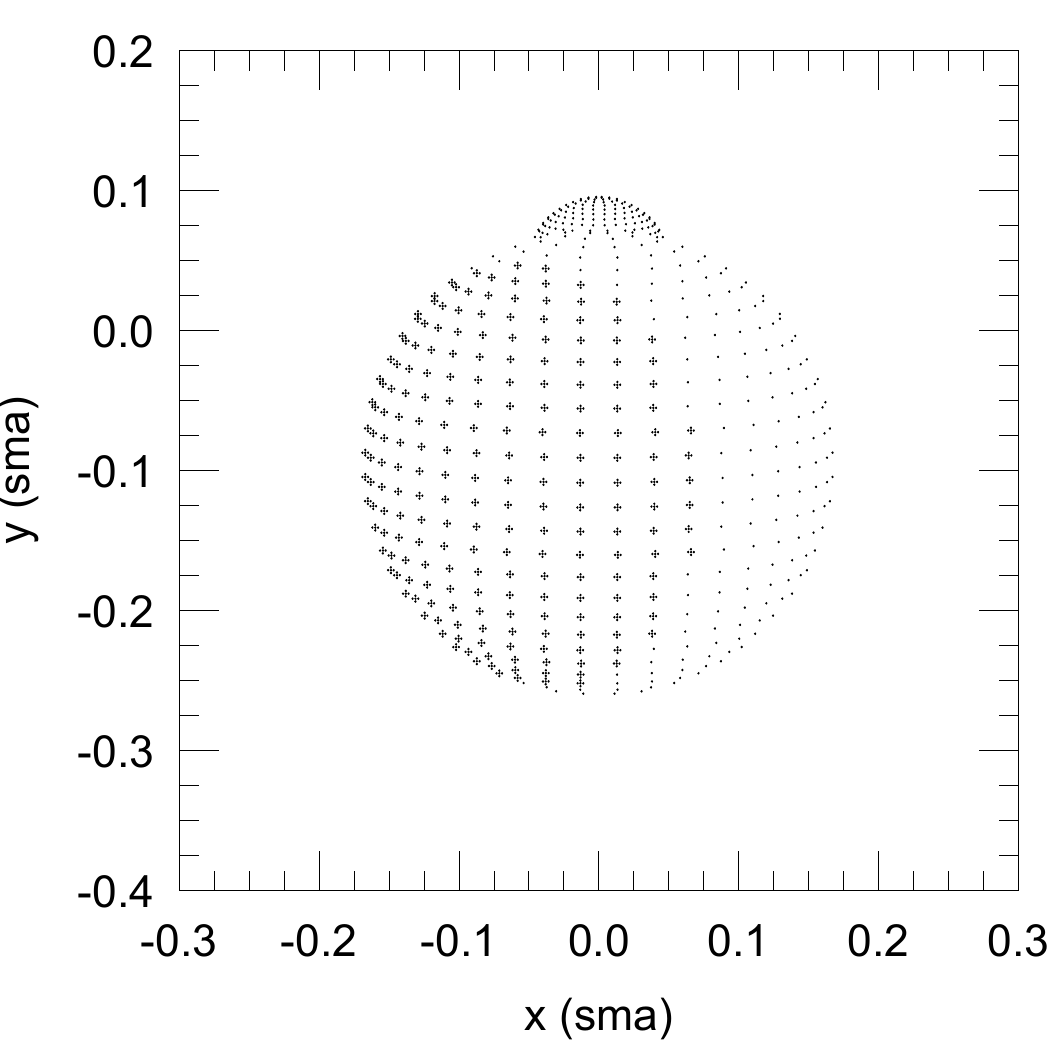}
    \includegraphics[width=0.24\linewidth]{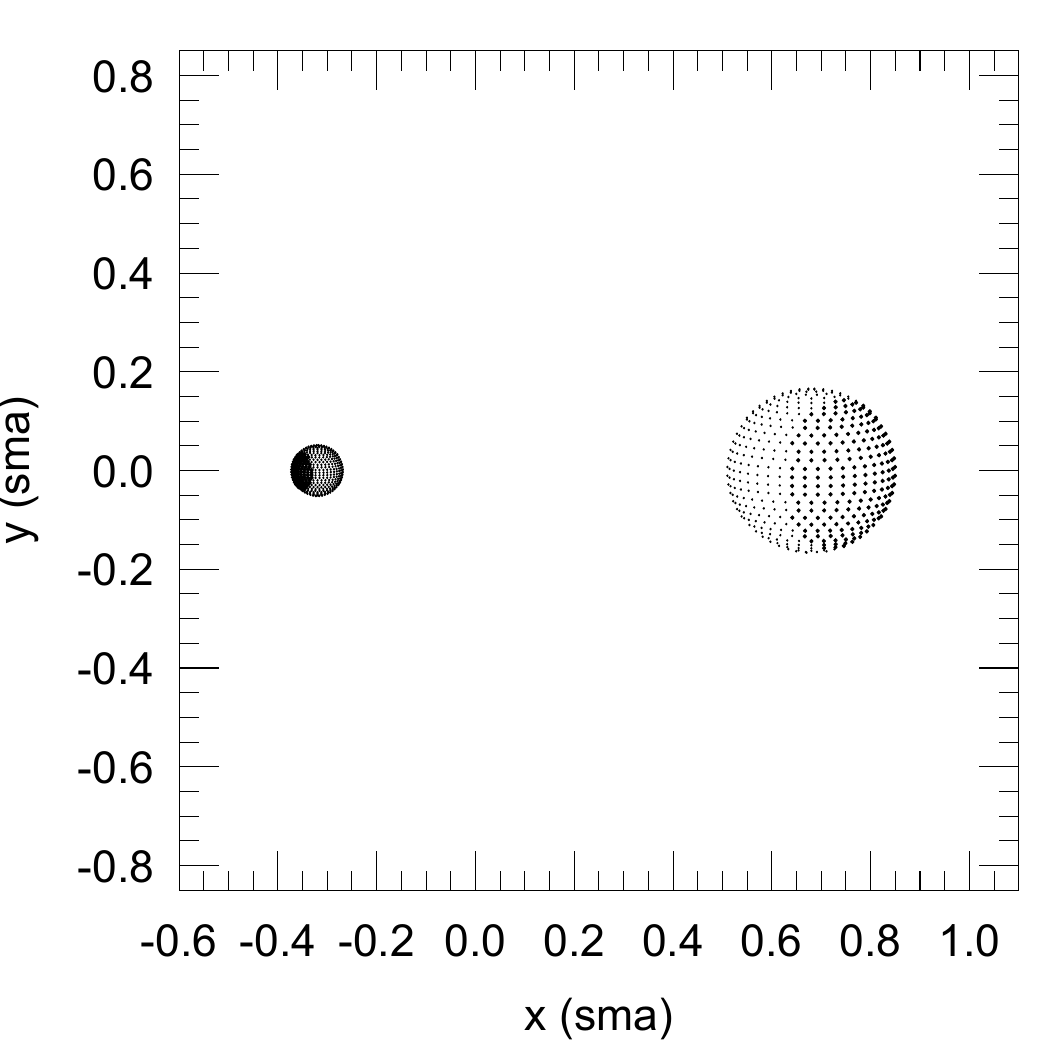}
    \includegraphics[width=0.24\linewidth]{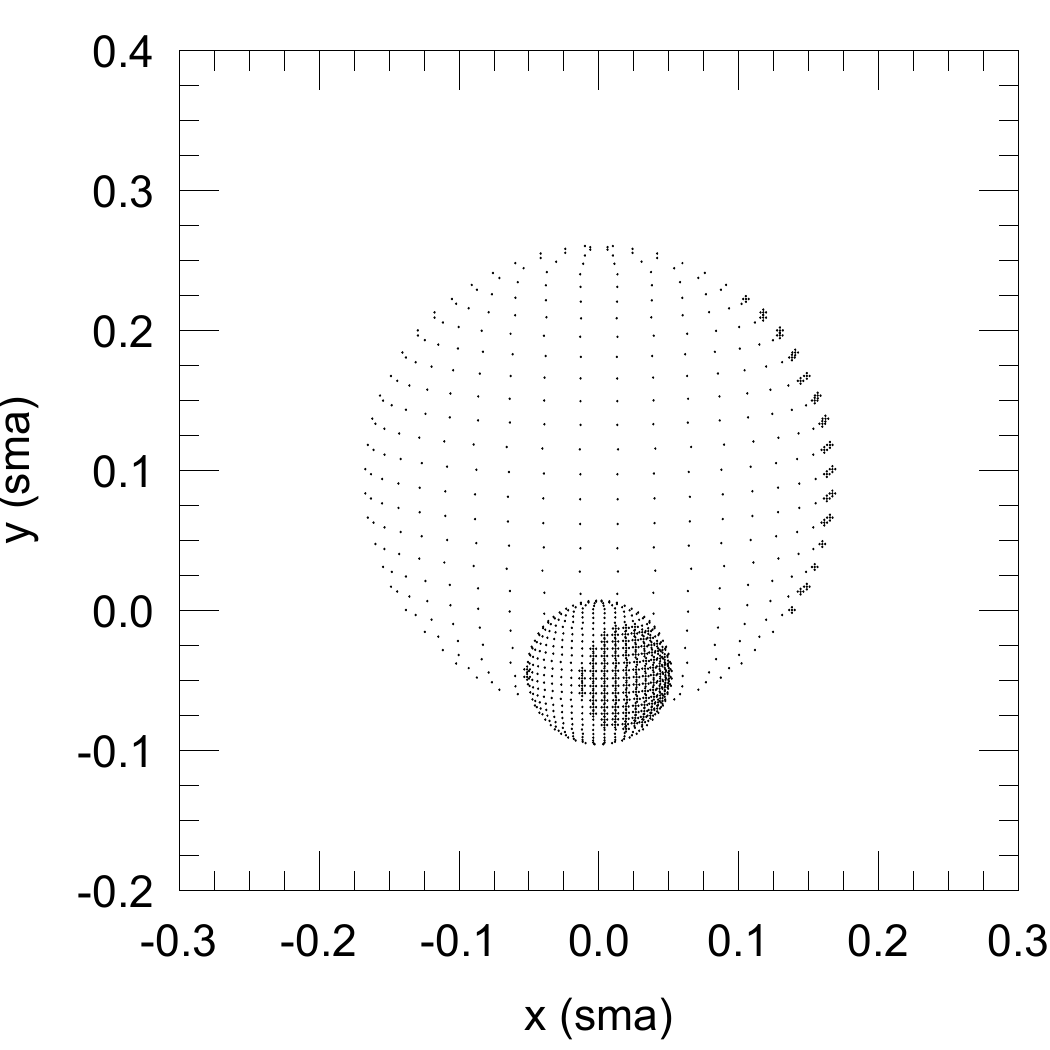}
    \includegraphics[width=0.24\linewidth]{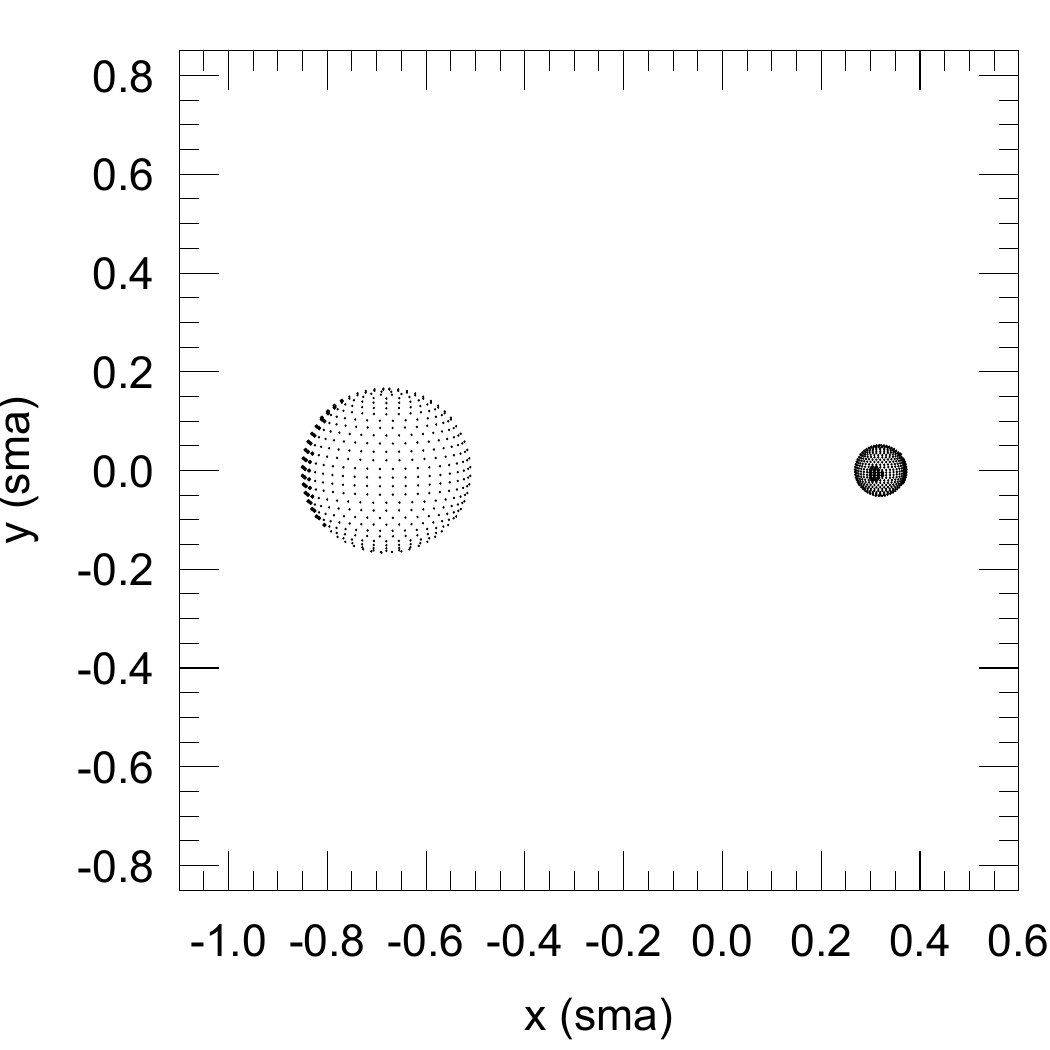}
    
    
    \includegraphics[width=0.24\linewidth]{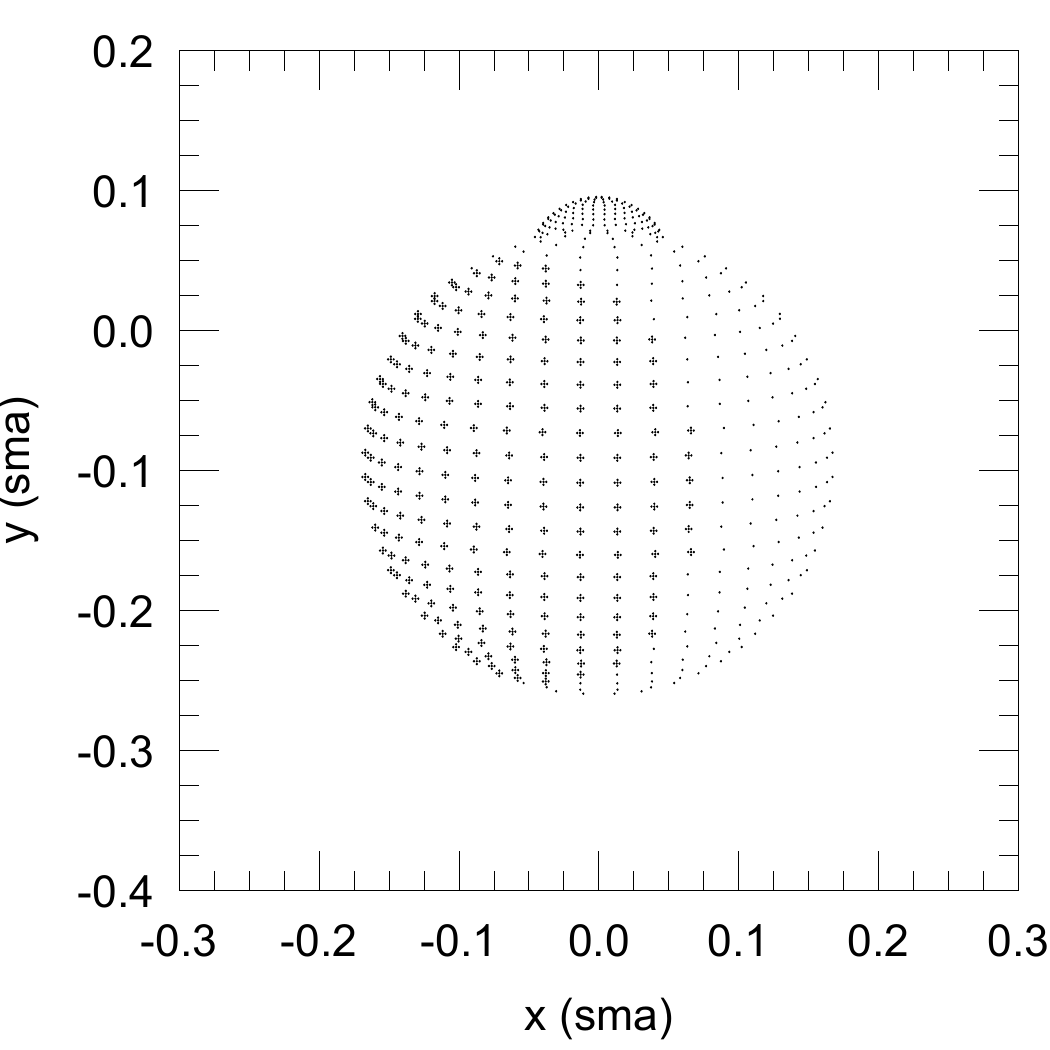}
    \includegraphics[width=0.24\linewidth]{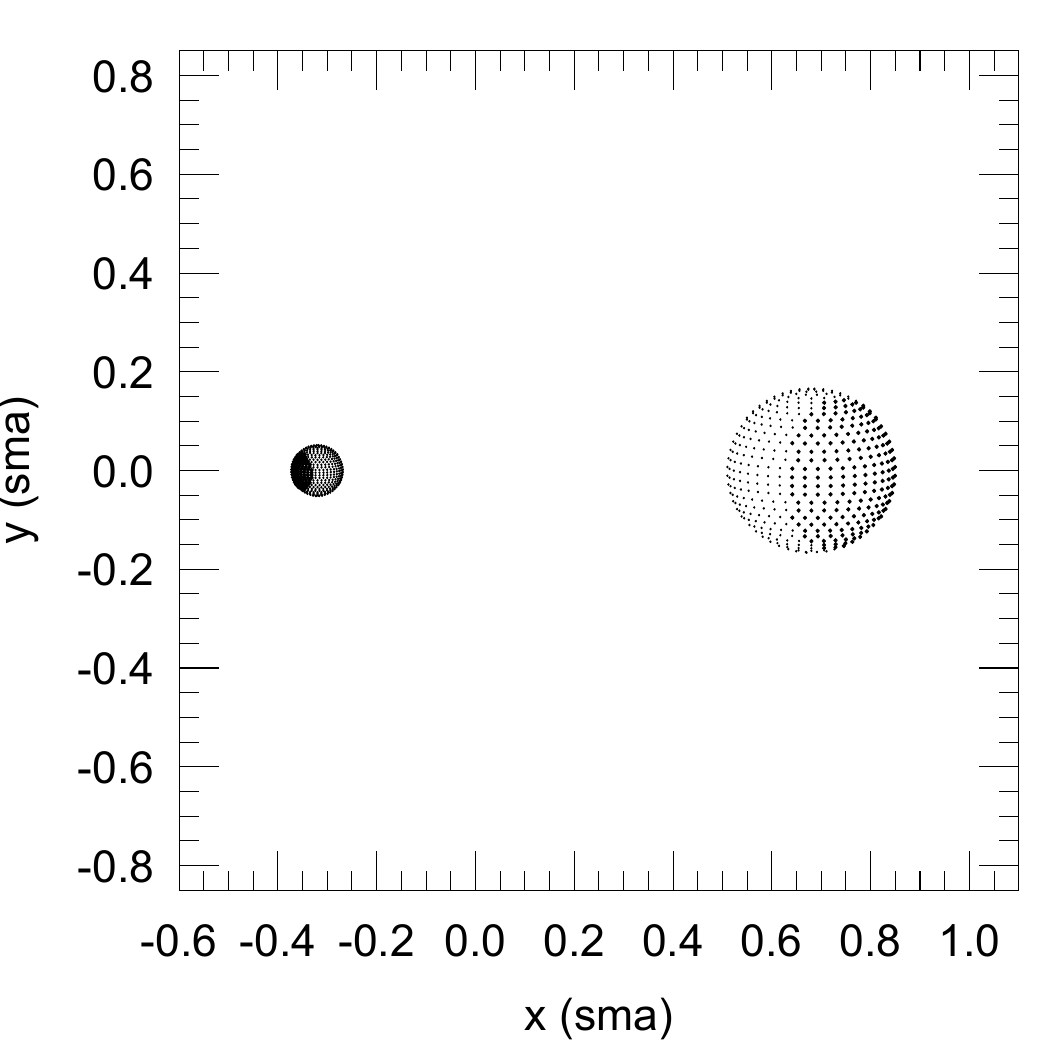}
    \includegraphics[width=0.24\linewidth]{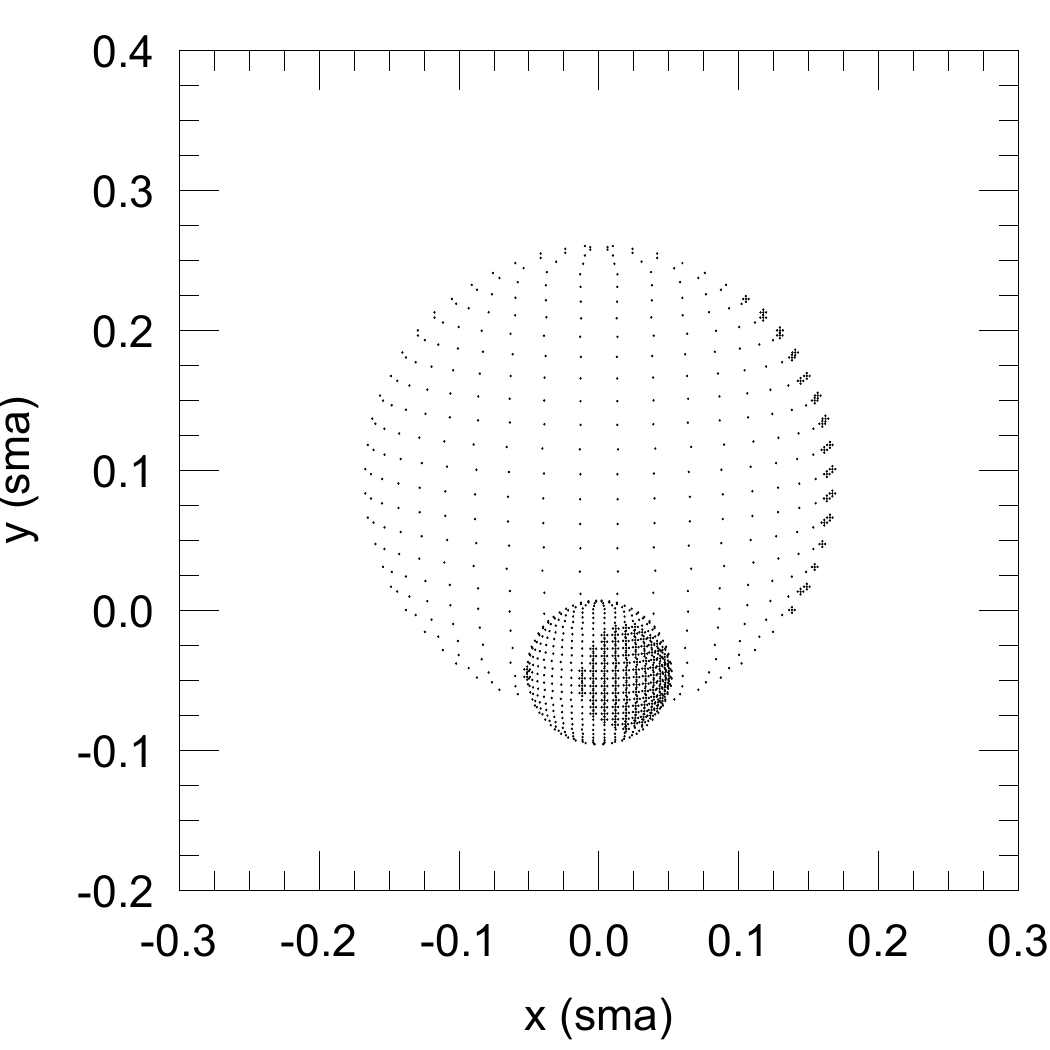}
    \includegraphics[width=0.24\linewidth]{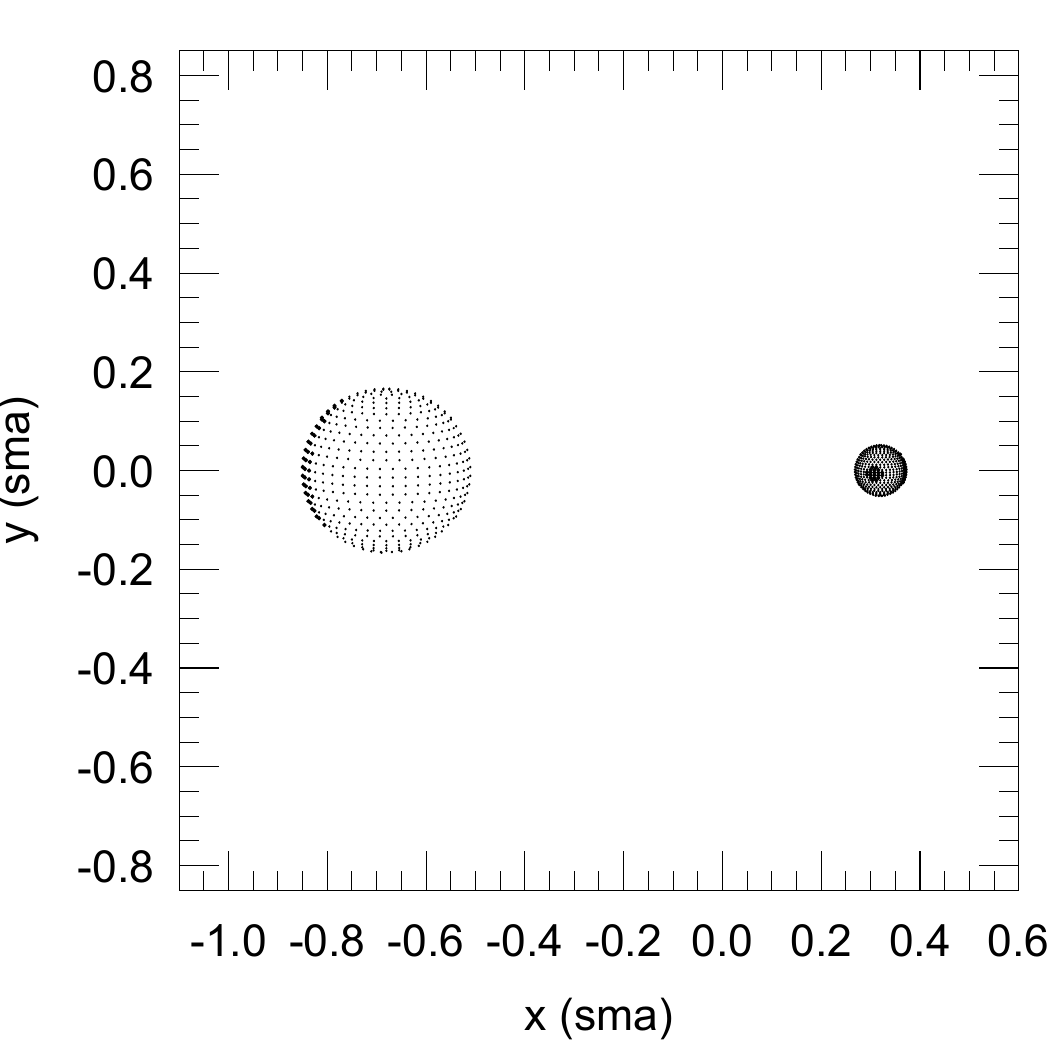}
    
    \caption{Same as Figure~\ref{fig:mesh1}, but for \textit{TESS} Sectors 65, 66, 93, and 94 (from top to bottom, respectively).}
    \label{fig:mesh2}
\end{figure*}

\begin{figure*}[htbp!]
    \centering
    \includegraphics[width=\linewidth]{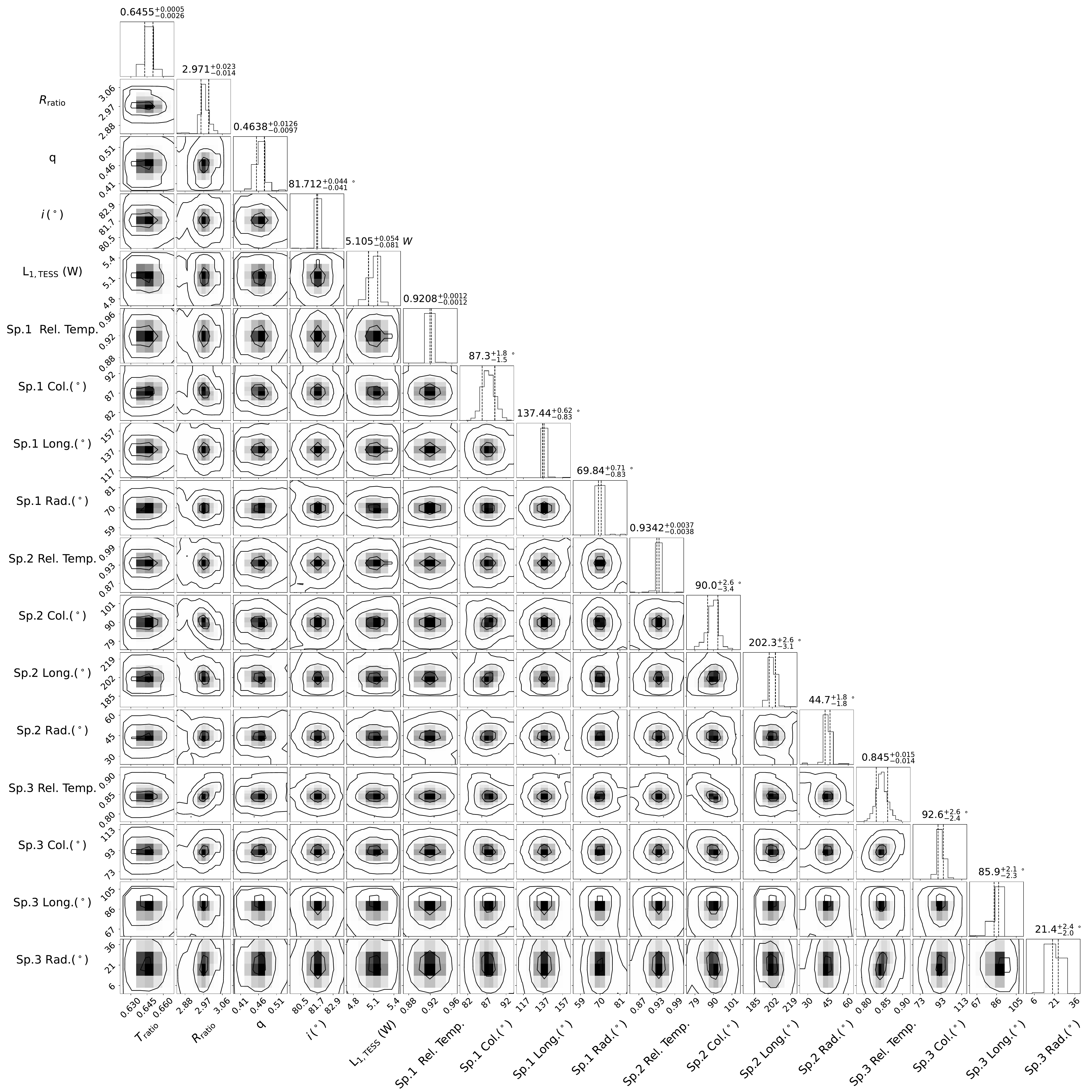}
    \caption{
    Posterior distributions and parameter correlations for the light curve model obtained from the MCMC analysis. The diagonal panels show the marginalized distributions, while the off-diagonal panels illustrate parameter covariances. Contours indicate the highest posterior density regions at 68\%, 95\%, and 99.7\%.
    }
    \label{fig:corner_lc}
\end{figure*}

\subsection{Supplementary SED Model Diagnostics}
\label{app:sed_diagnostics}

This appendix presents supplementary diagnostics associated with
the conditional circumstellar-dust interpretation discussed in
Sect.~\ref{sec:sed_analysis}. These figures are provided to illustrate
the behaviour and parameter degeneracies of the adopted model and should
not be interpreted as independent evidence for the presence of a bound
circumbinary disk.

\begin{figure*}[htbp!]
    \centering
    \includegraphics[width=\linewidth]{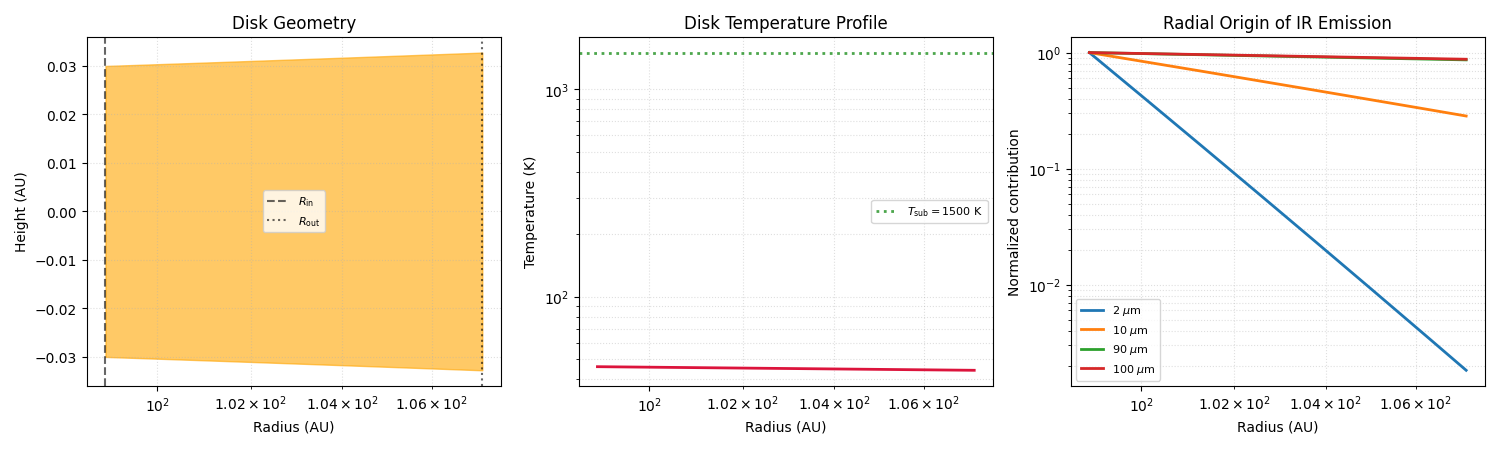}
    \caption{
    Disk-structure diagnostics for the adopted conditional
    circumstellar-dust model.
    \textit{Left:} Model representation of the radial--vertical
    geometry of the parametrized flared disk.
    \textit{Middle:} Temperature profile of the passively irradiated disk,
    following a power-law decrease with radius; the horizontal line
    indicates the approximate dust sublimation temperature.
    \textit{Right:} Predicted radial contribution to the thermal
    emission at representative wavelengths within the adopted model,
    illustrating that progressively longer wavelengths are predicted
    to receive relatively greater contributions from cooler material
    at larger radii. These diagnostics are entirely model-derived and
    do not constitute spatially resolved measurements or a unique
    reconstruction of the circumstellar structure.
    }
    \label{fig:sed_summary}
\end{figure*}

\begin{figure*}[htbp!]
    \centering
    \includegraphics[width=0.8\linewidth]{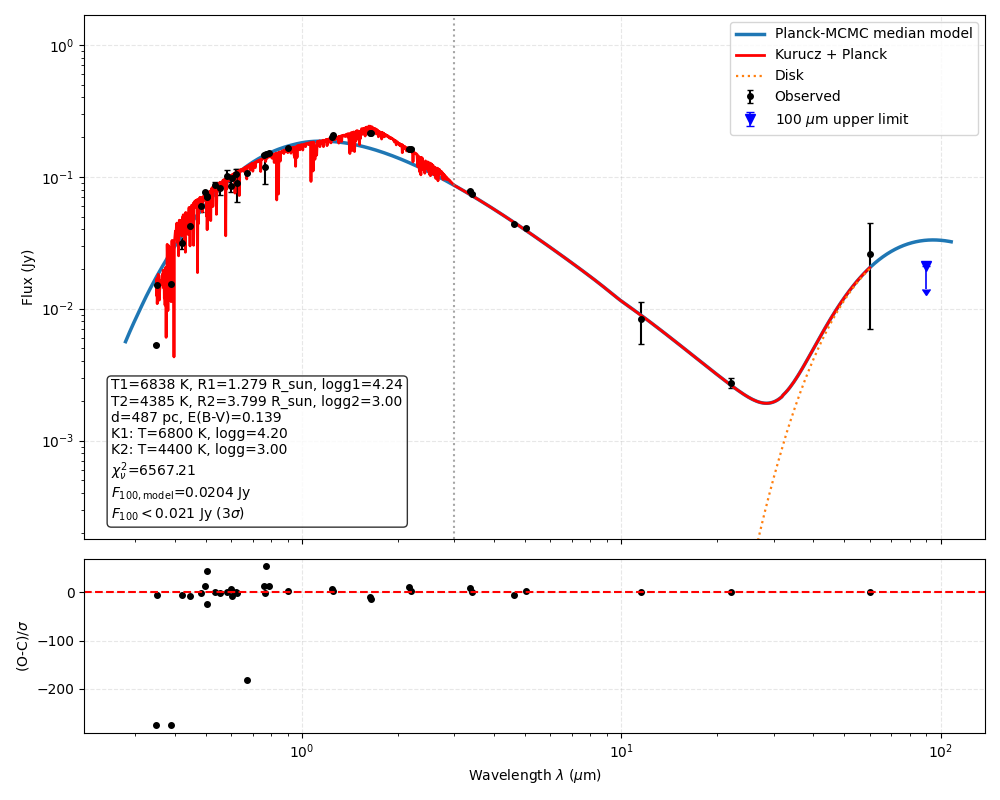}
    \caption{
    Supplementary comparison of the observed SED with alternative
    representations of the stellar photospheres.
    \textit{Top panel:} Black circles show the observed fluxes.
    The solid red curve corresponds to the hybrid model constructed using
    Kurucz stellar atmospheres combined with the
    conditional circumstellar-dust component, while the black
    curve shows the median blackbody-based MCMC model.
    The dotted line represents the dust contribution
    within the conditional model.
    The $90\,\mu$m upper limit is indicated by a blue downward arrow,
    whereas the low-significance
    $60\,\mu$m measurement provides the only positive far-infrared
    deviation from the stellar photospheres.
    The vertical dotted line marks the wavelength at which
    the dust contribution becomes dominant in the conditional
    model and should not be interpreted as an observationally determined
    transition wavelength.
    \textit{Bottom panel:} Normalized residuals $(O-C)/\sigma$ for the
    hybrid model.
    The Kurucz-atmosphere calculation is included only as a
    supplementary consistency check and was not used to derive the SED
    posterior parameters quoted in the main analysis. Both photospheric
    representations support the conclusion that no significant infrared
    excess is required through the WISE W4 band.
    }
    \label{fig:hybrid_sed}
\end{figure*}

The corner plot below shows the posterior distributions obtained
when the circumstellar-dust component is retained as a conditional part
of the SED model. Strong correlations are present among extinction,
stellar temperature, and stellar radius, as expected from broadband SED
fitting. The circumstellar parameters are more weakly constrained, with
broad and asymmetric posterior distributions reflecting the very limited
far-infrared information available for this system. In particular, the
low-significance $60\,\mu$m measurement and the $90\,\mu$m upper limit
do not uniquely constrain the radial structure or dust mass of the
conditional model. The light-curve constraints on the stellar
temperature and radius ratios, together with the \textit{Gaia} distance,
reduce the freedom in the stellar component of the model, but they do
not remove the substantial uncertainties in the circumstellar
parameters. Consequently, these posterior distributions should be
interpreted as parameter constraints conditional on the adopted dust
model rather than as evidence that such a component is required by the
data.

\begin{figure*}[htbp!]
    \centering
    \includegraphics[width=\linewidth]{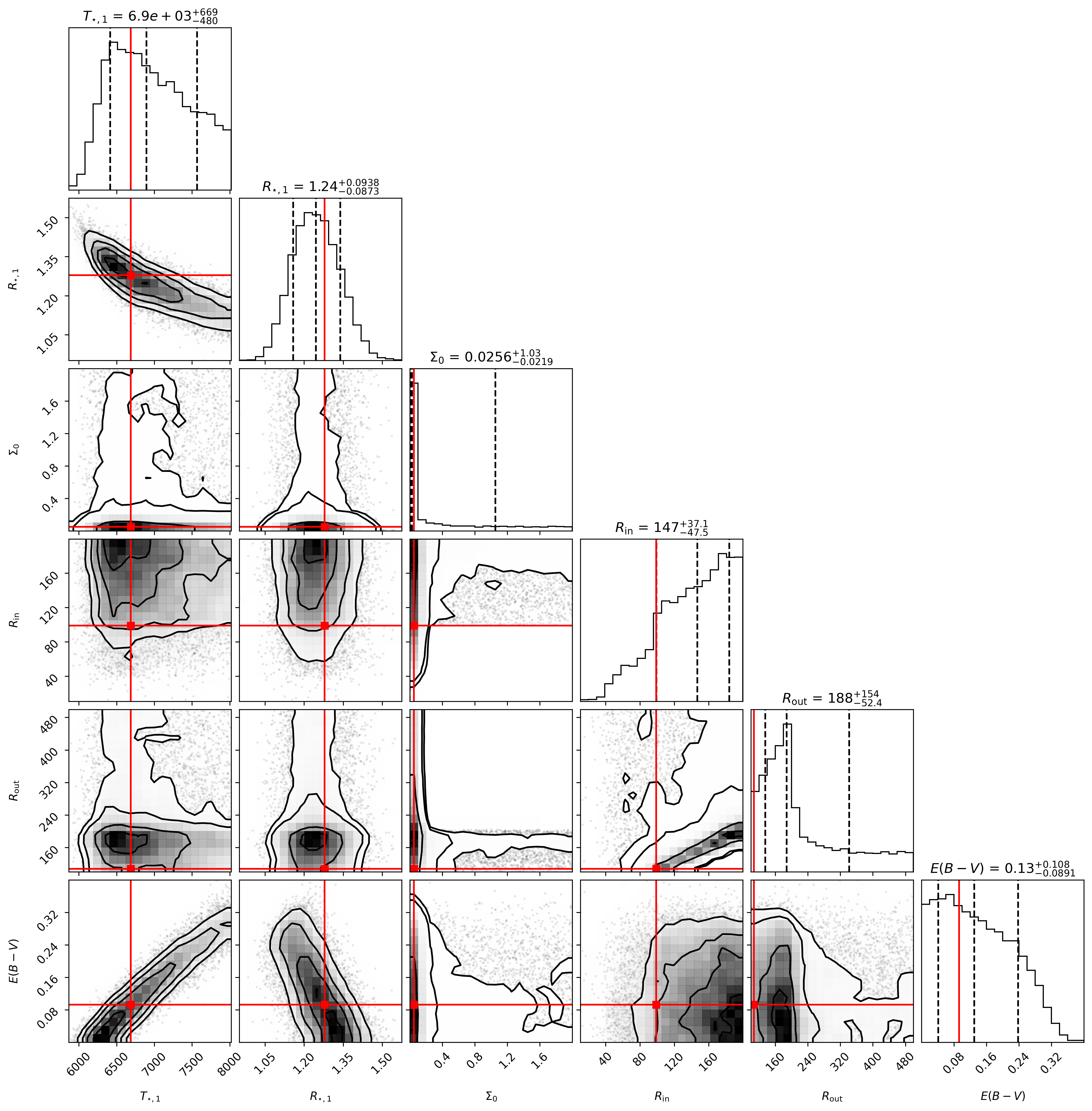}
    \caption{
    Posterior distributions and parameter correlations for the
    conditional circumstellar-dust SED model obtained from the MCMC
    analysis. The diagonal panels show the marginalized distributions,
    while the off-diagonal panels illustrate parameter covariances.
    Contours indicate the highest posterior density regions at 68\%,
    95\%, and 99.7\%.
    The broad and correlated circumstellar posteriors reflect the
    weak far-infrared constraints and should not be interpreted as a
    unique determination of the disk properties.
    }
    \label{fig:corner_sed}
\end{figure*}

\bibliography{CDS_bibfile}{}

@article{Anders2022,
  author  = {Anders, F. and Khalatyan, A. and Queiroz, A. B. A. and
             Chiappini, C. and Ard\`evol, J. and Casamiquela, L. and
             Figueras, F. and Jim\'enez-Arranz, O. and Jordi, C. and
             Mongui\'o, M. and Romero-G\'omez, M. and Altamirano, D. and
             Antoja, T. and Assaad, R. and Cantat-Gaudin, T. and
             Castro-Ginard, A. and Enke, H. and Girardi, L. and
             Guiglion, G. and Khan, S. and Luri, X. and Miglio, A. and
             Minchev, I. and Ramos, P. and Santiago, B. X. and
             Steinmetz, M.},
  title   = {Photo-astrometric distances, extinctions, and astrophysical
             parameters for Gaia EDR3 stars brighter than G = 18.5},
  journal = {Astronomy \& Astrophysics},
  volume  = {658},
  pages   = {A91},
  year    = {2022},
  doi     = {10.1051/0004-6361/202142369}
}

@ARTICLE{TovarMendoza2022,
       author = {{Tovar Mendoza}, Guadalupe and {Davenport}, James R.~A. and {Agol}, Eric and {Jackman}, James A.~G. and {Hawley}, Suzanne L.},
        title = "{Llamaradas Estelares: Modeling the Morphology of White-light Flares}",
      journal = {\aj},
         year = 2022,
        month = jul,
       volume = {164},
       number = {1},
          eid = {17},
        pages = {17},
          doi = {10.3847/1538-3881/ac6fe6},
archivePrefix = {arXiv},
       eprint = {2205.05706},
 primaryClass = {astro-ph.SR},
       adsurl = {https://ui.adsabs.harvard.edu/abs/2022AJ....164...17T}
}

@ARTICLE{Kwee1956,
       author = {{Kwee}, K.~K. and {van Woerden}, H.},
        title = "{A method for computing accurately the epoch of minimum of an eclipsing variable}",
      journal = {\bain},
         year = 1956,
        month = may,
       volume = {12},
        pages = {327},
       adsurl = {https://ui.adsabs.harvard.edu/abs/1956BAN....12..327K}
}

@article{Roccatagliata2018,
  author  = {Roccatagliata, V. and Sacco, G. G. and Franciosini, E. and Randich, S.},
  title   = {The double population of Chamaeleon I detected by Gaia DR2},
  journal = {Astronomy \& Astrophysics},
  year    = {2018},
  volume  = {617},
  pages   = {L4},
  doi     = {10.1051/0004-6361/201833890},
  eprint  = {1808.06931},
  archivePrefix = {arXiv},
  primaryClass  = {astro-ph.SR}
}

@ARTICLE{Carpenter2009,
       author = {{Carpenter}, John M. and {Bouwman}, Jeroen and {Mamajek}, Eric E. and {Meyer}, Michael R. and {Hillenbrand}, Lynne A. and {Backman}, Dana E. and {Henning}, Thomas and {Hines}, Dean C. and {Hollenbach}, David and {Kim}, Jinyoung Serena and {Moro-Martin}, Amaya and {Pascucci}, Ilaria and {Silverstone}, Murray D. and {Stauffer}, John R. and {Wolf}, Sebastian},
        title = "{Formation and Evolution of Planetary Systems: Properties of Debris Dust Around Solar-Type Stars}",
      journal = {\apjs},
         year = 2009,
        month = mar,
       volume = {181},
       number = {1},
        pages = {197-226},
          doi = {10.1088/0067-0049/181/1/197},
archivePrefix = {arXiv},
       eprint = {0810.1003},
 primaryClass = {astro-ph},
       adsurl = {https://ui.adsabs.harvard.edu/abs/2009ApJS..181..197C}
}

@ARTICLE{Christy2022,
       author = {{Christy}, C.~T. and {Jayasinghe}, T. and {Stanek}, K.~Z. and {Kochanek}, C.~S. and {Way}, Z. and {Prieto}, J.~L. and {Shappee}, B.~J. and {Holoien}, T.~W.-S. and {Thompson}, T.~A. and {Schneider}, A.},
        title = "{Citizen ASAS-SN Data Release. I. Variable Star Classification Using Citizen Science}",
      journal = {\pasp},
         year = 2022,
        month = feb,
       volume = {134},
       number = {1032},
          eid = {024201},
        pages = {024201},
          doi = {10.1088/1538-3873/ac44f0},
archivePrefix = {arXiv},
       eprint = {2111.02415},
 primaryClass = {astro-ph.SR},
       adsurl = {https://ui.adsabs.harvard.edu/abs/2022PASP..134b4201C}
}

@ARTICLE{Artymowicz1994,
       author = {{Artymowicz}, Pawel and {Lubow}, Stephen H.},
        title = "{Dynamics of Binary-Disk Interaction. I. Resonances and Disk Gap Sizes}",
      journal = {ApJ},
         year = 1994,
        month = feb,
       volume = {421},
        pages = {651},
          doi = {10.1086/173679},
       adsurl = {https://ui.adsabs.harvard.edu/abs/1994ApJ...421..651A}
}

@ARTICLE{Baraffe2015,
       author = {{Baraffe}, Isabelle and {Homeier}, Derek and {Allard}, France and {Chabrier}, Gilles},
        title = "{New evolutionary models for pre-main sequence and main sequence low-mass stars down to the hydrogen-burning limit}",
      journal = {A\&A},
         year = 2015,
        month = may,
       volume = {577},
          eid = {A42},
        pages = {A42},
          doi = {10.1051/0004-6361/201425481},
archivePrefix = {arXiv},
       eprint = {1503.04107},
 primaryClass = {astro-ph.SR},
       adsurl = {https://ui.adsabs.harvard.edu/abs/2015A&A...577A..42B}
}

@ARTICLE{Stassun2022,
       author = {{Stassun}, Keivan G. and {Torres}, Guillermo and {Kounkel}, Marina and {Feliz}, Dax L. and {Bouma}, Luke G. and {Howell}, Steve B. and {Gnilka}, Crystal L. and {Furlan}, E.},
        title = "{A Low-mass Pre-main-sequence Eclipsing Binary in Lower Centaurus Crux Discovered with TESS}",
      journal = {ApJ},
         year = 2022,
        month = dec,
       volume = {941},
       number = {2},
          eid = {125},
        pages = {125},
          doi = {10.3847/1538-4357/aca32e},
archivePrefix = {arXiv},
       eprint = {2211.07899},
 primaryClass = {astro-ph.SR},
       adsurl = {https://ui.adsabs.harvard.edu/abs/2022ApJ...941..125S}
}

@ARTICLE{David2019,
       author = {{David}, Trevor J. and {Hillenbrand}, Lynne A. and {Gillen}, Edward and {Cody}, Ann Marie and {Howell}, Steve B. and {Isaacson}, Howard T. and {Livingston}, John H.},
        title = "{Age Determination in Upper Scorpius with Eclipsing Binaries}",
      journal = {ApJ},
         year = 2019,
        month = feb,
       volume = {872},
       number = {2},
          eid = {161},
        pages = {161},
          doi = {10.3847/1538-4357/aafe09},
archivePrefix = {arXiv},
       eprint = {1901.05532},
 primaryClass = {astro-ph.SR},
       adsurl = {https://ui.adsabs.harvard.edu/abs/2019ApJ...872..161D}
}

@ARTICLE{GomezMaqueoChew2019,
       author = {{G{\'o}mez Maqueo Chew}, Y. and {Hebb}, L. and {Stempels}, H.~C. and {Paat}, A. and {Stassun}, K.~G. and {Faedi}, F. and {Street}, R.~A. and {Rohn}, G. and {Hellier}, C. and {Anderson}, D.~R.},
        title = "{Fundamental properties of the pre-main sequence eclipsing stars of MML 53 and the mass of the tertiary}",
      journal = {A\&A},
         year = 2019,
        month = mar,
       volume = {623},
          eid = {A23},
        pages = {A23},
          doi = {10.1051/0004-6361/201833299},
archivePrefix = {arXiv},
       eprint = {1901.10611},
 primaryClass = {astro-ph.SR},
       adsurl = {https://ui.adsabs.harvard.edu/abs/2019A&A...623A..23G}
}

@ARTICLE{KounkelStassun2024,
       author = {{Kounkel}, Marina and {Stassun}, Keivan G.},
        title = "{Two Young Eclipsing Binaries in Orion with Temperatures and Radii Affected by Spots and Third Bodies}",
      journal = {AJ},
         year = 2024,
        month = sep,
       volume = {168},
       number = {3},
          eid = {134},
        pages = {134},
          doi = {10.3847/1538-3881/ad6a17},
archivePrefix = {arXiv},
       eprint = {2408.00886},
 primaryClass = {astro-ph.SR},
       adsurl = {https://ui.adsabs.harvard.edu/abs/2024AJ....168..134K}
}

@ARTICLE{Murphy2020,
       author = {{Murphy}, Simon J. and {Lawson}, Warrick A. and {Onken}, Christopher A. and {Yong}, David and {Da Costa}, Gary S. and {Zhou}, George and {Mamajek}, Eric E. and {Bell}, Cameron P.~M. and {Bessell}, Michael S. and {Feinstein}, Adina D.},
        title = "{THOR 42: A touchstone {\ensuremath{\sim}}24 Myr-old eclipsing binary spanning the fully convective boundary}",
      journal = {MNRAS},
         year = 2020,
        month = feb,
       volume = {491},
       number = {4},
        pages = {4902-4924},
          doi = {10.1093/mnras/stz3198},
archivePrefix = {arXiv},
       eprint = {1911.05925},
 primaryClass = {astro-ph.SR},
       adsurl = {https://ui.adsabs.harvard.edu/abs/2020MNRAS.491.4902M}
}

@ARTICLE{Gillen2017,
       author = {{Gillen}, Edward and {Hillenbrand}, Lynne A. and {David}, Trevor J. and {Aigrain}, Suzanne and {Rebull}, Luisa and {Stauffer}, John and {Cody}, Ann Marie and {Queloz}, Didier},
        title = "{New Low-mass Eclipsing Binary Systems in Praesepe Discovered by K2}",
      journal = {ApJ},
         year = 2017,
        month = nov,
       volume = {849},
       number = {1},
          eid = {11},
        pages = {11},
          doi = {10.3847/1538-4357/aa84b3},
archivePrefix = {arXiv},
       eprint = {1706.03084},
 primaryClass = {astro-ph.SR},
       adsurl = {https://ui.adsabs.harvard.edu/abs/2017ApJ...849...11G}
}

@ARTICLE{Stassun2014,
       author = {{Stassun}, Keivan G. and {Feiden}, Gregory A. and {Torres}, Guillermo},
        title = "{Empirical tests of pre-main-sequence stellar evolution models with eclipsing binaries}",
      journal = {NewAR},
         year = 2014,
        month = jun,
       volume = {60},
        pages = {1-28},
          doi = {10.1016/j.newar.2014.06.001},
archivePrefix = {arXiv},
       eprint = {1406.3788},
 primaryClass = {astro-ph.SR},
       adsurl = {https://ui.adsabs.harvard.edu/abs/2014NewAR..60....1S}
}

@ARTICLE{Grossschedl2019,
       author = {{Gro{\ss}schedl}, Josefa Elisabeth and {Alves}, Jo{\~a}o and {Teixeira}, Paula S. and {Bouy}, Herv{\'e} and {Forbrich}, Jan and {Lada}, Charles J. and {Meingast}, Stefan and {Hacar}, {\'A}lvaro and {Ascenso}, Joana and {Ackerl}, Christine and {Hasenberger}, Birgit and {K{\"o}hler}, Rainer and {Kubiak}, Karolina and {Larreina}, Irati and {Linhardt}, Lorenz and {Lombardi}, Marco and {M{\"o}ller}, Torsten},
        title = "{VISION - Vienna survey in Orion. III. Young stellar objects in Orion A}",
      journal = {A\&A},
         year = 2019,
        month = feb,
       volume = {622},
          eid = {A149},
        pages = {A149},
          doi = {10.1051/0004-6361/201832577},
archivePrefix = {arXiv},
       eprint = {1810.00878},
 primaryClass = {astro-ph.SR},
       adsurl = {https://ui.adsabs.harvard.edu/abs/2019A&A...622A.149G}
}

@ARTICLE{vizier2000,
       author = {{Ochsenbein}, F. and {Bauer}, P. and {Marcout}, J.},
        title = "{The VizieR database of astronomical catalogues}",
      journal = {A\&AS},
         year = 2000,
        month = apr,
       volume = {143},
        pages = {23-32},
          doi = {10.1051/aas:2000169},
archivePrefix = {arXiv},
       eprint = {astro-ph/0002122},
 primaryClass = {astro-ph},
       adsurl = {https://ui.adsabs.harvard.edu/abs/2000A&AS..143...23O}
}

@ARTICLE{Gillen2020,
       author = {{Gillen}, Edward and {Hillenbrand}, Lynne A. and {Stauffer}, John and {Aigrain}, Suzanne and {Rebull}, Luisa and {Cody}, Ann Marie},
        title = "{Mon-735: a new low-mass pre-main-sequence eclipsing binary in NGC 2264}",
      journal = {MNRAS},
         year = 2020,
        month = jun,
       volume = {495},
       number = {2},
        pages = {1531-1548},
          doi = {10.1093/mnras/staa1016},
archivePrefix = {arXiv},
       eprint = {2004.04753},
 primaryClass = {astro-ph.SR},
       adsurl = {https://ui.adsabs.harvard.edu/abs/2020MNRAS.495.1531G}
}

@ARTICLE{Tofflemire2023,
       author = {{Tofflemire}, Benjamin M. and {Kraus}, Adam L. and {Mann}, Andrew W. and {Newton}, Elisabeth R. and {Gully-Santiago}, Michael A. and {Vanderburg}, Andrew and {Waalkes}, William C. and {Berta-Thompson}, Zachory K. and {Collins}, Kevin I. and {Collins}, Karen A. and {Nielsen}, Louise D. and {Bouchy}, Fran{\c{c}}ois and {Ziegler}, Carl and {Brice{\~n}o}, C{\'e}sar and {Law}, Nicholas M.},
        title = "{A Low-mass, Pre-main-sequence Eclipsing Binary in the 40 Myr Columba Association-Fundamental Stellar Parameters and Modeling the Effect of Star Spots}",
      journal = {AJ},
         year = 2023,
        month = feb,
       volume = {165},
       number = {2},
          eid = {46},
        pages = {46},
          doi = {10.3847/1538-3881/aca60f},
archivePrefix = {arXiv},
       eprint = {2210.10789},
 primaryClass = {astro-ph.SR},
       adsurl = {https://ui.adsabs.harvard.edu/abs/2023AJ....165...46T}
}

@ARTICLE{Bakis2026,
       author = {{Bak{\i}{\textcommabelow s}}, Volkan and {Habal{\i}}, Ay{\textcommabelow s}e Yadikar},
        title = "{Parametric SED Modelling of Protoplanetary Discs: Validation and Application to An Unstudied YSO}",
      journal = {Physics and Astronomy Reports},
         year = 2026,
        month = jun,
       volume = {4},
       number = {1},
        pages = {29-39},
          doi = {10.26650/PAR.2026.00005},
archivePrefix = {arXiv},
       eprint = {2604.03211},
 primaryClass = {astro-ph.SR},
       adsurl = {https://ui.adsabs.harvard.edu/abs/2026PARep...4...29B}
}

@ARTICLE{Gordon2023,
       author = {{Gordon}, Karl D. and
                 {Clayton}, Geoffrey C. and
                 {Decleir}, Marjorie and
                 {Fitzpatrick}, E. L. and
                 {Massa}, Derck and
                 {Misselt}, Karl A. and
                 {Tollerud}, Erik J.},
        title = "{One Relation for All Wavelengths: The Far-Ultraviolet
                  to Mid-Infrared Milky Way Spectroscopic
                  R(V)-dependent Dust Extinction Relationship}",
      journal = {\apj},
         year = 2023,
        month = jun,
       volume = {950},
       number = {2},
          eid = {86},
        pages = {86},
          doi = {10.3847/1538-4357/accb59},
archivePrefix = {arXiv},
       eprint = {2304.01991},
 primaryClass = {astro-ph.GA}
}

@ARTICLE{Huang2020,
       author = {{Huang}, Chelsea X. and {Vanderburg}, Andrew and {P{\'a}l}, Andras and {Sha}, Lizhou and {Yu}, Liang and {Fong}, Willie and {Fausnaugh}, Michael and {Shporer}, Avi and {Guerrero}, Natalia and {Vanderspek}, Roland and {Ricker}, George},
        title = "{Photometry of 10 Million Stars from the First Two Years of TESS Full Frame Images: Part I}",
      journal = {Research Notes of the American Astronomical Society},
         year = 2020,
        month = nov,
       volume = {4},
       number = {11},
          eid = {204},
        pages = {204},
          doi = {10.3847/2515-5172/abca2e},
archivePrefix = {arXiv},
       eprint = {2011.06459},
 primaryClass = {astro-ph.EP},
       adsurl = {https://ui.adsabs.harvard.edu/abs/2020RNAAS...4..204H}
}

@ARTICLE{Jayasinghe2019,
       author = {{Jayasinghe}, T. and {Stanek}, K.~Z. and {Kochanek}, C.~S. and {Shappee}, B.~J. and {Holoien}, T.~W.-S. and {Thompson}, Todd A. and {Prieto}, J.~L. and {Dong}, Subo and {Pawlak}, M. and {Pejcha}, O. and {Shields}, J.~V. and {Pojmanski}, G. and {Otero}, S. and {Hurst}, N. and {Britt}, C.~A. and {Will}, D.},
        title = "{The ASAS-SN catalogue of variable stars III: variables in the southern TESS continuous viewing zone}",
      journal = {MNRAS},
         year = 2019,
        month = may,
       volume = {485},
       number = {1},
        pages = {961-971},
          doi = {10.1093/mnras/stz444},
archivePrefix = {arXiv},
       eprint = {1901.00009},
 primaryClass = {astro-ph.SR},
       adsurl = {https://ui.adsabs.harvard.edu/abs/2019MNRAS.485..961J}
}

@INPROCEEDINGS{Shappee2014,
       author = {{Shappee}, Benjamin and {Prieto}, J. and {Stanek}, K.~Z. and {Kochanek}, C.~S. and {Holoien}, T. and {Jencson}, J. and {Basu}, U. and {Beacom}, J.~F. and {Szczygiel}, D. and {Pojmanski}, G. and {Brimacombe}, J. and {Dubberley}, M. and {Elphick}, M. and {Foale}, S. and {Hawkins}, E. and {Mullins}, D. and {Rosing}, W. and {Ross}, R. and {Walker}, Z.},
        title = "{All Sky Automated Survey for SuperNovae (ASAS-SN or ``Assassin'')}",
    booktitle = {American Astronomical Society Meeting Abstracts \#223},
         year = 2014,
       series = {American Astronomical Society Meeting Abstracts},
       volume = {223},
        month = jan,
          eid = {236.03},
        pages = {236.03},
       adsurl = {https://ui.adsabs.harvard.edu/abs/2014AAS...22323603S}
}

@ARTICLE{isophot1996,
       author = {{Lemke}, D. and {Klaas}, U. and {Abolins}, J. and {Abraham}, P. and {Acosta-Pulido}, J. and {Bogun}, S. and {Castaneda}, H. and {Cornwall}, L. and {Drury}, L. and {Gabriel}, C. and {Garzon}, F. and {Gemuend}, H.~P. and {Groezinger}, U. and {Gruen}, E. and {Haas}, M. and {Hajduk}, C. and {Hall}, G. and {Heinrichsen}, I. and {Herbstmeier}, U. and {Hirth}, G. and {Joseph}, R. and {Kinkel}, U. and {Kirches}, S. and {Koempe}, C. and {Kraetschmer}, W. and {Kreysa}, E. and {Krueger}, H. and {Kunkel}, M. and {Laureijs}, R. and {Luetzow-Wentzky}, P. and {Mattila}, K. and {Mueller}, T. and {Pacher}, T. and {Pelz}, G. and {Popow}, E. and {Rasmussen}, I. and {Rodriguez Espinosa}, J. and {Richards}, P. and {Russell}, S. and {Schnopper}, H. and {Schubert}, J. and {Schulz}, B. and {Telesco}, C. and {Tilgner}, C. and {Tuffs}, R. and {Voelk}, H. and {Walker}, H. and {Wells}, M. and {Wolf}, J.},
        title = "{ISOPHOT - capabilities and performance.}",
      journal = {A\&A},
         year = 1996,
        month = nov,
       volume = {315},
        pages = {L64-L70},
       adsurl = {https://ui.adsabs.harvard.edu/abs/1996A&A...315L..64L}
}

@MISC{lk,
   author = {{Lightkurve Collaboration} and {Cardoso}, J.~V.~d.~M. and
             {Hedges}, C. and {Gully-Santiago}, M. and {Saunders}, N. and
             {Cody}, A.~M. and {Barclay}, T. and {Hall}, O. and
             {Sagear}, S. and {Turtelboom}, E. and {Zhang}, J. and
             {Tzanidakis}, A. and {Mighell}, K. and {Coughlin}, J. and
             {Bell}, K. and {Berta-Thompson}, Z. and {Williams}, P. and
             {Dotson}, J. and {Barentsen}, G.},
    title = "{Lightkurve: Kepler and TESS time series analysis in Python}",
howpublished = {Astrophysics Source Code Library},
     year = 2018,
    month = dec,
archivePrefix = "ascl",
   eprint = {1812.013},
   adsurl = {http://adsabs.harvard.edu/abs/2018ascl.soft12013L},
}

@ARTICLE{MML48_2025,
  author  = {Y. Gomez Maqueo Chew and L. Hebb},
  title   = {Discovery of the pre-main-sequence eclipsing binary MML 48},
  journal = {Astronomy and Astrophysics},
  year    = {2025},
  volume  = {702},
  pages   = {17}
}

@ARTICLE{Serenelli2021,
       author = {{Serenelli}, Aldo and {Weiss}, Achim and {Aerts}, Conny and {Angelou}, George C. and {Baroch}, David and {Bastian}, Nate and {Beck}, Paul G. and {Bergemann}, Maria and {Bestenlehner}, Joachim M. and {Czekala}, Ian and {Elias-Rosa}, Nancy and {Escorza}, Ana and {Van Eylen}, Vincent and {Feuillet}, Diane K. and {Gandolfi}, Davide and {Gieles}, Mark and {Girardi}, L{\'e}o and {Lebreton}, Yveline and {Lodieu}, Nicolas and {Martig}, Marie and {Miller Bertolami}, Marcelo M. and {Mombarg}, Joey S.~G. and {Morales}, Juan Carlos and {Moya}, Andr{\'e}s and {Nsamba}, Benard and {Pavlovski}, Kre{\v{s}}imir and {Pedersen}, May G. and {Ribas}, Ignasi and {Schneider}, Fabian R.~N. and {Silva Aguirre}, Victor and {Stassun}, Keivan G. and {Tolstoy}, Eline and {Tremblay}, Pier-Emmanuel and {Zwintz}, Konstanze},
        title = "{Weighing stars from birth to death: mass determination methods across the HRD}",
      journal = {A\&ARev},
         year = 2021,
        month = dec,
       volume = {29},
       number = {1},
          eid = {4},
        pages = {4},
          doi = {10.1007/s00159-021-00132-9},
archivePrefix = {arXiv},
       eprint = {2006.10868},
 primaryClass = {astro-ph.SR},
       adsurl = {https://ui.adsabs.harvard.edu/abs/2021A&ARv..29....4S}
}

@ARTICLE{Spangler2001,
       author = {{Spangler}, C. and {Sargent}, A.~I. and {Silverstone}, M.~D. and {Becklin}, E.~E. and {Zuckerman}, B.},
        title = "{Dusty Debris around Solar-Type Stars: Temporal Disk Evolution}",
      journal = {ApJ},
         year = 2001,
        month = jul,
       volume = {555},
       number = {2},
        pages = {932-944},
          doi = {10.1086/321490},
archivePrefix = {arXiv},
       eprint = {astro-ph/0103185},
 primaryClass = {astro-ph},
       adsurl = {https://ui.adsabs.harvard.edu/abs/2001ApJ...555..932S}
}

@ARTICLE{Torres2010,
       author = {{Torres}, G. and {Andersen}, J. and {Gim{\'e}nez}, A.},
        title = "{Accurate masses and radii of normal stars: modern results and applications}",
      journal = {A\&ARev},
         year = 2010,
        month = feb,
       volume = {18},
       number = {1-2},
        pages = {67-126},
          doi = {10.1007/s00159-009-0025-1},
archivePrefix = {arXiv},
       eprint = {0908.2624},
 primaryClass = {astro-ph.SR},
       adsurl = {https://ui.adsabs.harvard.edu/abs/2010A&ARv..18...67T}
}

@ARTICLE{phoebe1,
       author = {{Pr{\v{s}}a}, A. and {Zwitter}, T.},
        title = "{A Computational Guide to Physics of Eclipsing Binaries. I. Demonstrations and Perspectives}",
      journal = {\apj},
         year = 2005,
        month = jul,
       volume = {628},
       number = {1},
        pages = {426-438},
          doi = {10.1086/430591},
archivePrefix = {arXiv},
       eprint = {astro-ph/0503361},
 primaryClass = {astro-ph},
       adsurl = {https://ui.adsabs.harvard.edu/abs/2005ApJ...628..426P}
}

@ARTICLE{phoebe2,
       author = {{Pr{\v{s}}a}, A. and {Conroy}, K.~E. and {Horvat}, M. and {Pablo}, H. and {Kochoska}, A. and {Bloemen}, S. and {Giammarco}, J. and {Hambleton}, K.~M. and {Degroote}, P.},
        title = "{Physics Of Eclipsing Binaries. II. Toward the Increased Model Fidelity}",
      journal = {\apjs},
         year = 2016,
        month = dec,
       volume = {227},
       number = {2},
          eid = {29},
        pages = {29},
          doi = {10.3847/1538-4365/227/2/29},
archivePrefix = {arXiv},
       eprint = {1609.08135},
 primaryClass = {astro-ph.SR},
       adsurl = {https://ui.adsabs.harvard.edu/abs/2016ApJS..227...29P}
}

@ARTICLE{phoebe3,
       author = {{Horvat}, Martin and {Conroy}, Kyle E. and {Pablo}, Herbert and {Hambleton}, Kelly M. and {Kochoska}, Angela and {Giammarco}, Joseph and {Pr{\v{s}}a}, Andrej},
        title = "{Physics of Eclipsing Binaries. III. Spin-Orbit Misalignment}",
      journal = {\apjs},
         year = 2018,
        month = aug,
       volume = {237},
       number = {2},
          eid = {26},
        pages = {26},
          doi = {10.3847/1538-4365/aacd0f},
archivePrefix = {arXiv},
       eprint = {1806.07680},
 primaryClass = {astro-ph.SR},
       adsurl = {https://ui.adsabs.harvard.edu/abs/2018ApJS..237...26H}
}

@ARTICLE{phoebe4,
       author = {{Jones}, David and {Conroy}, Kyle E. and {Horvat}, Martin and {Giammarco}, Joseph and {Kochoska}, Angela and {Pablo}, Herbert and {Brown}, Alex J. and {Sowicka}, Paulina and {Pr{\v{s}}a}, Andrej},
        title = "{Physics of Eclipsing Binaries. IV. The Impact of Interstellar Extinction on the Light Curves of Eclipsing Binaries}",
      journal = {\apjs},
         year = 2020,
        month = apr,
       volume = {247},
       number = {2},
          eid = {63},
        pages = {63},
          doi = {10.3847/1538-4365/ab7927},
archivePrefix = {arXiv},
       eprint = {1912.09474},
 primaryClass = {astro-ph.SR},
       adsurl = {https://ui.adsabs.harvard.edu/abs/2020ApJS..247...63J}
}

@ARTICLE{phoebe5,
       author = {{Conroy}, Kyle E. and {Kochoska}, Angela and {Hey}, Daniel and {Pablo}, Herbert and {Hambleton}, Kelly M. and {Jones}, David and {Giammarco}, Joseph and {Abdul-Masih}, Michael and {Pr{\v{s}}a}, Andrej},
        title = "{Physics of Eclipsing Binaries. V. General Framework for Solving the Inverse Problem}",
      journal = {\apjs},
         year = 2020,
        month = oct,
       volume = {250},
       number = {2},
          eid = {34},
        pages = {34},
          doi = {10.3847/1538-4365/abb4e2},
archivePrefix = {arXiv},
       eprint = {2006.16951},
 primaryClass = {astro-ph.SR},
       adsurl = {https://ui.adsabs.harvard.edu/abs/2020ApJS..250...34C}
}

@ARTICLE{mcmc,
       author = {{Foreman-Mackey}, Daniel and {Hogg}, David W. and {Lang}, Dustin and {Goodman}, Jonathan},
        title = "{emcee: The MCMC Hammer}",
      journal = {\pasp},
         year = 2013,
        month = mar,
       volume = {125},
       number = {925},
        pages = {306},
          doi = {10.1086/670067},
archivePrefix = {arXiv},
       eprint = {1202.3665},
 primaryClass = {astro-ph.IM},
       adsurl = {https://ui.adsabs.harvard.edu/abs/2013PASP..125..306F}
}

@ARTICLE{astro1,
       author = {{Astropy Collaboration} and {Robitaille}, Thomas P. and {Tollerud}, Erik J. and {Greenfield}, Perry and {Droettboom}, Michael and {Bray}, Erik and {Aldcroft}, Tom and {Davis}, Matt and {Ginsburg}, Adam and {Price-Whelan}, Adrian M. and {Kerzendorf}, Wolfgang E. and {Conley}, Alexander and {Crighton}, Neil and {Barbary}, Kyle and {Muna}, Demitri and {Ferguson}, Henry and {Grollier}, Fr{\'e}d{\'e}ric and {Parikh}, Madhura M. and {Nair}, Prasanth H. and {Unther}, Hans M. and {Deil}, Christoph and {Woillez}, Julien and {Conseil}, Simon and {Kramer}, Roban and {Turner}, James E.~H. and {Singer}, Leo and {Fox}, Ryan and {Weaver}, Benjamin A. and {Zabalza}, Victor and {Edwards}, Zachary I. and {Azalee Bostroem}, K. and {Burke}, D.~J. and {Casey}, Andrew R. and {Crawford}, Steven M. and {Dencheva}, Nadia and {Ely}, Justin and {Jenness}, Tim and {Labrie}, Kathleen and {Lim}, Pey Lian and {Pierfederici}, Francesco and {Pontzen}, Andrew and {Ptak}, Andy and {Refsdal}, Brian and {Servillat}, Mathieu and {Streicher}, Ole},
        title = "{Astropy: A community Python package for astronomy}",
      journal = {\aap},
         year = 2013,
        month = oct,
       volume = {558},
          eid = {A33},
        pages = {A33},
          doi = {10.1051/0004-6361/201322068},
archivePrefix = {arXiv},
       eprint = {1307.6212},
 primaryClass = {astro-ph.IM},
       adsurl = {https://ui.adsabs.harvard.edu/abs/2013A&A...558A..33A}
}

@ARTICLE{astro2,
       author = {{Astropy Collaboration} and {Price-Whelan}, A.~M. and {Sip{\H{o}}cz}, B.~M. and {G{\"u}nther}, H.~M. and {Lim}, P.~L. and {Crawford}, S.~M. and {Conseil}, S. and {Shupe}, D.~L. and {Craig}, M.~W. and {Dencheva}, N. and {Ginsburg}, A. and {VanderPlas}, J.~T. and {Bradley}, L.~D. and {P{\'e}rez-Su{\'a}rez}, D. and {de Val-Borro}, M. and {Aldcroft}, T.~L. and {Cruz}, K.~L. and {Robitaille}, T.~P. and {Tollerud}, E.~J. and {Ardelean}, C. and {Babej}, T. and {Bach}, Y.~P. and {Bachetti}, M. and {Bakanov}, A.~V. and {Bamford}, S.~P. and {Barentsen}, G. and {Barmby}, P. and {Baumbach}, A. and {Berry}, K.~L. and {Biscani}, F. and {Boquien}, M. and {Bostroem}, K.~A. and {Bouma}, L.~G. and {Brammer}, G.~B. and {Bray}, E.~M. and {Breytenbach}, H. and {Buddelmeijer}, H. and {Burke}, D.~J. and {Calderone}, G. and {Cano Rodr{\'\i}guez}, J.~L. and {Cara}, M. and {Cardoso}, J.~V.~M. and {Cheedella}, S. and {Copin}, Y. and {Corrales}, L. and {Crichton}, D. and {D'Avella}, D. and {Deil}, C. and {Depagne}, {\'E}. and {Dietrich}, J.~P. and {Donath}, A. and {Droettboom}, M. and {Earl}, N. and {Erben}, T. and {Fabbro}, S. and {Ferreira}, L.~A. and {Finethy}, T. and {Fox}, R.~T. and {Garrison}, L.~H. and {Gibbons}, S.~L.~J. and {Goldstein}, D.~A. and {Gommers}, R. and {Greco}, J.~P. and {Greenfield}, P. and {Groener}, A.~M. and {Grollier}, F. and {Hagen}, A. and {Hirst}, P. and {Homeier}, D. and {Horton}, A.~J. and {Hosseinzadeh}, G. and {Hu}, L. and {Hunkeler}, J.~S. and {Ivezi{\'c}}, {\v{Z}}. and {Jain}, A. and {Jenness}, T. and {Kanarek}, G. and {Kendrew}, S. and {Kern}, N.~S. and {Kerzendorf}, W.~E. and {Khvalko}, A. and {King}, J. and {Kirkby}, D. and {Kulkarni}, A.~M. and {Kumar}, A. and {Lee}, A. and {Lenz}, D. and {Littlefair}, S.~P. and {Ma}, Z. and {Macleod}, D.~M. and {Mastropietro}, M. and {McCully}, C. and {Montagnac}, S. and {Morris}, B.~M. and {Mueller}, M. and {Mumford}, S.~J. and {Muna}, D. and {Murphy}, N.~A. and {Nelson}, S. and {Nguyen}, G.~H. and {Ninan}, J.~P. and {N{\"o}the}, M. and {Ogaz}, S. and {Oh}, S. and {Parejko}, J.~K. and {Parley}, N. and {Pascual}, S. and {Patil}, R. and {Patil}, A.~A. and {Plunkett}, A.~L. and {Prochaska}, J.~X. and {Rastogi}, T. and {Reddy Janga}, V. and {Sabater}, J. and {Sakurikar}, P. and {Seifert}, M. and {Sherbert}, L.~E. and {Sherwood-Taylor}, H. and {Shih}, A.~Y. and {Sick}, J. and {Silbiger}, M.~T. and {Singanamalla}, S. and {Singer}, L.~P. and {Sladen}, P.~H. and {Sooley}, K.~A. and {Sornarajah}, S. and {Streicher}, O. and {Teuben}, P. and {Thomas}, S.~W. and {Tremblay}, G.~R. and {Turner}, J.~E.~H. and {Terr{\'o}n}, V. and {van Kerkwijk}, M.~H. and {de la Vega}, A. and {Watkins}, L.~L. and {Weaver}, B.~A. and {Whitmore}, J.~B. and {Woillez}, J. and {Zabalza}, V. and {Astropy Contributors}},
        title = "{The Astropy Project: Building an Open-science Project and Status of the v2.0 Core Package}",
      journal = {\aj},
         year = 2018,
        month = sep,
       volume = {156},
       number = {3},
          eid = {123},
        pages = {123},
          doi = {10.3847/1538-3881/aabc4f},
archivePrefix = {arXiv},
       eprint = {1801.02634},
 primaryClass = {astro-ph.IM},
       adsurl = {https://ui.adsabs.harvard.edu/abs/2018AJ....156..123A}
}

@ARTICLE{astro3,
       author = {{Astropy Collaboration} and {Price-Whelan}, Adrian M. and {Lim}, Pey Lian and {Earl}, Nicholas and {Starkman}, Nathaniel and {Bradley}, Larry and {Shupe}, David L. and {Patil}, Aarya A. and {Corrales}, Lia and {Brasseur}, C.~E. and {N{\"o}the}, Maximilian and {Donath}, Axel and {Tollerud}, Erik and {Morris}, Brett M. and {Ginsburg}, Adam and {Vaher}, Eero and {Weaver}, Benjamin A. and {Tocknell}, James and {Jamieson}, William and {van Kerkwijk}, Marten H. and {Robitaille}, Thomas P. and {Merry}, Bruce and {Bachetti}, Matteo and {G{\"u}nther}, H. Moritz and {Aldcroft}, Thomas L. and {Alvarado-Montes}, Jaime A. and {Archibald}, Anne M. and {B{\'o}di}, Attila and {Bapat}, Shreyas and {Barentsen}, Geert and {Baz{\'a}n}, Juanjo and {Biswas}, Manish and {Boquien}, M{\'e}d{\'e}ric and {Burke}, D.~J. and {Cara}, Daria and {Cara}, Mihai and {Conroy}, Kyle E. and {Conseil}, Simon and {Craig}, Matthew W. and {Cross}, Robert M. and {Cruz}, Kelle L. and {D'Eugenio}, Francesco and {Dencheva}, Nadia and {Devillepoix}, Hadrien A.~R. and {Dietrich}, J{\"o}rg P. and {Eigenbrot}, Arthur Davis and {Erben}, Thomas and {Ferreira}, Leonardo and {Foreman-Mackey}, Daniel and {Fox}, Ryan and {Freij}, Nabil and {Garg}, Suyog and {Geda}, Robel and {Glattly}, Lauren and {Gondhalekar}, Yash and {Gordon}, Karl D. and {Grant}, David and {Greenfield}, Perry and {Groener}, Austen M. and {Guest}, Steve and {Gurovich}, Sebastian and {Handberg}, Rasmus and {Hart}, Akeem and {Hatfield-Dodds}, Zac and {Homeier}, Derek and {Hosseinzadeh}, Griffin and {Jenness}, Tim and {Jones}, Craig K. and {Joseph}, Prajwel and {Kalmbach}, J. Bryce and {Karamehmetoglu}, Emir and {Ka{\l}uszy{\'n}ski}, Miko{\l}aj and {Kelley}, Michael S.~P. and {Kern}, Nicholas and {Kerzendorf}, Wolfgang E. and {Koch}, Eric W. and {Kulumani}, Shankar and {Lee}, Antony and {Ly}, Chun and {Ma}, Zhiyuan and {MacBride}, Conor and {Maljaars}, Jakob M. and {Muna}, Demitri and {Murphy}, N.~A. and {Norman}, Henrik and {O'Steen}, Richard and {Oman}, Kyle A. and {Pacifici}, Camilla and {Pascual}, Sergio and {Pascual-Granado}, J. and {Patil}, Rohit R. and {Perren}, Gabriel I. and {Pickering}, Timothy E. and {Rastogi}, Tanuj and {Roulston}, Benjamin R. and {Ryan}, Daniel F. and {Rykoff}, Eli S. and {Sabater}, Jose and {Sakurikar}, Parikshit and {Salgado}, Jes{\'u}s and {Sanghi}, Aniket and {Saunders}, Nicholas and {Savchenko}, Volodymyr and {Schwardt}, Ludwig and {Seifert-Eckert}, Michael and {Shih}, Albert Y. and {Jain}, Anany Shrey and {Shukla}, Gyanendra and {Sick}, Jonathan and {Simpson}, Chris and {Singanamalla}, Sudheesh and {Singer}, Leo P. and {Singhal}, Jaladh and {Sinha}, Manodeep and {Sip{\H{o}}cz}, Brigitta M. and {Spitler}, Lee R. and {Stansby}, David and {Streicher}, Ole and {{\v{S}}umak}, Jani and {Swinbank}, John D. and {Taranu}, Dan S. and {Tewary}, Nikita and {Tremblay}, Grant R. and {de Val-Borro}, Miguel and {Van Kooten}, Samuel J. and {Vasovi{\'c}}, Zlatan and {Verma}, Shresth and {de Miranda Cardoso}, Jos{\'e} Vin{\'\i}cius and {Williams}, Peter K.~G. and {Wilson}, Tom J. and {Winkel}, Benjamin and {Wood-Vasey}, W.~M. and {Xue}, Rui and {Yoachim}, Peter and {Zhang}, Chen and {Zonca}, Andrea and {Astropy Project Contributors}},
        title = "{The Astropy Project: Sustaining and Growing a Community-oriented Open-source Project and the Latest Major Release (v5.0) of the Core Package}",
      journal = {\apj},
         year = 2022,
        month = aug,
       volume = {935},
       number = {2},
          eid = {167},
        pages = {167},
          doi = {10.3847/1538-4357/ac7c74},
archivePrefix = {arXiv},
       eprint = {2206.14220},
 primaryClass = {astro-ph.IM},
       adsurl = {https://ui.adsabs.harvard.edu/abs/2022ApJ...935..167A}
}

@ARTICLE{matplotlib,
       author = {{Hunter}, John D.},
        title = "{Matplotlib: A 2D Graphics Environment}",
      journal = {Computing in Science and Engineering},
         year = 2007,
        month = may,
       volume = {9},
       number = {3},
        pages = {90-95},
          doi = {10.1109/MCSE.2007.55},
       adsurl = {https://ui.adsabs.harvard.edu/abs/2007CSE.....9...90H}
}

@ARTICLE{corner,
       author = {{Foreman-Mackey}, Daniel},
        title = "{corner.py: Scatterplot matrices in Python}",
      journal = {The Journal of Open Source Software},
         year = 2016,
        month = jun,
       volume = {1},
        pages = {24},
          doi = {10.21105/joss.00024},
       adsurl = {https://ui.adsabs.harvard.edu/abs/2016JOSS....1...24F}
}

@ARTICLE{numpy,
       author = {{Harris}, Charles R. and {Millman}, K. Jarrod and {van der Walt}, St{\'e}fan J. and {Gommers}, Ralf and {Virtanen}, Pauli and {Cournapeau}, David and {Wieser}, Eric and {Taylor}, Julian and {Berg}, Sebastian and {Smith}, Nathaniel J. and {Kern}, Robert and {Picus}, Matti and {Hoyer}, Stephan and {van Kerkwijk}, Marten H. and {Brett}, Matthew and {Haldane}, Allan and {del R{\'\i}o}, Jaime Fern{\'a}ndez and {Wiebe}, Mark and {Peterson}, Pearu and {G{\'e}rard-Marchant}, Pierre and {Sheppard}, Kevin and {Reddy}, Tyler and {Weckesser}, Warren and {Abbasi}, Hameer and {Gohlke}, Christoph and {Oliphant}, Travis E.},
        title = "{Array programming with NumPy}",
      journal = {\nat},
         year = 2020,
        month = sep,
       volume = {585},
       number = {7825},
        pages = {357-362},
          doi = {10.1038/s41586-020-2649-2},
archivePrefix = {arXiv},
       eprint = {2006.10256},
 primaryClass = {cs.MS},
       adsurl = {https://ui.adsabs.harvard.edu/abs/2020Natur.585..357H}
}
\bibliographystyle{aasjournalv7}



\end{document}